\documentclass[a4paper,11pt]{article}
\pdfoutput=1 
\usepackage{jcappub}
\usepackage{amsmath}
\usepackage{mathrsfs} 
\usepackage{amsfonts}
\usepackage{amssymb}
\usepackage{amsthm}
\usepackage{mathtools}
\usepackage{graphicx}
\usepackage{latexsym}
\usepackage{xspace}
\usepackage{hyperref} 
\usepackage{bm}
\usepackage{relsize}
\usepackage{tabularx}
\usepackage{multirow}
\usepackage{amssymb}
\usepackage{lscape}
\usepackage[most]{tcolorbox}
\usepackage{braket}
\usepackage{comment}
\usepackage[utf8]{inputenc}
\usepackage[english]{babel}
\usepackage{slashed}
\usepackage{empheq}
\usepackage[makeroom]{cancel}
\newcolumntype{P}[1]{>{\centering\arraybackslash}p{#1}}

\usepackage{floatrow}
\usepackage{graphicx}
\usepackage[label font=bf,labelformat=simple]{subfig}
\usepackage{upgreek}

\usepackage[margin=1in,includefoot]{geometry}
\usepackage{xfrac} 
\usepackage{tikz-feynman}
\usepackage{tikz}	
\tikzfeynmanset{compat=1.0.0}

\allowdisplaybreaks

\makeatletter
\gdef\@fpheader{}
\g@addto@macro\bfseries{\boldmath}
\makeatother

\makeatletter
\DeclareRobustCommand*\uell{{\mathpalette\@uell\relax}}
\newcommand*\@uell[2]{
\setbox0=\hbox{#1\ell#1\ell}
\setbox1=\hbox{\rotatebox{10}{#1\ell#1\ell}}
\dimen0=\wd0 \advance\dimen0 by -\wd1 \divide\dimen0 by 2
\mathord{\lower 0.1ex \hbox{\kern\dimen0\unhbox1\kern\dimen0}}
}

\def\uppiq{\uppi_{\text{q}}}
\def\uppicl{\uppi_{\text{cl}}}

\def\pia{\pi_{a}}
\def\pir{\pi_{r}}
\def\uppia{\uppi_{a}}
\def\uppir{\uppi_{r}}

\def\bfk{\textbf{k}}
\def\bfq{\textbf{q}}

\newcommand{\Os}{\delta \mathcal{O}_{S}}

\newcommand{\ie}{\textsl{i.e.}~}

\newcommand{\eg}{\textsl{e.g.}\xspace}

\newcommand{\Tr}{\mathrm{Tr}}

\newcommand{\dd}{\mathrm{d}}
\newcommand{\ee}{e}

\newcommand{\bmk}{\boldmathsymbol{k}}
\newcommand{\bmx}{\boldmathsymbol{x}}

\newcommand{\Ima}{\Im \mathrm{m}\,}

\newcommand{\beq}{\begin{equation}}
	\newcommand{\eeq}{\end{equation}}
\newcommand{\bea}{\begin{equation}\begin{aligned}}
		\newcommand{\eea}{\end{aligned}\end{equation}}

\newlength{\wsingfig}
\newlength{\wdblefig}
\newlength{\wquadfig}
\newlength{\wtriplefig}
\newcommand{\Eq}[1]{Eq.~(\ref{#1})}
\newcommand{\Eqs}[1]{Eqs.~(\ref{#1})}
\newcommand{\Fig}[1]{Fig.~{\ref{#1}}}
\newcommand{\Figs}[1]{Figs.~{\ref{#1}}}

\newcommand{\Sec}[1]{Sec.~\ref{#1}}

\newcommand{\App}[1]{Appendix~\ref{#1}}

\usepackage{dsfont}

\def\bmx{{\boldsymbol{x}}}

\def\bmk{{\boldsymbol{k}}}

\subheader{}

\title{The Open Effective Field Theory\\ of Inflation}

\author[a]{Santiago Ag\"u\'i Salcedo,}
\author[a]{Thomas Colas,}
\author[a]{and Enrico Pajer}

\affiliation[a]{Department of Applied Mathematics and Theoretical Physics, University of Cambridge, Wilberforce Road, Cambridge, CB3 0WA, UK}
\emailAdd{sa2013@cam.ac.uk}
\emailAdd{tc683@cam.ac.uk}
\emailAdd{enrico.pajer@gmail.com}

\begin{document}
\sloppy

\abstract{In our quest to understand the generation of cosmological perturbations, we face two serious obstacles: we do not have direct information about the environment experienced by primordial perturbations during inflation, and our observables are practically limited to correlators of massless fields, heavier fields and derivatives decaying exponentially in the number of e-foldings. The flexible and general framework of open systems has been developed precisely to face similar challenges. Building on previous work, we develop a Schwinger-Keldysh path integral description for an open effective field theory of inflation, describing the possibly dissipative and non-unitary evolution of the Goldstone boson of time translations interacting with an unspecified environment, under the key assumption of locality in space and time. Working in the decoupling limit, we study the linear and interacting theory in de Sitter and derive predictions for the power spectrum and bispectrum that depend on a finite number of effective couplings organised in a derivative expansion. The smoking gun of interactions with the environment is an enhanced but finite bispectrum close to the folded kinematical limit. We demonstrate the generality of our approach by matching our open effective theory to an explicit model. Our construction provides a standard model to simultaneously study phenomenological predictions as well as quantum information aspects of the inflationary dynamics.} 

\maketitle


\section{Introduction}

A cornerstone of our cosmological model is that cosmological observations on very large scales are related approximately linearly to correlators of perturbations at the beginning of the hot big bang. The initial conditions for cosmological perturbations must hence be determined by some yet unknown dynamics that took place before our universe reheated into a thermal bath of standard model particles. As the leading proposal for such unknown dynamics, inflation combines a prolonged phase of accelerated expansion with the quantum generation of vacuum fluctuations. This framework for the primordial universe forces us to grapple with a series of peculiar features. First, we are in practice not able to directly probe time evolution, but we only see its final outcome at the end of inflation. Even when focusing on late time observables, we find further severe restrictions to the quantities that are practically measurable. Indeed, we can only directly access observables related to fields that are massless or very light compared to the typical scale of the system, namely the Hubble scale of inflation. Light fields are not generic, and their existence can typically be traced back to some symmetry of the theory. The quintessential example is the graviton \cite{Starobinsky:1979ty}, whose masslessness is tied to general covariance. Because primordial perturbations have been measured to be scalar in nature, at least to a few percent precision \cite{BICEPKeck:2022mhb}, we know there must be more an additional massless scalar active during inflation. This can be interpreted as curvature perturbations or as the Goldstone boson of spontaneously broken time translations \cite{Cheung:2007st}, whose approximate masslessness is tied to an internal shift symmetry crucially combined with an approximately linear background time evolution \cite{Finelli:2018upr}. Heavier fields must necessarily decay with time, and their existence can at best be established indirectly, e.g. through their effect on light fields. The same applies to derivatives of light fields, all of which decay exponentially in the number of e-foldings of inflation. In turn, this tells us that even for light fields, we are practically unable to probe correlators involving conjugate momenta. \\

The fact that we can only probe a restricted part of the whole system, namely the late-time limit of correlators of light fields, has important implications for the theoretical framework within which we should model inflation. We can choose to model the whole system, including all fields that might decay at late time, and use the framework appropriate for a closed system. Alternatively, it might be advantageous to focus on modelling the subsystem that we do observe, namely the massless curvature perturbations, and instead work within the framework of an \textit{open system}. The dichotomy between open and closed systems manifests itself in various ways. First, conserved quantities can be precisely tracked and leveraged in a closed system, but their constraints on the observable dynamics are much more relaxed for open systems. Second, in an open system it is essential to keep track of the correlation with the rest of the full system, henceforth referred to as the \textit{environment}. While these observations apply also to classical evolution, e.g. in a statistical or stochastic theory, they are especially important when dealing with a quantum system. The way a quantum system predicts probabilities is very different in a closed system, where all possible histories are accounted for, from an open system, where information can be lost into the environment. \\

This work stems from recognising that a satisfactory and comprehensive understanding of quantum dynamics during inflation, which necessarily relies on effective models, must begin in the general and flexible framework of \textit{open quantum systems}. This seems to be a necessary starting point both on the theoretical and the observational side. On the observational side, sizeable environmental effects on curvature perturbations lead to new signals to be searched for in the data, as we will discuss in the rest of this paper. On the theoretical side, we need to understand how the flat-space rules of Effective Field Theories (EFTs) and the Wilsonian paradigm translate to an accelerated spacetime where perturbations are created out of the vacuum, energy is not conserved, length and time scales are dynamical notions, and information is redistributed globally. \\

To be more specific, let us mention some concrete setups. First, let us consider the most vanilla model of inflation we can think of, with a single scalar field and a gentle slow-roll dynamics. Even in this case, due to the Hubble expansion, the curvature perturbations we measure do not experience the same vacuum as they would in flat space. Instead, they interact with an approximately thermal bath of field excitations at the de Sitter temperature $H/2\pi$. This bath contains all degrees of freedom in the theory and hence also heavy ones. To have control over the regime of validity of a proposed EFT and to correctly estimate corrections to our predictions, one would like to estimate the interactions with the environment and judge if and when they are negligible. In the simplest case, one expects that excitations of heavy fields are suppressed by a Boltzmann factor (see e.g. \cite{Arkani-Hamed:2015bza,Tong:2021wai}). However, a variety of possibilities have been considered and devised to counteract this suppression, e.g. through a large breaking of de Sitter boost invariance \cite{Lee:2016vti}, a chemical potential \cite{Chen:2018xck,Wang:2019gbi,Bodas:2020yho,Tong:2022cdz,Sou:2021juh}, or a tachyonic mass \cite{McCulloch:2024hiz}. More generally, large classes of models feature some additional mechanism for producing particles in excess of the thermal de Sitter radiation. An idea that goes back a long way is that of warm inflation \cite{Berera:1995wh,Berera:1995ie,Berera:2023liv}, where the presence of a radiation bath is ensured by a continuous source that counteracts the Hubble expansion. In many setups the environment is populated non-thermally. A much studied example is the tachyonic production of gauge fields of one chirality induced by a coupling $\phi F\tilde F$ to a slowly rolling field $\phi$ \cite{Anber:2009ua}. Here, the universe is filled with gauge field excitations of wavelength of order Hubble, which are diluted and redshifted away by the expansion but are also continuously produced. Tens of additional possibilities have been studied over the years, but we refrain from providing a comprehensive list. The upshot is that, for both minimal and non-minimal setups, one would like to have a description of the dynamics of curvature fluctuations that allows for possible interactions with the omnipresent environment. Given such a description one can then more accurately specify the regimes in which such interactions matter or can be safely neglected. \\

As the number of possible environments during inflation is clearly infinite and their properties are completely model dependent, one might wonder how to make concrete progress in devising a unified description and in identifying some minimal set of characteristic predictions. To this end, it is natural to employ the machinery of EFTs for open quantum systems. First, we recognise that EFTs have no magical powers: for a generic setup, there need not be a clear distinction between system of interest and environment, either because the two are strongly coupled to each other or because they cannot be cleanly separated in terms of the charges available in the problem. Rather, it is in the presence of a clear separation, for example dictated by symmetries, and a hierarchy of scales in the problem that EFTs become useful as an organising principle. Because of this, we would like to consider here cases where an unknown environment during inflation can be separated from the sector of curvature perturbations around Hubble crossing because the former is characterised by faster evolution, i.e. high frequency, and short distance perturbations, i.e. high momenta. In this regime, one expects that only a handful of properties of the environment need to be specified to accurately describe the evolution of curvature perturbations, in contrast to the much more fine-grained description that would be needed in the study of the whole system including environment. The handful of properties about the environment are specified via Wilsonian effective couplings in an \textit{open EFT} description of the system.  \\

An essential feature of an EFT is that only a finite number of interactions need to be considered for any given finite desired precision. This finiteness is very often\footnote{Other possibilities exists. Some non-local EFTs may have different organisational principles, such as the excitations of Fermi surfaces or in the partial re-summation of certain non-localities.} ensured by the assumption that the EFT dynamics is local in space and time. Here we make this assumption and our discussion and results are contingent on this crucial property. Models where the environment can mediate interactions around the Hubble scale are beyond the scope of our description. Notice that this of course is \textit{not} a peculiarity of open systems: also in standard EFTs, integrating out degrees of freedom that are active at the scales of interest leads to intractable non-local interactions and the EFT is powerless without the specific input of a high-energy completion. These situations simply lack a useful hierarchy of scales and there is no natural organisation to be provided by the EFT.\\

For inflation described as a closed system, a powerful organising principle has been identified in the Effective Field Theory of Inflation (EFToI) \cite{Cheung:2007st}. Two pioneering investigations have studied the generalisation of this approach to the case of an open system and our work builds upon them. The first one by \cite{LopezNacir:2011kk} studied the situation in which the Goldstone boson of time translations $\pi$ is coupled to noise fluctuations in the environment and additionally may experience dissipative evolution. The workhorse of their approach is a stochastic generalisation of the classical equations of motions known as the \textit{Langevin equation}. This allows one to model a very generic open dynamics and to identify characteristic signatures. An aspect of this approach is that functional methods related to Lagrangians and Hamiltonians are set aside in favour of working directly with the equations of motion. While this can provide a systematic description, it comes with a few shortcomings. First, symmetries and their associated charges and currents are naturally discussed in the context of actions and Lagrangians. Second, this approach might naively appear fundamentally different from the usual techniques used to study inflation in the context of closed systems, where one employs the operator or path integral description of the Schwinger-Keldysh or in-in formalism. The irony is that the Schwinger-Keldysh was developed to study open systems to begin with, and there is no need to abandon it if we wish account for dissipation, fluctuations and associated phenomena. \\ 

Indeed, recent work \cite{Crossley:2015evo,Glorioso:2017fpd,Liu:2018kfw} has generated renewed interest in the Schwinger-Keldysh formalism within the high-energy community, much of which in the context of hydrodynamics. This work has clarified some of the general rules needed to ensure that a putative open effective theory indeed arises by integrating out the environment in a consistent, i.e. unitary, local and causal theory for the full closed system. Similarly to hydrodynamics, for inflation too we are interested in identifying the system through the techniques of spontaneous symmetry breaking. Very generally, one can think of the hydrodynamical modes as describing the dynamics of small local departures from global thermodynamical equilibrium, with possibly some conserved charge. Similarly, the system of interest in cosmology is the Goldstone boson of time translations that are spontaneously broken by the time evolution of some inflaton sector, which selects a preferred foliation of spacetime. Investigation of the open EFT for the Goldstone boson in this context was initiated in \cite{Hongo:2018ant} (see also the related work \cite{Hongo:2019qhi}). There, quadratic theory for $\pi$ in flat spacetime was derived and studied. Our work continues their investigation by moving to cosmological spacetimes and including interactions and the production of non-Gaussianities. For the benefit of the busy reader we provide a summary of our main results in \Sec{summary}.\\

In it interesting to compare and contrast our approach here to previous works in the literature. We start with the recently proposed \textit{in-out formalism} for cosmological correlators proposed in \cite{Donath:2024utn}. In that work, it was noticed that for unitary Hamiltonian evolution of a closed system, cosmological correlators in de Sitter can be computed in the same in-out formalism used for scattering amplitude in flat space, namely using Feymann rules that only invoke a single time order Feynman propagator. This is in contrast to the more cumbersome in-in formalism where four different bulb-bulk propagators and two different bulk-boundary propagators appear in different combinations depending on the $2^V$ ways to label the $V$ vertices of a diagram as coming from the left time evolution or right anti-time evolution. For IR-finite interactions the in-out formalism provides a sizable simplification and facilitate the derivation of a series of results such as the correlators equivalent of the wavefunction recursion relations of \cite{Arkani-Hamed:2017fdk}, cutting rules for correlators completely analogous to those of Cutkosky and Veltman and a natural definition of a de Sitter S-matrix. The equality between in-in and in-out fails in the presence of an open system and therefore applies only to the complement of the theories we consider in this work. \\

Open quantum system techniques have been used for inflation also for a completely different reason, namely for their ability to re-sum certain secular divergences and to repair the validity of perturbation theory in the presence of time logarithms \cite{Burgess:2015ajz, Burgess:2024eng}. This has been used extensively to study the regime of stochastic inflation \cite{Starobinsky:1986fx,Starobinsky:1994bd} and associated IR issues arising in de Sitter \cite{Gorbenko:2019rza, Cespedes:2023aal}. We should stress that our motivation to study open system is completely independent and therefore the models we study are all IR finite and do not require any IR resumation. In this sense, it is remarkable that a curved spacetime has different inequivalent ways to favour an open system description. We will reference other studies employing open system techniques in cosmology in due course.  


\paragraph{Nomenclature} There are a lot of variations in the terminology used by different communities to describe open and non-equilibrium quantum dynamics.\footnote{We thank C. Burgess for discussions on this issue.} While the high-energy and particle physicists often define unitarity through conservation of the state normalisation, that is
\begin{align}
\frac{\dd }{\dd t} \mathrm{Tr}[\rho]  = 0,
\end{align}
where $\rho$ is the density matrix of the system, quantum optics and condensed matter communities might prefer to consider unitary evolution as\footnote{The above two definitions of (non-)unitarity can be reconciled by absorbing the eventual change in the state's normalisation unveiling non-unitary dynamics into a non-Hamiltonian evolution, leading to the complete-positive and trace preserving dynamical maps used in quantum optics \cite{breuerTheoryOpenQuantum2002}. Ultimately, the common property is the loss of probability in the observed sector and the associated entropy production of open dynamics.} 
\begin{align}\label{eq:OQS}
\frac{\dd \rho }{\dd t}  = - i \left[H, \rho\right] \qquad \mathrm{with} \qquad H^\dag = H. 
\end{align}
In this article, to avoid confusion, we define unitary time evolutions through the existence of a unitary evolution operator $\mathcal{U}(t,t_0)$ such that
\begin{align}\label{eq:unitevol}
\rho(t_0) \rightarrow \rho(t) = \mathcal{U}(t,t_0)  \rho(t_0) \mathcal{U}^\dag(t,t_0),
\end{align}
with 
\begin{align}
\mathcal{U}^\dag(t,t_0) \mathcal{U}(t,t_0)  = \mathcal{U}(t,t_0)    \mathcal{U}^\dag(t,t_0) =  \mathrm{Id},
\end{align}
and initial condition $\mathcal{U}(t_0,t_0) = \mathrm{Id}$. We further consider density matrices which are (i) normalised $\mathrm{Tr} \rho =1$, (ii) Hermitian $\rho^\dag = \rho$ and (iii) positive definite $\rho > 0$.

Not all physical evolution can be written under the form of \Eq{eq:unitevol}. Whenever some degrees of freedom are experimentally inaccessible, initially available information can get distributed in the unknown environment and eventually lost. Such a possibility is studied in the context of open quantum systems (see \cite{breuerTheoryOpenQuantum2002} for a reference textbook) and maps pure to mixed states. While information quantifiers such as purity $\gamma \equiv \mathrm{Tr} [\rho^2]$ or entanglement entropy $S_{\mathrm{ent}} = - \mathrm{Tr} [\rho \log \rho]$  are conserved under \Eq{eq:unitevol}, they can vary over time when considering open dynamics. These variations precisely encode the lack of probability conservation in open systems due to the leakage of information between observed degrees of freedom and their surroundings, which is the object of the current article. In this case, the dynamical evolution of the open system is given in terms of an \textit{open effective functional} which denotes the full functional of the fields $\uppi_\pm$ that weighs the in-in path integral,
\begin{align}\label{eq:Seffintro}
S_{\mathrm{eff}}\left[ \uppi_+, \uppi_-\right] = S_\uppi\left[\uppi_+\right] - S_\uppi\left[\uppi_-\right] + S_{\mathrm{IF}}\left[\uppi_+,\uppi_- \right],
\end{align}
where $S_\uppi$ is the unitary action and non-unitary dissipative evolution is attributed to the influence functional $S_{\mathrm{IF}}$ \cite{FEYNMAN1963118}.


\paragraph{Notation and conventions} We employ the Keldysh rotation of the doubled fields defined by
\begin{align}
\uppi_r &= \frac{\uppi_+ + \uppi_-}{2} \quad \mathrm{and} \quad \uppi_a = \uppi_+ - \uppi_-\, \qquad \Leftrightarrow \qquad  \uppi_\pm =\uppir\pm \frac12 \uppia\,.
\end{align}  
The fonts $(\uppi, \upalpha, \upbeta, \upgamma, \updelta)$ and $(\pi, \alpha, \beta, \gamma, \delta)$ denote fields and EFT parameters before and after canonical normalisation, respectively, as discussed in \Sec{subsec:energy}. The mapping from one convention to the other is made through
\begin{align}
\pi &\equiv f_\pi^2 \uppi\,, \quad f_\pi^4 \equiv \upalpha_0 -2 \upalpha_1
\end{align}
and
\begin{align}
c_s^{2} &\equiv \frac{\upalpha_{0}}{f_\pi^4}\;,\quad\; \gamma \equiv
 \frac{2\upgamma_{1}}{f_\pi^4}\;,\quad\;\beta_{i} \equiv \frac{\upbeta_{i}}{f_{\pi}^{4}}\;\quad\mathrm{for}\quad i = 1 ~\mathrm{to}~8, \\
\alpha_2 &\equiv \frac{\upalpha_2}{f_\pi^4}\;,\quad\; \gamma_2 \equiv \frac{\upgamma_{4}}{f_\pi^4} \;,\quad\; \delta_i \equiv \frac{\updelta_{i}}{f_\pi^4} \quad ~~\mathrm{for}\quad i = 1 ~\mathrm{to}~6.
\end{align}
The mass dimensions of these parameters are
\begin{align}
[\pi] = E, &\quad [f_\pi] = E, \quad [c_s] = E^0, \quad [\gamma] = E, \quad [\beta_1] = E^2, \quad [\beta_2] = [\beta_4] = E^0, \\
[\alpha_2] = E^{0}, &\quad [\gamma_2] = E, \quad [\beta_6] =  [\beta_8] = E^0, \quad [\beta_3] = [\beta_7] = E, \quad [\beta_5] = E^2, \\
& \qquad  [\delta_1] = E^3, \quad [\delta_5] = [\delta_2] = E, \quad [\delta_4] = [\delta_6] = E^0  . 
\end{align}


\subsection{Summary of the main results} \label{summary}

Starting from \Eq{eq:Seffintro}, we construct the most generic open effective functional $S_{\mathrm{eff}}\left[ \uppi_r, \uppi_r\right]$ compatible with (1) \textit{unitarity of the UV theory}; (2) \textit{the spontaneous symmetry breaking of time-translations}; and (3) \textit{locality} in time and space of the open effective theory. Condition (1) provides a set of non-perturbative relations known as \textit{non-equilibrium constraints} \cite{Crossley:2015evo, Glorioso:2016gsa, Liu:2018kfw} 
\begin{align}
S_{\mathrm{eff}} \left[\uppi_r,\uppi_a = 0\right] &= 0\,, \label{eq:normintro} \\
S_{\mathrm{eff}} \left[\uppi_r,\uppi_a\right] &= - 	S^*_{\mathrm{eff}} \left[\uppi_r,-\uppi_a\right]\,, \label{eq:hermintro} \\
\Ima S_{\mathrm{eff}} \left[\uppi_r,\uppi_a\right] &\geq 0. \label{eq:posintro}
\end{align} 
The symmetry requirement further restricts the available dynamics. Breaking the time-translation symmetry leads to two Stueckelberg fields $\uppi_+$ and $\uppi_-$ in the unitary case. Non-unitary effects further break the symmetry group explicitly to its diagonal subgroup, such that only $\uppi_{r}$ transforms non-linearly under time-translations and boosts \cite{Hongo:2018ant, Akyuz:2023lsm}. More in detail, for $\epsilon_{r} \in \mathbb{R}$ and $\Lambda^\mu_{r~\nu} \in \mathrm{SO}(1,3)$
\begin{align}\label{eq:LorentzRsymintro}
\uppi_{r}(t,\bmx) \rightarrow \uppi'_{r}(t,\bmx) &= \uppi_{r}\left(\Lambda^0_{r~\mu}x^\mu+\epsilon_{r},\Lambda^i_{r~\mu}x^\mu \right)+\epsilon_{r}+\Lambda^0_{r~\mu}x^\mu - t\,, \\
\uppi_{a}(t,\bmx) \rightarrow \uppi'_{a}(t,\bmx) &= \uppi_{a}\left(\Lambda^0_{r~\mu}x^\mu+\epsilon_{r},\Lambda^i_{r~\mu}x^\mu \right).
\end{align}
Finally, locality in time and space ensures the existence of an IR-stable power counting scheme that can be truncated to the desired level of accuracy.

\paragraph{The Open EFT of Inflation} We work in de Sitter space in the \textit{decoupling limit} with at most one derivative per field. Using the canonically normalised fields $\pir$ and $\pia$, we obtain the most generic open effective functional. At leading order in the slow-roll expansion, the quadratic order reads
\begin{align}\label{eq:canonormintro} 
&\quad S_{\mathrm{eff}}^{(2)} = \int \dd^4 x \Big\{ a^2 \pi'_{r} \pi'_{a} - c_{s}^{2} a^2 \partial_i \pir  \partial^i \pia  \\
-&  a^3 \gamma \pi'_{r} \pia + i \left[\beta_{1} a^4 \pia^2 - \left(\beta_2 - \beta_4\right) a^2\pia^{\prime 2} + \beta_2 a^2 \left(\partial_i \pia \right)^2\right] \Big\}, \nonumber
\end{align}
and the cubic order
\begin{align}\label{eq:canonormcubintro}
S_{\mathrm{eff}}^{(3)} =   \frac{1}{f_\pi^2} \int & \dd^4 x \Big\{\Big[4 \alpha_2 -  \frac{3}{2} (c^2_s-1) \Big]  a \pir^{\prime2} \pi'_{a} +\frac{1}{2} (c^2_s-1) a \left[\left(\partial_i \pir \right)^2 \pi'_a + 2 \pi'_{r}	\partial_i \pir  \partial^i \pia \right] \\
& \qquad \quad + \left(4 \gamma_2 - \frac{\gamma}{2}\right)  a^2\pir^{\prime2} \pia	+ \frac{\gamma}{2} a^2	\left(\partial_i \pir \right)^2 \pia \nonumber \Big. \\
&+ i \Big[\left(2\beta_7-\beta_3\right) a^2 \pi'_{r} \pi'_{a} \pia + \beta_3 a^2 \partial_i \pir  \partial^i \pia \pia + 2(\beta_4+ \beta_6 - \beta_8) a \pi'_{r} \pia^{\prime2} \Big. \nonumber \\
& \qquad \quad - 2 \beta_4  a \partial_i \pir  \partial^i \pia \pi'_{a} - 2\beta_5 a^3 \pir^{\prime} \pia^2  - 2 \beta_6 a \pir^{\prime} (\partial_i\pia)^2 \Big]  \Big. \nonumber\\
&+\delta_1 a^4 \pia^3 + (\delta_5- \delta_2) a^2 \pia^{\prime2} \pia  + \delta_2 a^2 (\partial_i \pia)^2 \pia - \delta_4 a  (\partial_i \pia)^2 \pi'_a + (\delta_4-\delta_6) a \pia^{\prime3} \Big\}.  \nonumber
\end{align}
Primes denote time derivatives with respect to the conformal time $\eta = - 1/(aH)$ where $a$ is the scale factor and $H$ the Hubble parameter. While the standard effective field theory of inflation \cite{Cheung:2007st} is recovered in the unitary limit, the above open effective functional also captures non-unitary effects such as dissipation and diffusion of the pseudo-Goldstone boson in an unknown surrounding environment. For instance, the first line of \Eq{eq:canonormintro} corresponds to the usual unitary dynamics with the kinetic term and an effective speed of sound $c_s$, whereas the second line of \Eq{eq:canonormintro} captures dissipation (controlled by $\gamma$) and noise fluctuations (controlled by $\beta_{1}$, $\beta_2$ and $\beta_4$). Unitary time evolution of the EFT can be recovered as discussed in \Sec{subsubsec:unit}, where we develop a classification of the EFT operators. Just as in the usual EFToI \cite{Cheung:2007st}, non-linearly realised boosts relate non-unitary operators at different orders such as the dissipation parameter $\gamma [- a \pir^{\prime}  -\pir^{\prime2}/2 +\left(\partial_i \pir \right)^2/2] \pia $ \cite{LopezNacir:2011kk}. We discuss in \Sec{subsec:energy} the new energy scales that characterise non-unitary evolution and the associated heuristic estimate of primordial non-Gaussianities. 

\paragraph{The power spectrum} The open effective field theory of inflation provides theoretical predictions for standard cosmological observables such as the power spectrum (\Sec{sec:Pk}) and the bispectrum (\Sec{sec:bispec}). At leading order, curvature perturbations are related to the pseudo-Goldstone boson by $\zeta = - H\pi/ f_\pi^2$. Symmetry requirements ensure the existence of a nearly scale invariant power spectrum
\begin{align}
\Delta^2_\zeta(k) \equiv \frac{k^3}{2\pi^2}P_\zeta(k) \qquad \mathrm{with} \qquad \langle \zeta_\bmk \zeta_{-\bmk}\rangle = (2\pi)^3 \delta(\bmk + \bmk') P_\zeta(k).
\end{align}
Considering the first noise term of \Eq{eq:canonormintro}, which is controlled by $\beta_1$, we obtain the dissipative power spectrum and give an exact expression in \Eq{eq:PK1SH}. In the large and small dissipation regimes that result reduces to
\begin{align}
\Delta^2_\zeta(k)  &\propto  \begin{dcases}
\frac{\beta_1}{H^2} \frac{H^4}{f_\pi^4}   \sqrt{\frac{H}{\gamma}} + \mathcal{O}[(\gamma/H)^{-3/2}] , & \gamma \gg H,\\
\frac{\beta_1}{H^2} \frac{H^4}{f_\pi^4}  + \mathcal{O}(\gamma/H), & \gamma \ll H.
\end{dcases}
\end{align}
The observational constraint $\Delta^2_\zeta = 10^{-9}$ can be easily obeyed by imposing hierarchies between the various scales of the problem. While we never assume this in our discussion, one could further impose thermal equilibrium of the surrounding environment. Then, the power spectrum in the large dissipation regime scales as
\begin{align}
\Delta^2_\zeta \propto \frac{T}{H}\frac{H^4}{f_\pi^4}\, \sqrt{\frac{\gamma}{H}},
\end{align}
which reproduces the warm inflation results \cite{Berera:1995ie, Berera:1995wh,  Berera:2008ar, Ballesteros:2023dno, Montefalcone:2023pvh}. Analytical results for the two other noise directions $\pia^{\prime 2}$ and $\left(\partial_i \pia \right)^2$  can be found in \Sec{sec:Pk}. 

\begin{figure}[tbp]
 \begin{minipage}{6in}
    \centering
    \raisebox{-0.5\height}{\includegraphics[width=.48\textwidth]{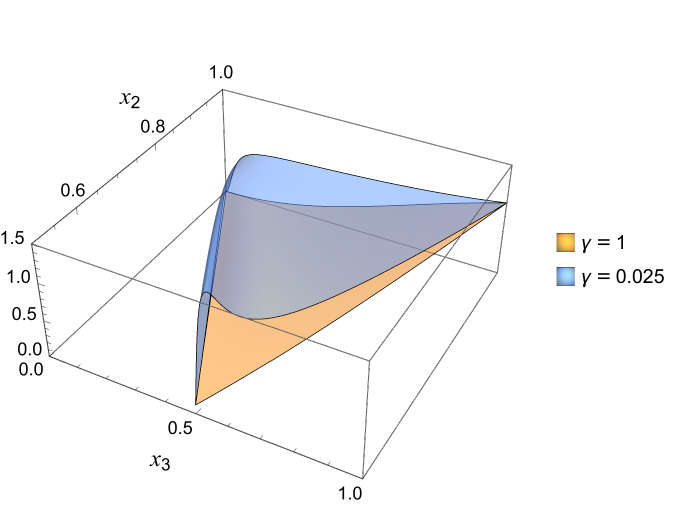}}
    \raisebox{-0.5\height}{\includegraphics[width=.48\textwidth]{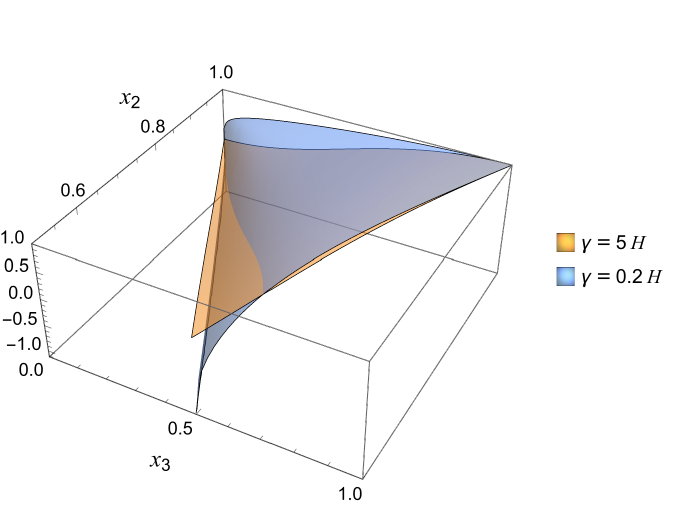}}
\end{minipage}
\caption{\label{fig:foldedintro} Shape function $S(x_2,x_3) \equiv (x_2 x_3)^2 [B(k_1, x_2 k_1, x_3 k_1)/B(k_1,k_1,k_1)]$ for the contact bispectrum generated by the operator $\pi_a^3$ in Minkowski (\textit{left}) and in de Sitter (\textit{right}). \textit{Left}: the bispectrum is given by \Eq{eq:genintro} with singularities controlled by $\mathrm{Sing}_\gamma$ in \Eq{eq:singdissipintro} (details in \Sec{subsec:flatB}). The singularity is resolved such that the bispectrum remains finite for any physical configuration. We observe the equilateral (\textit{orange}) to folded (\textit{blue}) transition of the shape function as the dissipation parameter $\gamma$ decreases. There is an enhancement of the signal close to the isofolded configuration in the low dissipation regime. \textit{Right}: the qualitive features of the signal remain the same in de Sitter. At large dissipation (\textit{orange}), the signal peaks in the equilateral configuration. At small dissipation (\textit{blue}), it is enhanced near the folded region. Two aspects are confirmed by analytical arguments but are not clearly visible in the plot: (i) dissipation regulates the folded divergence and (ii) consistency relations still hold in the squeezed limit $x_3 \ll x_2 = 1$.}
\end{figure}

\paragraph{The bispectrum} To discuss interactions, one can use the same treatment as in the standard in-in formalism \cite{Chen:2017ryl}, and we derive Feynman rules in \Sec{sec:bispec}. The rest of this section is devoted to the computation of the contact bispectrum 
\begin{align}
\langle \zeta_{\bmk_1}  \zeta_{\bmk_2}  \zeta_{\bmk_3} \rangle  = - \frac{H^3}{f_\pi^6}	\langle \pi_{\bmk_1}  \pi_{\bmk_2}  \pi_{\bmk_3} \rangle \equiv (2\pi)^3 \delta(\bmk_1 + \bmk_2 + \bmk_3) B(k_1,k_2,k_3).
\end{align} 
The bispectrum is generated by the cubic operators in \Eq{eq:canonormcubintro}, both in flat space and in de Sitter. The flat-space results are instructive as computations can easily be carried out analytically. The generic structure of the contact bispectrum is given by 
\begin{align}\label{eq:genintro}
B(k_1,k_2,k_3) =f(\mathrm{EFT}) \frac{  \mathrm{Poly}_n\left(e_1^\gamma, e_2^\gamma,e_3^\gamma \right)}{ \mathrm{Sing}_\gamma} \,,
\end{align}
where $f(\mathrm{EFT})$ is a rational function of the EFT coefficients (and possibly the kinematics for spatial derivative interactions), and $\mathrm{Poly}_n$ are elementary symmetric polynomials of the energy variables
\begin{align}\label{eq:varintro}
e_1^\gamma = E_1^{\gamma} + E_2^{\gamma} + E_3^{\gamma}, \quad e_2^\gamma = E_1^{\gamma} E_2^{\gamma} + E_2^{\gamma}  E_3^{\gamma} + E_1^{\gamma} E_3^{\gamma} \quad  e_3^\gamma = E_1^{\gamma} E_2^{\gamma} E_3^{\gamma}\,,
\end{align}
where we used the dispersion relation appropriate for this dissipative system, namely 
\begin{align}
E_k^{\gamma} \equiv \sqrt{c_s^2 k^2 - \gamma^2/4}\,.   
\end{align}
Moreover, $\mathrm{Sing}_\gamma$ is a place holder for the singularity structure
\begin{align}\label{eq:singdissipintro}
\mathrm{Sing}_\gamma =& \left| E_1^{\gamma} + E_2^{\gamma} + E_3^{\gamma} + \frac{3}{2}i \gamma\right|^2 \left| -E_1^{\gamma} + E_2^{\gamma} + E_3^{\gamma} + \frac{3}{2}i \gamma\right|^2 \nonumber\\
&\times \left| E_1^{\gamma} - E_2^{\gamma} + E_3^{\gamma} + \frac{3}{2}i \gamma\right|^2 \left| E_1^{\gamma} + E_2^{\gamma} - E_3^{\gamma} + \frac{3}{2}i \gamma\right|^2.
\end{align}
This singularity structure captures most of the specificities of the non-unitary dynamics. Physically, it represents $3 \leftrightarrow 0$ (all pluses) and $2 \leftrightarrow 1$ (mixed signs) interactions mediated by the real particles present in the environment. Fluctuations alone would generate \textit{folded singularities} because the state of the system differs from the Bunch Davies vacuum in that real particles are present also on sub-Hubble scales. On the other hand, dissipation induces a finite memory of the past. This regularises the folded divergences \cite{LopezNacir:2011kk, Green:2020whw}, by effectively moving the folded pole into complex kinematics. The singularity is not located in the physical plane and the bispectrum remains finite over the whole dynamical range. This is illustrated in the \textit{left} panel of \Fig{fig:foldedintro} where we observe an enhanced but finite signal near to folded region.  
The singularity structure $\mathrm{Sing}_\gamma$ exhibits two different behaviours depending on the magnitude of the dissipation coefficient $\gamma$: 
\begin{itemize}
\item In the strong dissipation regime, the $ 3 i \gamma/2$ term of \Eq{eq:singdissipintro} always dominates and the signal peaks in the \textit{equilateral} shape where $k_1 \simeq k_3 \simeq k_3$ (\textit{orange region} in \Fig{fig:foldedintro}). 
\item In the small dissipation regime, $\mathrm{Sing}_\gamma$ can become small in the folded region where $k_2 + k_3 \simeq k_1$ and the signal predominantly peaks in the \textit{isofolded}\footnote{Here we differentiate between folded singularities, which are a one-parameter family of configurations modulo rescaling, $k_1+k_2 = k_3$, and permutations thereof, from the isofolded configuration, which represents a single kinematic modulo rescaling, $k_2 \simeq k_3 \simeq k_1/2$ and permutations thereof.} configuration where $k_2 \simeq k_3 \simeq k_1/2$ (\textit{blue region} in \Fig{fig:foldedintro}).
\end{itemize} 
Beyond the flat space case, analytical results are hard to reach in full generality and we mostly rely on numerical results, as in \Sec{subsec:bispecdS}. Just as in flat space, different behaviours emerge in the large ($\gamma \gg H$) and small ($\gamma \ll H$) dissipation regime, see the \textit{right} panel of \Fig{fig:foldedintro}. For large dissipation the signal peaks in the equilateral configuration, as already noted in \cite{LopezNacir:2011kk}; for small dissipation the signal reaches an extremum near the folded region. At first sigh, this smoking gun of open dynamics might seem to be degenerate with other classes of models, which also lead to signal in the folded triangles, such as non-Bunch Davies initial states \cite{Holman:2007na, Chen:2006nt, Meerburg:2009ys, Agullo:2010ws, Ashoorioon:2010xg, Agarwal:2012mq, Ashoorioon:2013eia,  Albrecht:2014aga, Green:2020whw}. A crucial difference is that dissipation regulates the divergence by smoothing the peak and displacing it from the edge of the triangular configurations, leading to finite values of the bispectrum for any physical configuration. In particular, it implies no divergence in the squeezed limit of the bispectrum $k_1 \simeq k_2 \gg k_3$. Small values of $\gamma/H$ may eventually lead to an intermediate peak due to the regularised folded singularity, but consistency relations hold \cite{Maldacena:2002vr,Creminelli:2004yq, Cheung:2007sv, Creminelli:2012ed,Hinterbichler:2012nm,Assassi:2012zq,Pajer:2017hmb,Avis:2019eav} and the squeezed limit goes to zero due to the symmetries of the theory.

\paragraph{UV-models and matching} Our open effective field theory of inflation captures all single-clock models that display a local and possibly dissipative dynamics. One such explicit UV-model was recently studied in \cite{Creminelli:2023aly}. Here, in \Sec{sec:matching} we show that our formalism provides an accurate low-energy effective description of its dynamics. The model in question contains, in addition to the inflaton field $\phi$, a massive scalar field $\chi$ with a softly-broken $U(1)$ symmetry
\begin{align}\label{eq:actionmatchintro}
S= \int \dd^4 x &\sqrt{- g} \bigg[ \frac{1}{2} M_{\mathrm{Pl}}^2  R - \frac{1}{2} \left(\partial \phi \right)^2 - V(\phi) - \left|\partial \chi \right|^2 + M^2 \left| \chi \right|^2 \nonumber \\
- &\frac{\partial_\mu \phi}{f}\left( \chi \partial^\mu \chi^* - \chi^* \partial^\mu \chi\right) - \frac{1}{2}m^2\left(\chi^2 + \chi^{*2}\right)\bigg].
\end{align}
This model exhibits a narrow instability band in the sub-Hubble regime, during which particle production occurs. We demonstrate the equivalence of the non-linear Langevin equation 
\begin{align}\label{eq:Paolointro}
\pi'' + \left( 2H + \gamma\right)a\pi' - \partial_i^2 \pi &\simeq \frac{\gamma}{2\rho f} \left[ \left( \partial_i \pi\right)^2 - 2 \pi \xi \pi^{\prime2}\right] - \frac{a^2m^2}{f} \left(1+2\pi\xi \frac{\pi'}{a \rho f} \right) \delta \mathcal{O}_S\,,
\end{align}
used in \cite{Creminelli:2023aly} to the open effective field theory of inflation we discuss here, which for this UV model reduces to
\begin{align}\label{eq:Seffmatchintro}
\quad S_{\mathrm{eff}} &= \int \dd^4 x \Big[ a^2 \pi'_{r} \pi'_{a} - c_{s}^{2} a^2 \partial_i \pir  \partial^i \pia   -  a^3 \gamma \pi'_{r} \pia + i \beta_1 a^4 \pia^2 \nonumber \\
+ &\frac{ \left(8\gamma_2 - \gamma\right)}{2f^{2}_{\pi}}  a^2\pir^{\prime2} \pia	+ \frac{\gamma}{2f_{\pi}^{2}} a^2	\left(\partial_i \pir \right)^2 \pia - 2i\frac{\beta_{5}}{f_\pi^2} a^3 \pir^{\prime} \pia^2 +\frac{\delta_1}{f_\pi^2} a^4 \pia^3  \Big]. 
\end{align}

\paragraph{Outline} The rest of this paper is organized as follows. In \Sec{sec:1}, we develop the bottom-up construction of an open effective field theory for single-clock inflation working from the beginning in the decoupling limit. In \Sec{sec:Pk}, we derive the propagators of the theory and extract the power spectrum. \Sec{sec:bispec} is devoted to the perturbative treatment of interactions through the computation of the contact bispectrum. Lastly, in \Sec{sec:matching}, we perform an explicit matching of our open EFT to a specific UV completion. Conclusions are gathered in \Sec{sec:disc} followed by a series of appendices collecting further details of our calculations. 


\section{The Open Effective Field Theory of Inflation}\label{sec:1}

The starting point to study an open system is a choice of the degrees of freedom that we want to describe, usually known as the \textit{system}, and those that we consider unobservable, usually known as the \textit{environment}. Given our focus on single-clock inflation, our system will consists of a scalar field $\uppi$ to be identified with a Goldstone boson of time translations, to be defined shortly. We aim to construct an EFT for $\uppi$ in the presence of non-unitary time evolution induced by interactions with the environment. A pure state evolving unitarily admits a wavefunction representation
\begin{align}
\Psi\left[\uppi; \eta_0 \right] = \int_{\Omega}^{\uppi} \mathcal{D} \uppi_+ \ee^{i S_{\mathrm{eff}}\left[ \uppi_+\right]},
\end{align}
This is the common starting point of most studies of inflation and the field-theoretic wavefunction $\Psi$ has been the object of intense recent scrutiny \cite{Anninos:2014lwa,Arkani-Hamed:2017fdk,Arkani-Hamed:2018kmz,Pajer:2020wnj,Goodhew:2020hob,Cespedes:2020xqq,Goodhew:2021oqg,Melville:2021lst,Salcedo:2022aal,Baumann:2021fxj,Albayrak:2023hie,Stefanyszyn:2023qov,Sleight:2021plv,Celoria:2021vjw,Creminelli:2024cge,Chakraborty:2023yed} (see \cite{Baumann:2022jpr,Benincasa:2022gtd} for reviews and further references).
In contrast, a mixed state is described by a density matrix $\widehat{\rho}_{\mathrm{red}}$, whose elements can be determined by a path integral along the in-in contour, a.k.a. the \textit{closed-time path}. A useful bookkeeping trick to work with an in-in contour is to double the number of fields to $\uppi_+$ and $\uppi_-$, such that
\begin{align}
\rho_{\mathrm{red}}\left[\uppi, \uppi';\eta_0\right] = \int_{\Omega}^\uppi \mathcal{D}\uppi_+ \int_{\Omega}^{\uppi'} \mathcal{D}\uppi_- \ee^{i S_{\mathrm{eff}}\left[ \uppi_+, \uppi_-\right]}.\label{eq:densitymatrix}
\end{align}
Here $\Omega$ represents the choice of initial state and can be in principle an initial mixed density matrix. In the rest of this work we will assume that in the infinite past the system started in the Bunch-Davies state, and hence was a pure state even when reduced to the $\uppi$ sector. The functional $S_{\mathrm{eff}}$ contains several contributions. First one can identify a contribution representing time evolution according to a Hermitian Hamiltonian. This corresponds to interactions in $S_{\mathrm{eff}}$ that do not mix the two branches of the path integral, which take the general form $S_\uppi\left[\uppi_+\right] - S_\uppi\left[\uppi_-\right]$. Second, we can identify non-unitary (non-Hamiltonian) effects, which encode energy and information losses and gains. These appear in the interactions across the two branches of the path integral. The part of $S_{\mathrm{eff}}$ representing these contributions is sometimes referred to as the Feynman-Vernon influence functional $S_{\mathrm{IF}} \left[\uppi_+,\uppi_- \right]$. In summary, the \textit{open effective functional} $S_{\mathrm{eff}}\left[ \uppi_+, \uppi_-\right]$ admits a Hermitian and a non-Hermitian part
\begin{align}\label{eq:Seffref}
S_{\mathrm{eff}}\left[ \uppi_+, \uppi_-\right] = S_\uppi\left[\uppi_+\right] - S_\uppi\left[\uppi_-\right] + S_{\mathrm{IF}}\left[\uppi_+,\uppi_- \right].
\end{align}
Building the recent literature on non-equilibrium open EFTs \cite{Crossley:2015evo, Boyanovsky:2015xoa, Glorioso:2016gsa, Liu:2018kfw, Hongo:2018ant, Oppenheim:2023izn, Akyuz:2023lsm}, our goal here is to construct and study $S_{\mathrm{eff}}\left[ \uppi_+, \uppi_-\right]$ for single-clock inflation. 


\subsection{EFT construction}\label{subsec:EFT}

In this subsection, we introduce and discuss a series of constraints on the open effective functional $S_{\mathrm{eff}}\left[ \uppi_+, \uppi_-\right]$. Some of these constraints follow from general principles, such as conservation of probabilities, others are specific to systems with a hierarchy of scales and to the study of single-clock inflation.

    \subsubsection{Non-equilibrium constraints}
    
    A first set of constraints follow from requiring that an open quantum system arises from a closed system undergoing Hermitian time evolution upon tracing over the environment \cite{breuerTheoryOpenQuantum2002,Crossley:2015evo, Glorioso:2016gsa, Liu:2018kfw}
    \begin{align}
S_{\mathrm{eff}} \left[\uppi_+,\uppi_+\right] &= 0\,,  \\
    S_{\mathrm{eff}} \left[\uppi_+,\uppi_-\right] &= - 	S^*_{\mathrm{eff}} \left[\uppi_-,\uppi_+\right] \,, \\
    \Ima S_{\mathrm{eff}} \left[\uppi_+,\uppi_-\right] &\geq 0 \,.
\end{align}
These conditions follow respectively from the defining conditions of a density matrix:
\begin{enumerate}
\item Normalisation, $\Tr \widehat{\rho}_{\mathrm{red}}=1$,
\item Hermiticity, $\widehat{\rho}_{\mathrm{red}}^\dagger=\widehat{\rho}_{\mathrm{red}}$,
\item Positivity, $\bra{\phi}\widehat{\rho}_{\mathrm{red}}\ket{\phi}\geq 0$ for all $\ket{\phi}$ in the Hilbert space.
\end{enumerate}
A derivation of these constraints following \cite{Glorioso:2016gsa} is given in \App{app:deriv}. The induced structure holds non-perturbatively and, in particular, remains valid at all loop orders. As we will see below, these conditions strongly reduce the number of operators one may include in the EFT. 

As the number of terms in $S_{\mathrm{eff}} \left[\uppi_+,\uppi_-\right]$ grows rapidly with the number of fields and derivatives, it is convenient to choose an organising principle. A convenient basis to work with\footnote{Hermitian time-evolution is most manifest in the $+/-$ basis, where it corresponds to the absence of mix terms between the $+$ and $-$ branches. Conversely, the causality structure simplifies in the $r$-$a$ basis, where the $\uppi_a$ fields don't propagate \cite{kamenev_2011}. In the literature, sometimes, the following equivalent notation is used where $\uppir \equiv \uppicl$ and $\uppia \equiv \uppiq$ \cite{kamenev_2011, Radovskaya:2020lns}. While in some physical situations, this ``classical-quantum" notation can help counting powers of $\hbar$, this is not the case for inflation and so we prefer to avoid this potentially misleading nomenclature.} is known as the \textit{Keldysh} basis (sometimes referred to as the \textit{retarded/advanced} basis) 
\begin{align}\label{eq:basis}
 \uppi_r &= \frac{\uppi_+ + \uppi_-}{2} \quad \mathrm{and} \quad \uppi_a = \uppi_+ - \uppi_-\, \qquad \Leftrightarrow \qquad  \uppi_\pm =\uppir\pm \frac12 \uppia\,.
\end{align}            
In the $r$-$a$ basis, the conditions on the open effective functional $S_{\mathrm{eff}}$ read
\begin{align}
S_{\mathrm{eff}} \left[\uppi_r,\uppi_a = 0\right] &= 0 \,,\label{eq:norm} \\
S_{\mathrm{eff}} \left[\uppi_r,\uppi_a\right] &= - 	S^*_{\mathrm{eff}} \left[\uppi_r,-\uppi_a\right] \,,\label{eq:herm} \\
\Ima S_{\mathrm{eff}} \left[\uppi_r,\uppi_a\right] &\geq 0. \label{eq:pos}
\end{align} 


\subsubsection{Time-translation symmetry breaking}
    
Symmetries further restrict the number and structure of EFT operators. The general idea follows from the recent Schwinger-Keldysh coset construction \cite{Akyuz:2023lsm} which aims at studying symmetry breaking patterns in out-of-equilibrium systems. Let's consider a schematic microscopic theory $S_{\mathrm{UV}}[\uppi,\chi]$ that is invariant under a symmetry group $G$. Then, the Schwinger-Keldysh action $S_{\mathrm{UV}}[\uppi_+,\chi_+] - S_{\mathrm{UV}}[\uppi_-,\chi_-] $ for the closed system is invariant under $G_+ \times G_-$. Integrating out the $\chi$ field leads to a symmetry breaking pattern under which $S_{\mathrm{eff}} [\uppi_+,\uppi_-]$ becomes invariant under a smaller subgroup $G_{\mathrm{IR}}$ (which is often the diagonal subgroup transforming $\uppi_+$ and $\uppi_-$ simultaneously \cite{Akyuz:2023lsm}). Lastly, one might consider spontaneous symmetry breaking associated to the choice of state, such as the dynamical Kubo-Martin-Schwinger (KMS) symmetry for thermal states \cite{Liu:2018kfw}. We don't discuss this possibility here because cosmological environments are often nonthermal \cite{Colas:2022hlq, Hsiang:2021vgx}.

\begin{figure}[tbp]
    \centering
    \includegraphics[width=1\textwidth]{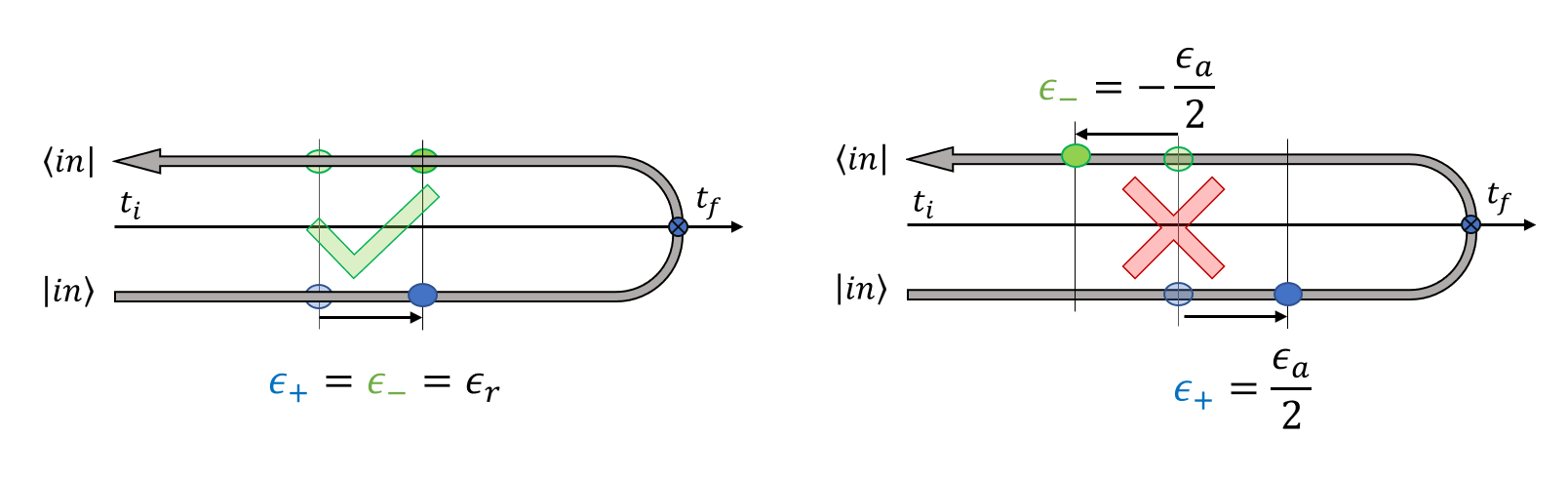}
\caption{The $\epsilon_{r}$ and $\epsilon_{a}$ transformations on the closed time path, where time is running from left to right in both contours and the arrow represent path ordering (time ordering in $\ket{in}$ and anti-time-ordering in $\bra{in}$). The $\epsilon_{r}$ transformation translates the $``+"$ and $``-"$ variables in the same direction while $\epsilon_{a}$ transformation does it in the opposite directions. While the $\epsilon_{r}$ transformation is preserved, the $\epsilon_{a}$ one is explicitly broken due to dissipative effects \cite{Hongo:2018ant}.}
    \label{fig:sym}
\end{figure}

As in the EFToI in the decoupling limit \cite{Creminelli:2006xe, Cheung:2007st, LopezNacir:2011kk}, our open EFT has a single degree of freedom, the Nambu-Goldstone boson of the spontaneous breaking of time translation by the inflaton background. We follow the pioneering work of \cite{Hongo:2018ant}, which first studied this symmetry breaking pattern. Consider a UV-action that is invariant under independent time-translations labelled by $\epsilon_\pm$ realised non-linearly as
\begin{align}
        \uppi_+(t) \rightarrow \uppi'_+(t) = \uppi_+(t+\epsilon_+)+\epsilon_+,\quad\quad\uppi_-(t) \rightarrow \uppi'_-(t) = \uppi_-(t+\epsilon_-)+\epsilon_-,  
    \end{align}
and acting linearly on additional environment fields. Upon tracing over the environment, the open effective functional $S_{\mathrm{eff}}\left[\uppi_+,\uppi_-\right]$ is not in general invariant under general $\epsilon_\pm$ translations. More precisely, the effective functional remains invariant under the translation (see the left-hand panel of \Fig{fig:sym}) \cite{Hongo:2018ant}
    \begin{align}
        \epsilon_+ = \epsilon_- =\epsilon_{r}\,,
    \end{align}
    while the translations
    \begin{align}
        \epsilon_+ =  - \epsilon_- =\frac{\epsilon_{a}}{2}
    \end{align}
    are explicitly broken (see the right-hand panel of \Fig{fig:sym}). In this way, out of two time-translational symmetries of the microscopic action, we are left with a single diagonal subgroup $\epsilon_+ = \epsilon_-$.

Let us consider the transformation of $\uppi_r$ and $\uppi_a$ under the diagonal subgroup of time translations and boosts. In the Keldysh basis, the $\epsilon_{r}$-transformations read \cite{Hongo:2018ant} (see left-hand panel of \Fig{fig:sym})\footnote{One can check that $S_{\mathrm{eff}}[\uppi_r,\uppi_a]$ is left invariant by the transformation given in \Eq{eq:translRsym}. To see it explicitly, one can consider the expansion
    \begin{align}\label{eq:translRsymexp}
        \uppi_{r}(t,\bmx) &\rightarrow \uppi_{r}(t,\bmx) + \epsilon_r\left[1+\dot{\uppi}_{r}(t,\bmx)\right] + \mathcal{O}\left[(\epsilon_r)^2\right],\\
        \uppi_{a}(t,\bmx) &\rightarrow  \uppi_{a}(t,\bmx) + \epsilon_r\dot{\uppi}_{a}(t,\bmx) + \mathcal{O}\left[(\epsilon_r)^2\right].
\end{align} } 
\begin{align}\label{eq:translRsym} 
    \uppi_{r}(t,\bmx) \rightarrow \uppi'_{r}(t,\bmx) &= \uppi_{r}(t+\epsilon_{r},\bmx) + \epsilon_{r},\\
    \uppi_{a}(t,\bmx) \rightarrow \uppi'_{a}(t,\bmx) &= \uppi_{a}(t+\epsilon_{r},\bmx),
\end{align} 
whereas the $\Lambda_{r}$-transformations follow 
\begin{align}\label{eq:LorentzRsym}
\uppi_{r}(t,\bmx) \rightarrow \uppi'_{r}(t,\bmx) &= \uppi_{r}\left(\Lambda^0_{r~\mu}x^\mu,\Lambda^i_{r~\mu}x^\mu \right)+\Lambda^0_{r~\mu}x^\mu - t, \\
\uppi_{a}(t,\bmx) \rightarrow \uppi'_{a}(t,\bmx) &= \uppi_{a}\left(\Lambda^0_{r~\mu}x^\mu,\Lambda^i_{r~\mu}x^\mu \right),
\end{align}
where we introduced $\Lambda^\mu_{r~\nu} \in \mathrm{SO}(1,3)$. The important point is that $\uppi_r$ \textit{non-linearly realises time-translations and boosts} whereas $\uppi_{a}$ \textit{transforms linearly}, just as ordinary matter \cite{Hongo:2018ant}. 


\subsubsection{Locality} 

In this subsection, we discuss \textit{locality} of $S_{\mathrm{eff}}$ in time and space. In the most general case, tracing over the environment yields an unwieldy non-local effective functional that is intractable unless one knows the exact UV-completion. Instead, just like for standard EFTs, a dramatic simplification takes place in the presence of a \textit{separation of scales}. Here we focus on precisely this possibility: we envisage that the typical length and time scales characterising the environment are much shorter than the Hubble time and the Hubble radius at which we compute cosmological correlators. This hierarchy ensures that our open EFT is local in space and time, i.e. it features operators that are the product of fields at the same spacetime point and a finite but arbitrary number of derivatives thereof. Not all UV-models display such hierarchy. For example, the much studied models of inflation featuring the tachyonic production of gauge modes engendered by a $\phi F\tilde F$ coupling \cite{Anber:2009ua} would give a non-local open EFT for $\uppi$ because the gauge fields are mostly produced around Hubble crossing. See also \cite{breuerColloquiumNonMarkovianDynamics2016, Prudhoe:2022pte, Shandera:2017qkg, Colas:2022hlq, Colas:2022kfu, Colas:2024xjy} for additional examples where the open EFT could be non-local and/or non-Markovian. The fact that our open EFT is not useful to describe these models is nothing new or specific to open systems: the same would happen for a standard EFT if one tried to integrate out very light or massless fields.\\

So we want to fucus on models that exhibit a hierarchy of scales. One such model was constructed and studied in detail recently in \cite{Creminelli:2023aly} and we were able to match our EFT description to that UV completion, as we will discuss in \Sec{sec:matching}. We further discuss the physical implications of locality in \Sec{sec:disc}.  

Now that we motivated a local open EFT, we want to write down all possible local operators compatible with the non-equilibrium constraints and symmetries. We can do this by using invariant combinations as fundamental building blocks \cite{Akyuz:2023lsm}. The open effective functional can be constructed out of $\uppi_{a}$, $t+ \uppi_{r}$ and their derivatives $\partial_\mu \uppi_a$ and
\begin{align}\label{eq:Pmu}
    P_\mu = \partial_\mu (t + \uppi_{r}) = \delta^0_{~\mu}+ \partial_\mu \uppi_{r}
\end{align}
just as one usually do in e.g. the EFToI \cite{Hongo:2018ant}. Working at leading order in the derivative expansion, we now restrict ourselves to operators with at most one derivative per field. At last, following \cite{LopezNacir:2011kk}, we work in the decoupling limit where there is a unique dynamical degree of freedom $\uppi$ up to slow-roll corrections (which remains a valid description despite the presence of the environment, see Appendix D of \cite{LopezNacir:2011kk}). Notice that one might also like to derive our open EFToI starting in unitary gauge and specifying a theory of gravity that is invariant under spatial diffeomorphisms but not time diffeomorphisms, as in the original derivation \cite{Cheung:2007st}. While doing this order by order in metric perturbations is straightforward, it is not clear to us how the non-perturbative and diffeomorphism invariant structure of gravity is organised when fields are doubled. For example, one seems to have infinitely many choices to covariantise indices using $g_{\mu\nu}^+$, $g_{\mu\nu}^-$ or combinations thereof. We hope to come back to this issue in future research. 


\subsection{Open effective functional}

We now aim at writing the most generic local open effective functional compatible with the non-equilibrium constraints and the above symmetries, with at most one derivative per field. In addition to the derivative expansion, a useful way to organise the open effective functional is in powers of $\uppi_a$ such that $ S_{\mathrm{eff}} = \int \dd^4 x \sqrt{-g }  \mathcal{L}_{\mathrm{eff}}$ with 
\begin{align}\label{eq:expref}
    \mathcal{L}_{\mathrm{eff}} = \sum_{n=1}^\infty \mathcal{L}_n\quad\mathrm{with}\quad \mathcal{L}_n = \mathcal{O}(\uppi_{a}^n)
\end{align}
where we used the unitarity condition \eqref{eq:norm} to notice that $\mathcal{L}_{\mathrm{eff}}$ starts from the first order term in $\uppi_{a}$. Notice that we are \textit{not} saying that operators with more powers of $\uppi_a$ are suppressed, and we are not invoking any argument about the $\hbar$ expansion, which is not justified in general in this context \cite{Radovskaya:2020lns}. Instead we are just using the organisation in powers of $\uppi_a$ to neatly groups the many different terms. Restricting ourselves to cubic operators for practical applications, we focus on $\mathcal{L}_1$, $\mathcal{L}_2$ and $\mathcal{L}_3$ for which we illustrate the general procedure, the next order being at best quartic in $\uppi_a$. 

\paragraph{On total derivatives} Before developing the explicit construction,  we wish to discuss the status of total derivatives whose removal greatly simplifies the discussion \cite{Hongo:2018ant}. In the inflationary context, working with the conformal time $\eta \in (-\infty,0]$, it might be a concern that integration-by-part (IBP) might generate boundary terms \cite{Salcedo:2022aal, Braglia:2024zsl} and might not be as harmless as in flat space. A way to avoid this issue is obviously to avoid IBP. Yet, it is instructive to see how this issues manifests itselft in our open EFT and contrast it to the usual EFToI. The boundary conditions of the in-in path integral impose $\uppi_a =0$ at the boundary. Consequently, any boundary term proportional to $\uppi_a$ vanishes. Non-vanishing boundary terms must contain at least one time derivative acting on $\uppi_a$. They correspond to operators with more than one derivative acting on the field variables in the bulk, which we neglected by focusing on the leading EFT operators. For this reason, we safely discard total derivatives in this work. 
    
\subsubsection{An effective functional for the Goldstone boson}

Let us construct $\mathcal{L}_{n}$ order by order in $\uppi_a$ following the above principles. 
    
\paragraph{$\mathcal{L}_1$ functional}

Let us illustrate the procedure by first considering $\mathcal{L}_1$. We aim at using the building blocks defined around \Eq{eq:Pmu} to construct invariant combinations that are linear in $\uppi_a$. The only option consists in multiplying $\uppi_a$ and $P^\mu \partial_\mu \uppi_{a} = \left(-\dot{\uppi}_{a} + \partial^\mu \uppi_{r} \partial_\mu \uppi_{a}\right)$ by powers of
    \begin{align}
        \left( P_\mu P^\mu  +1\right) = -2 \dot{\uppi}_{r} + \left( \partial_\mu \uppi_{r}\right)^2,
    \end{align}
    leading to \cite{Hongo:2018ant}\footnote{The sign in front of the $\upgamma_n$ term is chosen for later convenience.}
    \begin{align}
        \mathcal{L}_1 = & \sum_{n=0}^\infty \left( P_\mu P^\mu  +1\right)^{n} \left[ \upgamma_n \uppi_{a} -\upalpha_n P^\mu \partial_\mu \uppi_{a}  \right].
    \end{align}
    The EFT coefficients $\upgamma_n$ and $\upalpha_n$ are in general functions of $t + \uppi_{r}$ which have to be real because of the conjugate condition \eqref{eq:herm}.  In the slow-roll regime, one can assume time-independence for the EFT coefficients at leading order in slow-roll \cite{Hongo:2018ant}. 

     \begin{tcolorbox}[%
enhanced, 
breakable,
skin first=enhanced,
skin middle=enhanced,
skin last=enhanced,
before upper={\parindent15pt},
]{}
\paragraph{Background evolution and tadpoles}

Before describing the dynamics of the fluctuations, let us connect with the standard background evolution of the EFToI by discussing tadpole cancellation \cite{Cheung:2007st, Collins:2012nq}. As mentioned below \Eq{eq:expref}, the unitarity condition \eqref{eq:norm} imposes that $\mathcal{L}_{\mathrm{eff}}$ starts linear in $\uppi_{a}$. Therefore, the only available tadpoles are \cite{LopezNacir:2011kk,Hongo:2018ant}
\begin{align}\label{eq:tad1}
            S_{\mathrm{eff}}  \supset  \int \dd^4 x  \sqrt{-g} \left[ \upgamma_0(t) \uppi_a  + \upalpha_0(t) \dot{\uppi}_a \right],
\end{align}
        which leads to the tadpole cancellation 
        \begin{align}
           - \upgamma_0 + \dot{\upalpha}_0 + 3H \upalpha_0 = 0.
        \end{align}
        Let us compare this result to the usual construction. In the standard EFToI approach, the background FLRW evolution is obtained from varying  
        \begin{align}\label{eq:SEFToI}
            S_{\mathrm{EFToI}} = \int \dd^4x \sqrt{-g} \left[\frac{1}{2} M_{\mathrm{Pl}}^2 R - \Lambda(t) - c(t) g^{00} \right]
        \end{align}
        with respect to the $g^{00}$ and $g^{ii}$ components of the metric, leading to 
        \begin{align}\label{comb1}
            3 M_{\mathrm{Pl}}^2  H^2 &= c(t) + \Lambda(t)\,, \\
            - M_{\mathrm{Pl}}^2 \dot{H} &= c(t). \label{comb2}
        \end{align}
One can also reintroduce diffeomorphism invariance through the St\"uckelberg trick $t\rightarrow t + \uppi$ and then vary the action \Eq{eq:SEFToI} with respect to $\uppi$. This leads to the tadpole cancellation \cite{Cheung:2007st}
        \begin{align}\label{eq:tad2}
            \dot{\Lambda} + \dot{c} + 6 H c = 0\,.
        \end{align}
We see that this is not a new equation but instead follows from combining \Eqs{comb1} and \eqref{comb2}. This is to be expected because Einstein's equation imply the conservation of the total stress-energy tensor $\nabla^\mu T_{\mu \nu} = 0$. Rotated in the Keldysh basis, the above two tadpoles lead to\footnote{Notice that these terms are manifestly unitary. They obviously pass the unitary test described below in \Sec{subsubsec:unit} if one accounts for the slow-roll suppressed terms such as $\Ddot{\Lambda} \uppi_r \uppi_a$.} 
        \begin{align}
              -\left.\Lambda(t)\right|_{t\rightarrow t + \uppi_r + \frac{\uppi_a}{2}} + \left.\Lambda(t)\right|_{t\rightarrow t + \uppi_r - \frac{\uppi_a}{2}}  &= - \dot{\Lambda}(t) \uppi_a + \mathcal{O}(\uppi_a^2),
        \end{align}
        and 
        \begin{align}
            &-\left.c(t) g^{00}\right|_{t\rightarrow t + \uppi_r + \frac{\uppi_a}{2}} + \left.c(t) g^{00}\right|_{t\rightarrow t + \uppi_r - \frac{\uppi_a}{2}}    =  \dot{c}(t) \uppi_a + 2 c(t) \dot{\uppi}_a + \mathcal{O}(\uppi_a^2).
        \end{align}
        We identify 
        \begin{align}\label{eq:tadpole}
            \upalpha_0(t) \equiv 2 c(t) \qquad \mathrm{and} \qquad \upgamma_0 (t)\equiv  \dot{c}(t) - \dot{\Lambda}(t),
        \end{align}
        such that \Eqs{eq:tad1} and \eqref{eq:tad2} are equivalent. 
   \end{tcolorbox}

    The main physical outcome of the above discussion is the following. As noticed in \cite{LopezNacir:2011kk}, there is no new tadpole for this class of local dissipative models of inflation. The background evolution is fixed by the slicing and probes the global energy density, which does not distinguish the contributions of the inflaton from those of the unknown environment.\footnote{It would be interesting to further investigate if there exists a slicing where system and environment are distinguishable from the background dynamics, for instance through different charges.} It is only at the level of the fluctuations that the distinction between system and environment becomes relevant, as we can disentangle observable degrees of freedom associated to the hydrodynamical direction $\uppi$ and unobservable degrees of freedom that have been integrated out. 
    
    Now we have fixed the background dynamics and removed the tadpoles, $\mathcal{L}_1$ takes the explicit form
    \begin{align}
        \mathcal{L}_1 = - \upalpha_0 \partial^\mu \uppi_{r} \partial_\mu \uppi_{a} &-\sum_{n=1}^\infty \upalpha_n \left[-2 \dot{\uppi}_{r} + \left( \partial_\mu \uppi_{r}\right)^2\right]^{n}\left(-\dot{\uppi}_{a} + \partial^\mu \uppi_{r} \partial_\mu \uppi_{a}\right) \nonumber \\
        +& \sum_{n=1}^\infty \upgamma_n \left[-2 \dot{\uppi}_{r} + \left( \partial_\mu \uppi_{r}\right)^2\right]^n\uppi_{a}.
    \end{align}
    Notice that only $\upalpha_0$, $\upalpha_1$ and $\upgamma_1$ provide quadratic terms in $\uppi$ relevant for the dispersion relation of the Goldstone mode \cite{Hongo:2018ant}. The $\upalpha_0$ term is the usual kinetic term written in the Keldysh basis, see the basis transformation \eqref{eq:basis}. The tadpole cancellation discussed above leads to the relation $\upalpha_0(t) = 2 c(t) = 2M_{\mathrm{Pl}}^2 |\dot{H}|$. The $\upalpha_1$ term generates a non-trivial speed of sound accompanied by higher order operators controlling the appearance of equilateral non-Gaussianities \cite{Cheung:2007st}. 
    In contrast to the $\upalpha_0$ and $\upalpha_1$ terms, the $\upgamma_1$ term has no unitary counterpart and leads to a dissipative term in the $\uppi_{r}$ equation of motion, as we will see in \Sec{sec:matching}. Interestingly, the dissipation term $\dot{\uppi}_{r}\uppi_{a}$ is accompanied by a cubic interaction $\left( \partial_\mu \uppi_{r}\right)^2 \uppi_{a}$, as first noted in \cite{LopezNacir:2011kk}, such that the combination is invariant under Lorentz boosts. 

    Let us explicitly consider the quadratic and cubic contributions. Indeed, in addition to the expansion in powers of $\uppi_a$, one can classify
    \begin{align}
        \mathcal{L}_{\mathrm{eff}} = \sum_{n,m=1}^\infty  \mathcal{L}^{(m)}_n \quad\mathrm{with}\quad \mathcal{L}^{(m)}_n = \mathcal{O}(\uppi^m,~\uppi_{a}^n,~ \uppi_r^{m-n})
    \end{align}
    where $m$ labels the number of field operators. The quadratic contributions are
    \begin{align}
        \mathcal{L}_1^{(2)} &= \left( \upalpha_0 -2 \upalpha_1\right) \dot{\uppi}_{r} \dot{\uppi}_{a} - \upalpha_0 \partial_i \uppi_{r}  \partial^i \uppi_{a} - 2 \upgamma_1  \dot{\uppi}_{r} \uppi_a
    \end{align}
    where $\upalpha_0$ and $\upalpha_1$ control the unitary kinetic term with a non-trivial speed of sound and $\upgamma_1$ encodes the linear dissipation of the system onto its environment. At cubic order, $\mathcal{L}_1$ reads
    \begin{align}\label{eq:cubicinter}
        \mathcal{L}_1^{(3)} =  &\left(4 \upalpha_2 -3\upalpha_1 \right) \dot{\uppi}^2_{r} \dot{\uppi}_{a} + \upalpha_1 \left(\partial_i \uppi_{r} \right)^2 \dot{\uppi}_a + 2 \upalpha_1  \dot{\uppi}_{r}	\partial_i \uppi_{r}  \partial^i \uppi_{a} \Big.\nonumber \\
        &\qquad + \left(4 \upgamma_2 -\upgamma_1 \right)  \dot{\uppi}^2_{r} \uppi_{a}	+ \upgamma_1 	\left(\partial_i \uppi_{r} \right)^2 \uppi_a 
    \end{align} 
    where the first line corresponds to parts of the unitary operators $\dot{\uppi}^3$ and $(\partial_i \uppi)^2 \dot{\uppi}$ and the second line to the non-linear dissipation induced by the non-linearly realised symmetries, as discussed in \cite{LopezNacir:2011kk}.
    
    \paragraph{$\mathcal{L}_2$ functional} 

    Following \cite{Hongo:2018ant}, let us now construct $\mathcal{L}_2$, which is quadratic in $\uppi_a$. Just as above, working with operators containing at most one derivative, in the slow-roll limit, we obtain  
\begin{align}\label{eq:diffquad}
    \mathcal{L}_2 &= i \Big[\upbeta_1 \uppi_{a}^2 + \upbeta_2 \left( \partial_\mu \uppi_{a}\right)^2 +\upbeta_3 \left(-\dot{\uppi}_{a} + \partial^\mu \uppi_{r} \partial_\mu \uppi_{a}\right)\uppi_{a} +\upbeta_4 \left(-\dot{\uppi}_{a} + \partial^\mu \uppi_{r} \partial_\mu \uppi_{a}\right)^2 + \cdots \Big]
\end{align}
where the third and fourth terms are obtained from $P^\mu \partial_\mu \uppi_{a} \uppi_a$ and $(P^\mu \partial_\mu \uppi_{a})^2$ respectively and the dots represent higher order terms obtained by multiplying the first four terms by arbitrary powers of $(P^\mu P_\mu + 1) = -2 \dot{\uppi}_{r} + \left( \partial_\mu \uppi_{r}\right)^2$. The action being at least quadratic in $\uppi_a$, there is no tadpole contribution. The $i$ in front directly follows from the conjugate condition \eqref{eq:herm}. While the term proportional to $\upbeta_1$ is the standard noise term appearing in the Langevin equation, see \Sec{sec:matching}, we observe the presence of derivative corrections such as $\left(\partial_\mu \uppi_{a} \right)^2$ and $\dot{\uppi}_{a}^2 $ in the $\upbeta_2$ and $\upbeta_4$ terms which make the noise scale-dependent. 

There exists a positivity condition on the $\upbeta$'s coefficients due to \Eq{eq:pos} which imposes $\Ima S_{\mathrm{eff}} [\uppi_r,\uppi_a] \geq 0$. In flat space, making use of the derivative expansion which tells us that $\omega^2, k^2 \ll |\upbeta_1/\upbeta_{2,4}|$ (the quadratic term in $\upbeta_3$ can be written as a total derivative and removed), the authors of \cite{Hongo:2018ant} concluded that $\upbeta_1$ dominates in $\mathcal{L}_2$, such that the positivity constraint imposes 
\begin{align}
    \upbeta_1 > 0.
\end{align}
We will see this constraint emerging from a different point of view in \Sec{sec:Pk}, where it follows from requiring the positivity of the power spectrum. This different requirement further imposes constraints on $\upbeta_2$ and $\upbeta_4$. This positivity constraint on the noise kernel directly translates into consequences for the non-Gaussian signal if we multiply this operator by higher powers of $(P^\mu P_\mu + 1) = -2 \dot{\uppi}_{r} + \left( \partial_\mu \uppi_{r}\right)^2$. 

Note that if the positivity constraints $\Ima S_{\mathrm{eff}} [\uppi_r,\uppi_a] \geq 0$ seems easy to satisfy for odd powers of $\uppi_a$ as long as the Wilsonian coefficients are real, the case of even powers is much less trivial. For even powers of $\uppi_a$, one can ask if there exists combination of operators which are not positive definite or total derivatives. So far, we have not identified any of these terms even considering higher-derivative operators, which may guarantee, at least in principle, the boundedness of even powers of $\uppi_a$. The second concern is related to the presence of non-linear interactions involving $\uppi_r$. If one considers the simplest noise term $i \upbeta_1 \uppi_a^2$, one can multiply this operator by $(P^\mu P_\mu + 1) = -2 \dot{\uppi}_{r} + \left( \partial_\mu \uppi_{r}\right)^2$ to obtain equally valid operators. The cubic term $\dot{\uppi}_{r} \uppi_a^2$ is unbounded and one must impose a (non-unitary) perturbativity bounds to ensure convergence. 

As above, let us explicitly consider the quadratic and cubic contributions for $\mathcal{L}_2$. The three quadratic noise are controlled by 
\begin{align}
\mathcal{L}_2^{(2)} &= i \left[\upbeta_1 \uppi_a^2 - \left(\upbeta_2 - \upbeta_4\right)\dot{\uppi}_a^2 + \upbeta_2 \left(\partial_i \uppi_{a} \right)^2\right]\,,\label{eq:piq2free}
\end{align}
where we removed the total derivative related to $\upbeta_3$ at quadratic order. At cubic order, we obtain 
\begin{align}\label{eq:cubicinter2}
\mathcal{L}_2^{(3)} &= i \Big[-(\upbeta_3 -2 \upbeta_7)\dot{\uppi}_{r} \dot{\uppi}_{a} \uppi_a + \upbeta_3 \partial_i \uppi_{r}  \partial^i \uppi_{a} \uppi_a +2 (\upbeta_4+\upbeta_6 - \upbeta_8) \dot{\uppi}_{r} \dot{\uppi}_a^2 \nonumber \\
& \qquad \quad - 2 \upbeta_4   \partial_i \uppi_{r}  \partial^i \uppi_{a} \dot{\uppi}_{a} - 2\upbeta_5 \dot{\uppi}_{r} \uppi_a^2  - 2 \upbeta_6 \dot{\uppi}_{r}(\partial_i\uppi_a)^2\Big],
\end{align}
where we needed to introduce the Wilsonian coefficients $\upbeta_5, \upbeta_6, \upbeta_7$ and $\upbeta_8$ associated to the higher-order operators included in the dots of \Eq{eq:diffquad}, obtained from multiplying the first four terms in \Eq{eq:diffquad} by $(P^\mu P_\mu + 1) = -2 \dot{\uppi}_{r} + \left( \partial_\mu \uppi_{r}\right)^2$. The physical interpretation of these terms is discussed below in \Sec{subsubsec:unit}.

    \paragraph{$\mathcal{L}_3$ functional}

    One can carry on this construction to access $\mathcal{L}_3$. We have
    \begin{align}
        \mathcal{L}_3 &= \updelta_1 \uppi_a^3 + \updelta_2 (\partial_\mu \uppi_a)^2 \uppi_a + \updelta_3 \left(-\dot{\uppi}_{a} + \partial^\mu \uppi_{r} \partial_\mu \uppi_{a}\right) \uppi_a^2 + \updelta_4 \left(-\dot{\uppi}_{a} + \partial^\mu \uppi_{r} \partial_\mu \uppi_{a}\right) (\partial_\nu \uppi_a)^2 \Big.  \nonumber\\
        &\qquad \qquad + \updelta_5 \left(-\dot{\uppi}_{a} + \partial^\mu \uppi_{r} \partial_\mu \uppi_{a}\right)^2 \uppi_a  + \updelta_6 \left(-\dot{\uppi}_{a} + \partial^\mu \uppi_{r} \partial_\mu \uppi_{a}\right)^3 + \cdots,
    \end{align}
    where the $\updelta_3$ term originates from $(P^\mu \partial_\mu \uppi_a) \uppi_a^2$, the $\updelta_4$ term from $(P^\mu \partial_\mu \uppi_a) (\partial_\nu \uppi_a)^2 $, the $\updelta_5$ from $\left(P^\mu \partial_\mu \uppi_a\right)^2 \uppi_a$ and the $\updelta_6$ term from $\left(P^\mu \partial_\mu \uppi_a\right)^3$. As above, the dots represent higher order terms obtained by multiplying the first terms by arbitrary powers of $(P^\mu P_\mu + 1) = -2 \dot{\uppi}_{r} + \left( \partial_\mu \uppi_{r}\right)^2$. As we will see below, the interpretation of these terms is ambiguous, as they can either be associated to unitary or non-unitary operators depending on how they relate to contributions from $\mathcal{L}_1$. For this reason, we develop in \Sec{subsubsec:unit} a classification of these terms. 
    
    As above, let us explicitly consider the quadratic and cubic contributions for $\mathcal{L}_3$. Since these contributions are at least cubic in $\uppi_a$, there is no quadratic contribution. If we restrict ourselves to the cubic order, we obtain 
     \begin{align}
        \mathcal{L}_3^{(3)} &=\updelta_1 \uppi_a^3 + (\updelta_5- \updelta_2)  \dot{\uppi}_{a}^2 \uppi_a  + \updelta_2  (\partial_i \uppi_{a})^2 \uppi_a - \updelta_4  (\partial_i \uppi_{a})^2 \dot{\uppi}_a + (\updelta_4-\updelta_6)\dot{\uppi}_{a}^3 ,
    \end{align}
    the $\updelta_3$ cubic contribution being a total derivative. 

    \begin{tcolorbox}[%
enhanced, 
breakable,
skin first=enhanced,
skin middle=enhanced,
skin last=enhanced,
before upper={\parindent15pt},
]{}
\paragraph{Summary}

Under the assumptions specified in \Sec{subsec:EFT}, the most generic second order open effective functional (up to total derivatives) is
\begin{align}\label{eq:L2ref}
    &\qquad \mathcal{L}^{(2)} =\left( \upalpha_0 -2 \upalpha_1\right) \dot{\uppi}_{r} \dot{\uppi}_{a} - \upalpha_0 \partial_i \uppi_{r}  \partial^i \uppi_{a} \nonumber   \\
    &- 2 \upgamma_1  \dot{\uppi}_{r} \uppi_a+i \left[\upbeta_1 \uppi_a^2 - \left(\upbeta_2 - \upbeta_4\right)\dot{\uppi}_a^2 + \upbeta_2 \left(\partial_i \uppi_{a} \right)^2\right],
\end{align}
where the EFT coefficients are chosen to match the notations of \cite{Hongo:2018ant}. The first line corresponds to the usual unitary dynamics which the kinetic term and an effective speed of sound. The first term of the second line controlled by $\upgamma_1$ corresponds to the dissipation due to the surrounding environment. At last, the $\upbeta_i$ coefficients control the diffusion (noise-induced) process.\footnote{Note that if the environment obeys thermal equilibrium, the KMS symmetry imposes a relation between the $\upgamma_i$ and the $\upbeta_j$ \cite{Liu:2018kfw}. 
}  

\indent At cubic order, the most generic open effective functional (up to total derivatives) writes
\begin{align}
    \mathcal{L}^{(3)} &= \left(4 \upalpha_2 -3\upalpha_1 \right) \dot{\uppi}^2_{r} \dot{\uppi}_{a} + \upalpha_1 \left(\partial_i \uppi_{r} \right)^2 \dot{\uppi}_a + 2 \upalpha_1  \dot{\uppi}_{r}	\partial_i \uppi_{r}  \partial^i \uppi_{a} \Big. \label{eq:L13} \\
    &\qquad \quad + \left(4 \upgamma_2 -\upgamma_1 \right)  \dot{\uppi}^2_{r} \uppi_{a}	+ \upgamma_1 	\left(\partial_i \uppi_{r} \right)^2 \uppi_a \nonumber \\ 
    +&i \Big[-(\upbeta_3 -2 \upbeta_7)\dot{\uppi}_{r} \dot{\uppi}_{a} \uppi_a + \upbeta_3 \partial_i \uppi_{r}  \partial^i \uppi_{a} \uppi_a +2 (\upbeta_4+\upbeta_6 - \upbeta_8) \dot{\uppi}_{r} \dot{\uppi}_a^2 \label{eq:L23} \\
    & \qquad \quad - 2 \upbeta_4   \partial_i \uppi_{r}  \partial^i \uppi_{a} \dot{\uppi}_{a} - 2\upbeta_5 \dot{\uppi}_{r} \uppi_a^2  - 2 \upbeta_6 \dot{\uppi}_{r}(\partial_i\uppi_a)^2\Big] \nonumber \\
    +&\updelta_1 \uppi_a^3 + (\updelta_5- \updelta_2)  \dot{\uppi}_{a}^2 \uppi_a  + \updelta_2  (\partial_i \uppi_{a})^2 \uppi_a - \updelta_4  (\partial_i \uppi_{a})^2 \dot{\uppi}_a + (\updelta_4-\updelta_6)\dot{\uppi}_{a}^3, \label{eq:L33}
\end{align}
where \ref{eq:L13} originates from $\mathcal{L}_1^{(3)}$, \ref{eq:L23} originates from $\mathcal{L}_2^{(3)}$ and \ref{eq:L33} from $\mathcal{L}_3^{(3)}$.

\end{tcolorbox}

In de Sitter, in terms of the conformal time and the scale factor $a = -1/(H\eta)$, the open effective functional up to cubic order reads
\begin{align}
&\quad S_{\mathrm{eff}}^{(2)} = \int \dd^4 x \Big\{\left( \upalpha_0 -2 \upalpha_1\right) a^2 \uppi'_{r} \uppi'_{a} - \upalpha_0 a^2 \partial_i \uppi_{r}  \partial^i \uppi_{a}   \\
-& 2 a^3 \upgamma_1 \uppi'_{r} \uppi_a + i \left[\upbeta_1 a^4 \uppi_a^2 - \left(\upbeta_2 - \upbeta_4\right) a^2\uppi_a^{\prime2} + \upbeta_2 a^2 \left(\partial_i \uppi_{a} \right)^2\right] \Big\}, \nonumber
\end{align}
and
\begin{align}
S_{\mathrm{eff}}^{(3)} =  \int &\dd^4 x \Big\{\left(4 \upalpha_2 -3\upalpha_1 \right) a \uppi_r^{\prime2} \uppi'_{a} + \upalpha_1 a \left(\partial_i \uppi_{r} \right)^2 \uppi'_a + 2 \upalpha_1 a \uppi'_{r}	\partial_i \uppi_{r}  \partial^i \uppi_{a} \\
& \qquad \quad + \left(4 \upgamma_2 -\upgamma_1 \right)  a^2\uppi_r^{\prime2} \uppi_{a}	+ \upgamma_1 a^2	\left(\partial_i \uppi_{r} \right)^2 \uppi_a \nonumber \Big. \\
&+ i \Big[-(\upbeta_3-2\upbeta_7) a^2 \uppi'_{r} \uppi'_{a} \uppi_a + \upbeta_3 a^2 \partial_i \uppi_{r}  \partial^i \uppi_{a} \uppi_a +2 (\upbeta_4+\upbeta_6 - \upbeta_8) a \uppi'_{r} \uppi_a^{\prime2} \nonumber \\
& \qquad \quad - 2 \upbeta_4  a \partial_i \uppi_{r}  \partial^i \uppi_{a} \uppi'_{a} - 2\upbeta_5 a^3 \uppi_r^{\prime} \uppi_a^2  - 2 \upbeta_6 a \uppi_r^{\prime} (\partial_i\uppi_a)^2 \Big] \nonumber\\
&+\updelta_1 a^4 \uppi_a^3 + (\updelta_5- \updelta_2) a^2 \uppi_a^{\prime2} \uppi_a  + \updelta_2 a^2 (\partial_i \uppi_{a})^2 \uppi_a - \updelta_4 a  (\partial_i \uppi_{a})^2 \uppi'_a + (\updelta_4-\updelta_6) a \uppi_a^{\prime3} \Big\}. \nonumber
\end{align}


\subsubsection{Classification of the EFT operators}\label{subsubsec:unit}

The above construction exhibits a wide zoology of terms compared to its unitary counterpart: $5$ free parameters in $\mathcal{L}^{(2)}$ compared to only $1$ in the standard EFToI \cite{Cheung:2007st}; $13$ free parameters in $\mathcal{L}^{(3)}$ compared to only $1$ in \cite{Cheung:2007st}. While some operators describe faithful non-unitary effects generated by the presence of additional degrees of freedom, others are simply a consequence of writing unitary interactions in the Keldysh basis. In this section, we develop a procedure to distinguish unitary from non-unitary operators.

\paragraph{Recovering the EFToI} In this work, we study the decoupling limit of $\pi$ interacting with an unknown environment. Therefore, we expect our open effective functional $S_{\mathrm{eff}}$ to be able to reproduce in a certain limit the EFToI \cite{Cheung:2007st}. This limit defines the unitary direction of the parameter space of the theory. Let us first consider the quadratic terms of the EFToI which reads in the Keldysh basis
\begin{equation}
\frac{1}{2}\left[\dot{\uppi}_{+}^{2}-c_{s}^{2}(\partial_{i}\uppi_{+})^{2}\right]-\frac{1}{2}\left[\dot{\uppi}_{-}^{2}-c_{s}^{2}(\partial_{i}\uppi_{-})^{2}\right]=\dot{\uppi}_r\dot{\uppi}_a-c_{s}^{2}\partial_{i}{\uppir}\partial_{i}{\uppia}.
\end{equation}
It can be matched with the first line of \Eq{eq:L2ref} for
\begin{equation}
c_{s}^{2} = \frac{\upalpha_0}{\upalpha_{0}-2\upalpha_1}.
\end{equation}
We can go to the next order in the EFToI and consider cubic operators. Starting with $\dot{\uppi}^{3}$, the unitary interaction in the Keldysh basis reads
\begin{equation}\label{eq:Unitarypiq3}
\dot{\uppi}^{3}_{+}-\dot{\uppi}^{3}_{-}=3\dot{\uppi}_{r}^{2}\dot{\uppi}_{a} + \frac{1}{4}\dot{\uppi}^{3}_{a}.
\end{equation}
Comparing with the terms in \Eqs{eq:L13} and \eqref{eq:L33} which include $\dot{\uppi}_{r}^{2}\dot{\uppi}_{a}$ and  $\dot{\uppi}^{3}_{a}$, the unitary combination in \Eq{eq:Unitarypiq3} imposes the relation among the EFT coefficients\footnote{This relation defines an orthogonal direction $\updelta_4-\updelta_6= - 12\left(4 \upalpha_2 -3\upalpha_1 \right)$      intrinsically related to non-unitary effects. It would be interesting to explore whether information  quantifiers such as the purity only depends on the value of the EFT coefficients along this orthogonal direction. }
\begin{equation}\label{eq:combin1}
\updelta_4-\updelta_6= \frac{1}{12}\left(4 \upalpha_2 -3\upalpha_1 \right).
\end{equation}
A similar procedure follows for $(\partial_{i}\uppi)^{2}\dot{\uppi}$. In the Keldysh basis, this vertex reads
\begin{equation}
(\partial_{i}\uppi_{+})^{2}\dot{\uppi}_{+}-(\partial_{i}\uppi_{-})^{2}\dot{\uppi}_{-}=(\partial_{i}\uppir)^{2}\dot{\uppi}_{a} + 2 \partial_{i}\uppir\partial_{i}\uppia\dot{\uppi}_r+\frac{1}{4}\dot{\uppi}_a(\partial_{i}\uppir)^{2}.
\end{equation}
Comparing it to the operators appearing in \Eqs{eq:L13} and \eqref{eq:L33}, it specifies the unitary direction\footnote{The orthogonal direction is then defined by $\delta_{4}=-4\alpha_1$. It would again be interesting to consider if the entanglement tracers depends on the combination $(4\updelta_{4}+\upalpha_1)$ alone.} 
\begin{equation}\label{eq:combin2}
\updelta_{4}=-\frac{1}{4}\upalpha_1.
\end{equation}
One can then be reassured that in a certain limit, the current constructions reduces to the usual EFToI of \cite{Cheung:2007st}.

\paragraph{Unitary and orthogonal directions}

What about the other operators? Our open effective functional has more Wilsonain coefficients than the EFToI. Does it imply that all the remaining operators are intrinsically related to non-unitary effects?  
The symmetry structure of the theory allows one to  answer this question is a systematic manner. In the limit where non-unitary effects are absent, \Eq{eq:Seffref} reduces to the unitary effective action 
\begin{align}\label{eq:Seffunit}
S_{\mathrm{eff}}\left[ \uppi_+, \uppi_-\right] = S_\uppi\left[\uppi_+\right] - S_\uppi\left[\uppi_-\right].
\end{align}
This restriction is obtained by restoring the $\epsilon_a$ symmetry explicitly broken by the non-unitary effects (see \textit{Right} panel of \Fig{fig:sym}) \cite{Hongo:2018ant}. Indeed, in the unitary limit, the two branches of the path integral must transform equally under $\epsilon_\pm$                               
\begin{align}\label{eq:qsym}
\uppi_{\pm}(t,\bmx) \rightarrow \uppi'_{\pm}(t,\bmx) &= \uppi_{\pm}(t+\epsilon_{\pm},\bmx) + \epsilon_{\pm}.
\end{align}
One can impose this by acting on $S_{\mathrm{eff}}$ with the $\epsilon_{a}$ symmetry given by $\epsilon^0_+ =  - \epsilon^0_- = \epsilon_{a}/2$. Expressing \Eq{eq:qsym} in the Keldysh basis and expanding linear order in $\epsilon_a$ we obtain
\begin{align}
\uppi_{r}(t,\bmx) \rightarrow \uppi'_{r}(t,\bmx) &= \uppi_{r}(t,\bmx) + \frac{\epsilon_a}{2}\dot{\uppi}_{a}(t,\bmx) + \mathcal{O}\left(\epsilon_a^2\right)\label{eq:translAsym} \\
\uppi_{a}(t,\bmx) \rightarrow \uppi'_{a}(t,\bmx) &= \uppi_{a}(t,\bmx) + \epsilon_a\left[1+\dot{\uppi}_{r}(t,\bmx)\right] + \mathcal{O}\left(\epsilon_a^2\right). \label{eq:translAsym2}
\end{align} 
Unitary combinations of operators must leave $S_{\mathrm{eff}}$ invariant under the above transformation.

Let us illustrate this procedure with the kinetic terms $\dot{\uppi}_r \dot{\uppi}_a$ and $\partial_i\uppi_r \partial^i \uppi_a $ appearing in \Eq{eq:L2ref} that have been identified as being unitary through the comparison with the EFToI. Under the $\epsilon_{a}$ transformation \Eqs{eq:translAsym} and \eqref{eq:translAsym2}, we notice they lead to total derivatives. Hence, $S_{\mathrm{eff}}$ made of these terms is $\epsilon_a$-invariant, indicating they can be encountered in a unitary theory as one would expect. On the contrary, the quadratic dissipative term $\dot{\uppi}_r \uppi_a$ and diffusive terms $\uppi_a^2$, $\dot{\uppi}_a^2$ and $(\partial_i \uppi_a)^2$ are not invariant. Consequently, they have no unitary counterpart and represent genuine non-unitary effects.

For cubic interactions the effective action is invariant under $\epsilon_a$ only for the specific combinations identified in \Eqs{eq:combin1} and \eqref{eq:combin2}. Indeed, one can check the combinations $3\dot{\uppi}_r^2\dot{\uppi}_a + \dot{\uppi}_a^3/4$ and $(\partial_i \uppi_r)^2\dot{\uppi}_a + 2 \dot{\uppi}_r\partial_i \uppi_r \partial^i \uppi_a  +  (\partial_i \uppi_a)^2\dot{\uppi}_a/4$ are invariant. From these, one recovers the usual cubic interactions $\dot{\uppi}^3$ and $(\partial_i \uppi)^2\dot{\uppi}$ expressed in the Keldysh basis. Any deviation from these fine-tuned combinations would be associated to non-unitary dynamics. In particular, notice that unitary combinations only involve \textit{odd} powers of $\uppi_a$. Hence, any \textit{even} powers of $\uppi_a$ always relate to diffusive/noise processes \cite{Hongo:2018ant}. 


\paragraph{Unitarization procedure}

It interesting to understand if we can \textit{complete} an operator is order to obtain a unitary combination. A proposal to unitarize an operator is the following. Let us consider an operator $\mathcal{O}\left(\uppi_r,\uppi_a\right)$ made of field insertions of $\uppi_r$, $\uppi_a$ and their (at most single) derivatives. The combination of operators
\begin{align}\label{eq:unitarize}
\mathcal{U}\left[\mathcal{O}\left(\uppi_r,\uppi_a\right)\right] = \mathcal{O}\left(\frac{\uppi_+}{2},\uppi_+\right) + \mathcal{O}(\frac{\uppi_-}{2},-\uppi_-) - \mathcal{O}\left(\frac{\uppi_-}{2},\uppi_-\right) - \mathcal{O}(\frac{\uppi_+}{2},-\uppi_+)
\end{align}
is unitary by construction. For operators such as $\uppi_a^2$ (the diffusion operators) that are intrinsically non-unitary, the above combination vanishes. Also note that some operators such as $\dot{\uppi}_r \uppi_a$ (the dissipation operators) are unitarized into total derivatives and so do not add any net contribution to the open effective functional. Lastly, some of the contributions obtained out of \Eq{eq:unitarize} eventually violate symmetries of the problem (\eg the fact that $\uppi_r$ non-linearly realises time-translations and boosts) and must then be discarded.
This classification allows us to rewrite Wilsonian coefficients in terms of unitary and orthogonal directions. An interesting avenue would consist of exploring which non-unitary directions maximise the entangling dynamics between system and environment.    


\subsection{Setting up expectations}\label{subsec:energy}

The EFT coefficients are dimensionful quantities, such that $[\upalpha_0] =[\upalpha_1] = E^{4}$, $[\upgamma_1] = E^{5}$, $[\upbeta_1] = E^{6}$ and $[\upbeta_2] =[\upbeta_4] = E^{4}$. One can then ask what are the relevant scales controlling the physics and the regime of validity of our EFT description. 

\subsubsection{Energy scales and canonical normalisation} 

To treat the problems of the scales, we first canonically normalise the fields such that they have dimension of energy $E$. This canonical normalisation relates the original Wilsonian coefficients to quantities of physical interest such as the speed of sound $c_s$, the dissipation scale $\gamma$ or the fluctuations of the environment $\beta_1$. We first define the energy scale
\begin{align}
f_\pi^4 \equiv \upalpha_0 -2 \upalpha_1\,.
\end{align}
From this we construct the canonically normalised field
\begin{equation}\label{eq:can}
\pi \equiv f_\pi^2 \uppi.
\end{equation}
Notice the use of different Greek fonts to distinguish normalised and un-normalised fields. It follows that at leading order, the curvature perturbations are given by the relation
\begin{align}\label{eq:zetarel}
\zeta =  - \frac{H}{f_\pi^2} \pi.
\end{align}
In terms of the canonically normalised variables, the quadratic action takes the form
\begin{align}\label{eq:canonorm} 
&\quad S_{\mathrm{eff}}^{(2)} = \int \dd^4 x \Big\{ a^2 \pi'_{r} \pi'_{a} - c_{s}^{2} a^2 \partial_i \pir  \partial^i \pia  \\
-&  a^3 \gamma \pi'_{r} \pia + i \left[\beta_{1} a^4 \pia^2 - \left(\beta_2 - \beta_4\right) a^2\pia^{\prime 2} + \beta_2 a^2 \left(\partial_i \pia \right)^2\right] \Big\}. \nonumber
\end{align}
The rescaled coefficients are 
\begin{align}
c_s^{2} \equiv \frac{\upalpha_{0}}{f_\pi^4}\;,\quad\; \gamma \equiv
\frac{2\upgamma_{1}}{f_\pi^4}\;,\quad\;\beta_{i} \equiv \frac{\upbeta_{i}}{f_{\pi}^{4}}\;\quad\mathrm{for}\quad i = 1 ~\mathrm{to}~8.
\end{align}
The dimensions of the parameters appearing above are 
\begin{align}
[\pi] = E, \quad [f_\pi] = E, \quad [c_s] = E^0, \quad [\gamma] = E, \quad [\beta_1] = E^2, \quad [\beta_2] = [\beta_4] = E^0.
\end{align}
Expressed in terms of the canonically normalised variables, the cubic action becomes
\begin{align}\label{eq:canonormcub}
S_{\mathrm{eff}}^{(3)} =   \frac{1}{f_\pi^2} \int & \dd^4 x \Big\{\Big[4 \alpha_2 -  \frac{3}{2} (c^2_s-1) \Big]  a \pir^{\prime2} \pi'_{a} +\frac{1}{2} (c^2_s-1) a \left[\left(\partial_i \pir \right)^2 \pi'_a + 2 \pi'_{r}	\partial_i \pir  \partial^i \pia \right] \\
& \qquad \quad + \left(4 \gamma_2 - \frac{\gamma}{2}\right)  a^2\pir^{\prime2} \pia	+ \frac{\gamma}{2} a^2	\left(\partial_i \pir \right)^2 \pia \nonumber \Big. \\
&+ i \Big[\left(2\beta_7-\beta_3\right) a^2 \pi'_{r} \pi'_{a} \pia + \beta_3 a^2 \partial_i \pir  \partial^i \pia \pia + 2(\beta_4+ \beta_6 - \beta_8) a \pi'_{r} \pia^{\prime2} \Big. \nonumber \\
& \qquad \quad - 2 \beta_4  a \partial_i \pir  \partial^i \pia \pi'_{a} - 2\beta_5 a^3 \pir^{\prime} \pia^2  - 2 \beta_6 a \pir^{\prime} (\partial_i\pia)^2 \Big]  \Big. \nonumber\\
&+\delta_1 a^4 \pia^3 + (\delta_5- \delta_2) a^2 \pia^{\prime2} \pia  + \delta_2 a^2 (\partial_i \pia)^2 \pia - \delta_4 a  (\partial_i \pia)^2 \pi'_a + (\delta_4-\delta_6) a \pia^{\prime3} \Big\},  \nonumber
\end{align}
where we defined the rescaled coefficients\footnote{Notice that $\alpha_2$ can be related to the EFToI \cite{Cheung:2007st} parameter $M_3^4 = - f_\pi^4 \alpha_2$.}
\begin{align}
\alpha_2 \equiv \frac{\upalpha_2}{f_\pi^4}\;,\quad\; \gamma_2 \equiv \frac{\upgamma_{4}}{f_\pi^4} \;,\quad\; \delta_i \equiv \frac{\updelta_{i}}{f_\pi^4} \quad\mathrm{for}\quad i = 1 ~\mathrm{to}~6\,,
\end{align}
with dimensions
\begin{align}
[\alpha_2] = E^{0}, &\quad [\gamma_2] = E, \quad [\beta_6] =  [\beta_8] = E^0, \quad [\beta_3] = [\beta_7] = E, \quad [\beta_5] = E^2, \nonumber \\
& \qquad  [\delta_1] = E^3, \quad [\delta_5] = [\delta_2] = E, \quad [\delta_4] = [\delta_6] = E^0  . 
\end{align}
From now on, we work in this canonical basis and use the rescaled action given in \Eqs{eq:canonorm} and \eqref{eq:canonormcub}. 


\subsubsection{Heuristic estimates} 

We are now in the position to carry out a rough estimate of the non-Gaussianities sourced by the cubic operators of \Eq{eq:canonormcub}. For simplicity, we set $\beta_2 = \beta_4 = 0$ and focus on the leading noise term of \Eq{eq:canonorm} controlled by $\beta_1$. The estimate relies on the following rules:
\begin{itemize}
    \item We can approximate $\pir$ from the amplitude of the primordial power spectrum $\Delta^2_{\zeta}$, accounting for the canonical normalisation and the leading-order relation with $\zeta$ such that
    \begin{equation}\label{eq:rule1}
        \pir\sim\frac{f_{\pi}^{2}}{H}\Delta_{\zeta}.
    \end{equation}
    \item  We can estimate spatial derivatives by the spatial momenta. The value of spatial derivatives in different directions is, on average, the same by isotropy:
    \begin{equation}\label{eq:rule2}
        \partial_{i}\pi_{r,a}\sim k\pi_{r,a}.
    \end{equation}
    Adiabatic perturbations of momentum $k$ freeze at a scale factor $a_*$ that depends on the dissipation parameter $\gamma$ \cite{LopezNacir:2011kk}. We derive this freezing time by comparing the early and the late time limit of the power spectrum in \Sec{sec:Pk}. This leads to the relation
    \begin{equation}\label{eq:rule3}
        c_s k\sim a_* H \sqrt{\frac{H+\gamma}{H}},
    \end{equation}
    where we use the expression $(H+\gamma)$ as a shorthand reminder of a quantity that scales to leading order in $\gamma\to\infty$ as $\gamma$ and to leading order in $\gamma \to 0$ as $H$. While freezing still occurs at wavelengths around (sound) horizon crossing at low dissipation, it is displaced to sub (sound) horizon wavelength at large dissipation.
    \item The characteristic frequencies of the retarded $\pir$ and advanced $\pia$ components are estimated in \Sec{sec:Pk} to be
    \begin{equation}\label{eq:rule4}
        \pi_{r,a}'\sim a H \pi_{r,a}.
    \end{equation}
    The fact that a derivative acts similarly on $\pir$ and $\pia$ follows from consistency under integration by part, for instance considering
    \begin{align}
        a^2 \pir'\pia' = - a^2 \left[\pir'' + 2 a H \pir' \right]  \pia.
    \end{align}
    Similarly consistency under integration by parts should apply to all other interactions.
    \item The retarded component $\pir$ evolves according to a dynamics sourced by the advanced component $\pia$ and controlled by the environment noise $\beta_1$ \cite{kamenev_2011}. This sourced dynamics implies that $\pir$ and $\pia$ are not of the same amplitude, their ratio being controlled by
    \begin{equation}\label{eq:rule5}
        \frac{\pir}{\pia} \sim \frac{\beta_1}{H(H+\gamma)}.
    \end{equation}
    This relation can be obtained from evaluating the equation of motion for $\pir$ at $a=a_{*}$ as done in \Sec{sec:Pk}.
\end{itemize}

These prescriptions imply some hierarchies among the quadratic operators. Comparing the kinetic terms $a^2 \pir'\pia'$ and $c_s^2 a^2 \partial_i \pir \partial^i \pia$ and the linear dissipation $a^3\gamma\pir' \pia$ with the noise $a^{4} \beta_1 \pia^2$, we observe that
\begin{align}\label{eq:quad1}
    \frac{a^2 \pir'\pia'}{a^{4} \beta_1 \pia^2} \sim \frac{H}{H+\gamma}, \qquad  \qquad  \frac{c_s^2 a^2 \partial_i \pir \partial^i \pia}{a^{4} \beta_1 \pia^2} \sim 1, \qquad \qquad \frac{a^3\gamma\pir' \pia}{a^{4} \beta_1 \pia^2} \sim \frac{\gamma}{H+\gamma}.
\end{align}
This illustrates the dynamical regimes of a driven-dissipative harmonic oscillator. At low dissipation, $a^3\gamma\pir' \pia$ is negligible and the system is mostly controlled by the other three operators. At large dissipation, we enter the overdamped regime in which $a^2 \pir'\pia'$ becomes subdominant compared to the other contributions. 

To estimate the size of non-Gaussianities, we can first approximate the ratio between the cubic operators in \Eq{eq:canonormcub} and the dominant quadratic terms in \Eq{eq:canonorm}. Based on the above estimate, a choice of operator that is valid both at large and small dissipation is $a^{4} \beta_1 \pia^2$ (or equivalently $c_s^2 a^2 \partial_i \pir \partial^i \pia$), leading to
\begin{equation}
    f_{\text{NL}}\Delta_{\zeta}\sim\frac{\mathcal{L}_{3}}{a^{4} \beta_1 \pia^2}\,.
\end{equation}
Then we can compare our analytical prediction for the dependence on $\gamma$ of the amplitude of non-Gaussianities in the equilateral configuration with numerical results obtained in \Sec{sec:bispec}, see e.g. \Fig{fig:fnleq}. 

Like in the EFToI, one of the amplitudes of the unitary vertices in \Eq{eq:canonormcub} is controlled by the speed of sound $c_s$. The associated non-Gaussianities can be estimated through the above prescriptions, leading for instance to
\begin{equation}
	\mathcal{L}_{3} \supset \frac{(c^2_s-1)}{2f_{\pi}^{2}}  a \left(\partial_i \pir \right)^2 \pi'_a \quad \rightarrow \quad f_{\text{NL}}\sim \frac{(c^2_s-1)}{c_{s}^{2}} .
\end{equation}
This matches the usual expectation from \cite{Cheung:2007st}. We can then consider the two dissipative vertices $(\partial_{i}\pir)^{2}\pia$ and $\pir^{\prime 2}\pia$ controlled by $\gamma$ and related to the quadratic dissipation $\pir^{\prime}\pia$ through non-linearly realised boosts. Using the above prescriptions, we find for the first operator the linear-in-$\gamma$ scaling
\begin{equation}
    \mathcal{L}_{3} \supset \frac{\gamma}{f_{\pi}^{2}}a^2(\partial_{i}\pir)^{2}\pia \quad \rightarrow \quad f_{\text{NL}}\sim \frac{1}{c_s^2} \frac{\gamma}{H} ,
\end{equation}
in agreement with the results of \cite{LopezNacir:2011kk, Creminelli:2023aly}. The second vertex can be estimated by
\begin{equation}
    \mathcal{L}_{3} \supset \frac{\gamma}{f_{\pi}^{2}}a^2 \pir^{\prime 2}\pia \quad \rightarrow \quad f_{\text{NL}}\sim \frac{\gamma}{H+\gamma},
\end{equation}
which correctly reproduces the numerical results of \Fig{fig:fnleq}. Noise terms quadratic in $\pia$ can also be estimated in the same manner, leading for instance to
\begin{equation}
 \mathcal{L}_{3} \supset \frac{i\beta_{5}}{f_{\pi}^{2}}a^{3}\pir'\pia^{2} \quad \rightarrow \quad  f_{\text{NL}}\sim \frac{\beta_{5}}{\beta_{1}}.
\end{equation}
Noticeably, this ratio of the noise amplitudes $\beta_1$ and $\beta_5$ is independent of the dissipation parameter $\gamma$ and leads to approximately constant $f_{\text{NL}}$ for any value of $\gamma/H$, as seen in the specific model studied in \cite{Creminelli:2023aly} and further discussed in \Sec{sec:matching}.  At last, cubic operators in $\pia$ also source a bispectrum signal such as
 \begin{equation}\label{eq:5thestimate}
     \mathcal{L}_{3} \supset \frac{\delta_{1}}{f_{\pi}^{2}}a^{4}\pia^{3} \quad \rightarrow \quad f_{\text{NL}}\sim \frac{\delta_{1}}{\beta^2_{1}} (H+\gamma).
 \end{equation}
Again our estimates correctly reproduce the numerical results of \Fig{fig:fnleq} which plateaus at low dissipation and grows linearly at large dissipation for fixed values of $\beta_{1}$ (\textit{red curve} in \Fig{fig:fnleq}). As we discussed in \Sec{sec:matching}, the ratio of $\delta_{1}$ over $\beta_1^2$ controls the amplitude of the non-Gaussian noise compared to its Gaussian counterpart. Similar estimates can be obtained for all cubic operators of \Eq{eq:canonormcub}. This heuristic derivation correctly accounts for all contributions of \Fig{fig:fnleq}, hence providing a valuable insight to the rich physics of the open EFT of inflation.


\section{Power spectrum}\label{sec:Pk}

Once the open effective functional is known, we can ask how one can derive a new set of Feynman rules. To this end, first we need to solve the free theory, as it is the one that determines the propagators. In this section, we obtain the generating functional of the correlation functions of the free theory of the canonically normalised fields $\pia$ and $\pir$. We start from the quadratic action \eqref{eq:canonorm} and introduce sources $J_a$ and $J_r$. The path integral over $\pia$ and $\pir$ results in a Gaussian partition function $\mathcal{Z}[J_{a},J_{r}]$. The functional derivatives of $\mathcal{Z}[J_{a},J_{r}]$ yield the correlators between $\pia$ and $\pir$. Of special interest are the two-point functions $\langle\pir(\eta_{1})\pir(\eta_{2})\rangle$ and $\langle\pir(\eta_{1})\pia(\eta_{2})\rangle$. The former is dubbed the \textit{Keldysh propagator} and encodes fluctuations of the system mediated by the presence of the environment \cite{kamenev_2011}. The latter is the \textit{retarded Green function} of the system, which encodes the propagation of perturbations sourced by the environment.


\subsection{Generating functional}

The open effective functional in \Eq{eq:canonorm} can be integrated by parts to
\begin{align}
S_{\text{eff}}^{(2)} =\int d^{4}x\bigg\{&-\frac{1}{2}\left(a^{2}\pir'\right)' \pia-\frac{1}{2} \left(a^{2}\pia'\right)' \pir+\frac{a^2 c_s^2}{2} \left(\pia\partial_{i}^{2}\pir+\pir\partial_{i}^{2}\pia\right)\nonumber\\
&\qquad \qquad -\frac{a^3\gamma}{2}\pir'\pia+\frac{a^3\gamma}{2}\pia'\pir+\frac{3a^3\mathcal{H}\gamma}{2} \pia\pir\nonumber\\
&+i \left[a^4\beta_1  \pia^2 + \left(\beta_2 - \beta_4\right)\left(a^{2}\pia'\right)' - a^2\beta_2 \pia\partial_i^{2} \pia \right]\bigg\},\label{eq:S2effparts}
\end{align}
where we defined $\mathcal{H} \equiv a' / a$. This integration by parts can be done thanks to the boundary condition on $\pia$. Even though some terms now seem to break invariance under the non-linear transformation of $\pir$, there is a balance between coefficients such that the action is still invariant. The form of \Eq{eq:S2effparts} allows us to write the action as a bilinear on the fields
\begin{equation}
S_{\text{eff}}^{(2)} = -\frac{1}{2}\int d^{4}x \left(\pir,\pia\right)\begin{pmatrix}
0 & \widehat{D}_{A} \\ \widehat{D}_{R} & -2i\widehat{D}_{K}
\end{pmatrix}\begin{pmatrix}
\pir \\ \pia
\end{pmatrix},
\end{equation}
the matrix being a second order differential operator acting to the right, which is made of
\begin{align}
\widehat{D}_{A}& \equiv a^2\partial_{\eta}^{2}+ a^3 \left(\frac{2\mathcal{H}}{a}- \gamma\right)\partial_{\eta}-a^2c_{s}^{2} \partial_{i}^{2}- 3a^3 \mathcal{H} \gamma,\\
\widehat{D}_{R}&\equiv a^2\partial_{\eta}^{2}+a^3\left(\frac{2\mathcal{H}}{a}+ \gamma\right)\partial_{\eta}- a^2 c_s^2\partial_{i}^{2},\\
\widehat{D}_{K}&\equiv a^4\beta_{1}+a^2(\beta_{2}-\beta_{4})\left(\partial^{2}_{\eta}+2\mathcal{H}\partial_{\eta}\right)-a^2 \beta_{2}\partial_{i}^{2}.
\end{align}

The following path integral of the free theory computes the diagonal of the density matrix of the system
\begin{align}
\rho_{\mathrm{red}}\left[\pi, \pi;\eta_0\right] = &\int_{\Omega}^\pi \mathcal{D}\pir  \int_{\Omega}^{0}\mathcal{D}\pia \exp\left\{i  S_{\text{eff}}^{(2)}[\pir,\pia]\right\}.
\end{align}
Upon the introduction of sources, we obtain the generating function
\begin{align}
\mathcal{Z}[J_{r},J_{a}]=\int_{\Omega}^\pi \mathcal{D}\pir  \int_{\Omega}^{0}\mathcal{D}\pia \exp\Big\{-\frac{i}{2}  &\int d^{4}x \left(\pir,\pia\right)\begin{pmatrix}
0 & \widehat{D}_{A} \\ \widehat{D}_{R} & -2i\widehat{D}_{K}
\end{pmatrix}\begin{pmatrix}
\pir \\ \pia
\end{pmatrix} \nonumber\\
&+\int d^{4}x \left(J_{r}\pir+J_{a}\pia\right)\Big\}
\end{align}
Completing the square for $\pia$ and $\pir$ allows us to factorise the dependence on the sources $J_{a}$ and $J_{r}$. This is done by introducing a shift in the path integral
\begin{equation}
\begin{pmatrix}
\pir \\ \pia
\end{pmatrix}=\begin{pmatrix}
\Pi_{r} \\ \Pi_{a}
\end{pmatrix}+\int d^{4}y \begin{pmatrix}
A_{11}(x,y) & A_{12}(x,y) \\  A_{21}(x,y) & 0
\end{pmatrix}\begin{pmatrix}
\pir \\ \pia
\end{pmatrix}
\end{equation}   
Demanding that the terms linear in $\Pi_{r/a}$ vanish, we find\footnote{There exists a similar equation where we integrate by parts on $\widehat{D}_{A/R/K}$ which yields the same information.}
\begin{equation}
- \frac{i}{2} \begin{pmatrix}
0 & \widehat{D}_{A} \\ \widehat{D}_{R} & -2i\widehat{D}_{K}
\end{pmatrix}\begin{pmatrix}
A_{11}(x,y) & A_{12}(x,y) \\  A_{21}(x,y) & 0
\end{pmatrix}+\frac{1}{2}\begin{pmatrix}
\delta(x-y) & 0 \\  0 & \delta(x-y)
\end{pmatrix}=\begin{pmatrix}
0 & 0 \\  0 & 0
\end{pmatrix}\,,
\end{equation}
where the Dirac delta $\delta(x-y)$ is a tensor density. These equations can be rewritten in a covariant form such that it becomes straightforward to recognise the advanced and retarded propagators
\begin{align}
\widehat{D}_{R}(x)A_{12}(x,y)&=-i\delta(x-y)\;,\quad \;A_{12}(x,y)=-iG^{R}(x,y)\;,\quad \; G^{R}(x^{0}<y^{0})=0,\\
\widehat{D}_{A}(x)A_{21}(x,y)&=-i\delta(x-y)\;,\quad \;A_{21}(x,y)=-iG^{A}(x,y)\;,\quad \; G^{A}(x^{0}>y^{0})=0.
\end{align}
A fundamental property of the retarded and advanced propagator is that they are mapped to each other under the exchange of the arguments:
\begin{equation}\label{eq:Rxy_Ayx}
G^{R}(x,y)=G^{A}(y,x)\,.
\end{equation}
The Keldysh propagator is obtained from the matrix element $A_{11}(x,y)=-iG^{K}(x,y)$, which obeys the differential equation
\begin{equation}
\widehat{D}_{R}(x)A_{11}(x,y)=2\widehat{D}_{K}(x)G^{A}(x,y).
\end{equation}
Notice that $G^{K}$ is thus not a Green's function of some equation of motion. We choose to symmetrise in $x\leftrightarrow y$ such that 
\begin{align}\label{eq:Keldyshreff}
G^{K}(x,y)=&i\int d^{4}z \sqrt{-g(z)} G^{R}(x,z)\widehat{D}_{K}(z)G^{A}(z,y) \nonumber \\
&+i\int d^{4}z \sqrt{-g(z)} G^{R}(y,z)\widehat{D}_{K}(z)G^{A}(z,x)
\end{align}
where we have used the property \eqref{eq:Rxy_Ayx} to write $A_{11}(x,y)$ in the most symmetric way\footnote{The generalisation to non-local noises is straightforward from here by upgrading the local $\widehat{D}_{K}(z)$ to a non-local $\widehat{D}_{K}(z_{1}-z_{2})$, where now we have the integrand to be $G^{R}(x,z_{1})\widehat{D}_{K}(z_{1}-z_{2})G^{A}(z_{2},y)$ with the appropriate factors of the square root of the metric.}. Note that causality is implemented in a natural way as the retarded propagator requires $x^{0}>z^{0}$ and the advanced propagator requires $z^{0}<y^{0}$.

This leaves the partition function to be 
\begin{equation}
\mathcal{Z}[J_{r},J_{a}]= \exp\left\{ - \frac{i}{2}\int d^{4}x \int d^{4}y \left(J_{r}(x),J_{a}(x)\right)\begin{pmatrix}
G^{K}(x,y) & G^{R}(x,y) \\ G^{A}(x,y) & 0
\end{pmatrix}\begin{pmatrix}
J_{r}(y) \\ J_{a}(y)
\end{pmatrix}\right\}
\end{equation}
where we used the fact that $\mathcal{Z}[0,0] = 1$ in the closed time contour from trace normalisation \cite{kamenev_2011}. The two-point function reduces to
\begin{align}\label{eq:extract}
\langle&\pir(x_{1})\pir(x_{2})\rangle=\left.\frac{\delta^{2}}{\delta J_{r}(x_{1})\delta J_{r}(x_{2})}\mathcal{Z}[J_{a},J_{r}]\right|_{J_{r,a}=0}=-i G^K(x_{1},x_{2}).
\end{align}
Evaluating the two-point function at coincident times in Fourier space provides an expression for the power spectrum of the system
\begin{align}
\langle&\pir(\bfk, \eta_{0})\pir(\bfk', \eta_{0})\rangle=P^\pi_{k}(\eta_{0})(2\pi)^{3}\delta(\bfk+\bfk') \quad \mathrm{with} \quad P^\pi_{k}(\eta_{0}) = -i G^K(k; \eta_0,\eta_0).\bigg.
\end{align}      


\subsection{Flat space intuition}\label{subsec:M4Pk}

At linear order, the dissipative theory introduces new ingredients compared to the unitary case, which are the dissipation coefficient $\gamma$ and the three noises controlled by $\beta_1, \beta_2$ and $\beta_4$. Before discussing the inflationary case, it is instructive to get familiar with the formalism through some explicit flat space computation. In flat space, the open effective functional reads 
\begin{align}
S^{(2)}_{\mathrm{eff}} \left[\pir,\pia\right] &= \int \dd^4 x  \begin{pmatrix}
\pir & \pia
\end{pmatrix} \begin{pmatrix}
0 & -\frac{1}{2}\left(\partial_t^2 + \gamma \partial_t - c_s^2 \partial_i^2  \right)\\
-\frac{1}{2}\left(\partial_t^2 - \gamma \partial_t - c_s^2 \partial_i^2\right)  & i\left[\beta_{1}-(\beta_{2}-\beta_{4})\partial^{2}_{t}+\beta_{2}\partial_{i}^{2}\right]
\end{pmatrix} \begin{pmatrix}
\pir \\ \pia
\end{pmatrix}\,.
\end{align}
The equations of motion for the propagators read  
\begin{align}
\left(\partial_t^2 \pm \gamma \partial_t + c_s^2 k^2 \right) G^{R/A} (k; t_1, t_2) =  \delta(t_1-t_2)\,,			 
\end{align}
and, from \Eq{eq:Keldyshreff},
\begin{align}
G^K (k;t_1,t_2) = 2 i \int \dd t' G^{R} (k;t_1,t') \left[\beta_1- (\beta_{2}-\beta_{4}) \partial_{t'}^2 + \beta_2 k^2 \right] G^{A} (k;t',t_2).
\end{align}
The dynamics is easily solved in frequency space, where 
\begin{align}
G^{R/A} (k; \omega) &= - \frac{1}{\omega^2 \pm i \gamma \omega - c_s^2k^2}= -\frac{1}{(\omega_- - \omega_+)} \left[ \frac{1}{\omega- \omega_-} - \frac{1}{\omega- \omega_+}   \right]
\end{align}
with 
\begin{align}
\omega_\pm \equiv - i \frac{\gamma}{2} \pm E_k^{\gamma} \qquad \qquad \mathrm{and} \qquad \qquad 	E_k^{\gamma} \equiv \sqrt{c_s^2 k^2 - \frac{\gamma^2}{4}}.
\end{align}
Two comments are in order. First, notice that dissipation is incompatible with linearly realised Lorentz boosts. Instead, boosts are non-linearly realised only on $\pi_r$. Second, there is no need to introduce any $i\epsilon$ in the retarded-advanced propagators because the correct position of the poles is already determined by the non-vanishing dissipation. Related to this, in the presence of dissipation the retarded Green's function has only a finite memory of the past, decaying exponentially in time. As a consequence, time integrals involving $G^{R/A}$ already converge without the need to specify any small rotation into the complex plane.\\

We assume $E_k^{\gamma}$ to be real (\textit{damped regime}), keeping in mind that the \textit{overdamped regime} for which $E_k^{\gamma} \in i\mathbb{R}$ can be obtained by analytic continuation. Performing the temporal inverse Fourier transform, we obtain 
\begin{align}\label{eq:retardedM4dissip}
G^{R/A} (k; \tau) &= \int \frac{\dd \omega}{2\pi} \ee^{i \omega \tau} G^{R/A} (k; \omega) = \mp \frac{\sin \left[E_k^{\gamma} \tau\right]}{E_k^{\gamma}} \ee^{- \frac{\gamma}{2}\tau}\theta\left(\pm \tau\right).
\end{align}
where for $G^{R/A}(k;t_1,t_2)$ we defined $\tau \equiv t_1-t_2$. Notice that the presence of dissipation automatically ensures convergence, without the need for an $i\epsilon$. The Keldysh component is also easily obtained in frequency space where
\begin{align}
G^{K} (k; \omega) &= 2 G^{R} (k; \omega) \widehat{D}_K (k; \omega) G^{A} (k; \omega) = 2 i \frac{\beta_1+ (\beta_{2}-\beta_{4}) \omega^2 + \beta_2 k^2 }{\left(\omega^2-c_s^2k^2\right)^2+\gamma^2 \omega^2}.
\end{align}
One can analyse the pole structure and obtain the real-space propagator from 
\begin{align}
G^{K} (k; \tau) &= 2 i \int \frac{\dd \omega}{2\pi} \frac{\left[\beta_1+ (\beta_{2}-\beta_{4}) \omega^2 + \beta_2 k^2 \right] \ee^{-i\omega \tau}}{\prod_{i=1}^{4}(\omega-\omega_i)}
\end{align}
with the poles satsfying $\omega_2 = - \omega_1$, $\omega_3 = - \omega^*_1$, $\omega_4 = -\omega^*_1$ and 
\begin{align}
\omega_1 = E_k^{\gamma} + i \frac{\gamma}{2}.
\end{align}
A long but straightforward derivation leads to 
\begin{align}\label{eq:distribM4dissip}
G^{K} (k; \tau) &= i \frac{\ee^{-\frac{\gamma}{2}\tau}}{c_s^2 k^2} \bigg[\frac{2}{\gamma} \left[\beta_1 +c_s^2 k^2  (\beta_{2}-\beta_{4})+ k^2 \beta_2\right]\cos\left(E_k^{\gamma} \tau\right) \nonumber \\
+& \left[\beta_1 - c_s^2 k^2 (\beta_{2}-\beta_{4})+ k^2 \beta_2\right]\frac{\sin\left(E_k^{\gamma} \tau\right)}{E_k^{\gamma}}\bigg]
\end{align}
In the coincident time limit, we obtain the dissipative power spectrum
\begin{align}
P^\pi_k = \frac{2 \beta_1}{\gamma c_s^2k^2} + \frac{2 (\beta_{2}-\beta_{4})}{\gamma} + \frac{2\beta_2}{c_s^2\gamma}.
\end{align}
Note that the vacuum contribution to the power spectrum, the usual $P_k^{\pi}=1/2E_k^{\gamma = 0}$ term, is absent because of the exponential decay in time caused by dissipation. Nevertheless, with hindsight, one could still recover the standard vacuum Minkowski power spectrum by taking both $\beta_1$ and $\gamma$ to zero while keeping their ratio fixed. From the fact that the equal-time power spectrum must be non-negative we conclude that the $\beta$'s must be positive. This computation highlights several features of the formalism. While $G^R(k, \tau)$ encodes the dynamics but is oblivious to the state of the system, $G^K(k, \tau)$ captures the state of the environment by probing the statistics of fluctuations \cite{kamenev_2011}. The final outcome is an interplay between the dissipation of the system into its surrounding and the fluctuations of the environment getting imprinted onto the observable sector. Crucially, these two effects cannot be easily disentangle from one another. 


\subsection{Dissipative and noise-induced power spectrum}\label{subsec:dSPk}

Back to cosmology, we now extend the previous computation to account for the expanding spacetime. Embedded in the inflating background, the retarded and Keldysh differential operators become
\begin{align}\label{eq:DRexp}
\widehat{D}_{R} &= \frac{1}{H^2\eta^2} \left[\partial_\eta^2 - \frac{2 + \frac{\gamma}{H}}{\eta}\partial_\eta - c_s^2 \partial_i^2 \right]
\end{align}
and 
\begin{align}\label{eq:GKexp}
\widehat{D}_K &=\frac{1}{H^2\eta^2}\left[\frac{\beta_1}{H^2\eta^2} - \left(\beta_2 - \beta_4\right) \left(\partial_\eta^2 -\frac{2}{\eta} \partial_\eta\right) + \beta_2 \partial_i^2  \right].
\end{align}
From now on, we set $c_s = 1$ for simplicity and discuss the effect of $c_s$ at the end of the computation.


\paragraph{Scaling dimensions} Before deriving the propagators of the theory, let us briefly comment on the two homogeneous solutions of $\widehat{D}_{R}$ in Fourier space (mode functions). Obeying the dynamical equation 
\begin{align}
\left(\partial_{\eta}^2 - \frac{2 + \frac{\gamma}{H}}{\eta}\partial_{\eta} +  k^2  \right) \pi_k =0,
\end{align}
the two homogeneous solutions are given by 
\begin{align}
\pi_k(\eta) \propto \eta^{\nu_\gamma}  H^{(1)}_{\nu_\gamma}(- k \eta) \qquad \mathrm{and} \qquad \propto \eta^{\nu_\gamma} H^{(2)}_{\nu_\gamma}(- k \eta) \,,
\end{align}
where 
\begin{align}
    \nu_\gamma \equiv \frac{3}{2}+ \frac{\gamma}{2H}\,,
\end{align}
and $H^{(1)}$ and $H^{(2)}$ are Hankel functions. Selecting positive frequency mode functions of the form $\ee^{- i k \eta}$ in the asymptotic past, $- k \eta \gg 1$, one can safely discard one of the two solutions. At late times, where $- k \eta \ll 1$, the dissipative mode functions acquire scaling solutions
\begin{align}
\pi_k(z = - k \eta) &=  \mathcal{O}_+  z^{\Delta_+}  \left[1 +\mathcal{O}(z^2) \right] + \mathcal{O}_- z^{\Delta_-}  \left[1 +\mathcal{O}(z^2) \right]\,,
\end{align}
where $\mathcal{O}_+$ and $\mathcal{O}_-$ depend on $\gamma$ and $H$, and the scaling dimensions are
\begin{align}
\Delta_+ = 0 \qquad \mathrm{and} \qquad \Delta_- = 3 + \frac{\gamma}{H}.
\end{align}
Comparing this result to the well-known relation for closed systems $\Delta_+ + \Delta_-= d$, involving representations related by a shadow transform, we note that dissipation acts in the same way as a continuation to an non-integer number of dimensions. It would be interesting to investigate if this suggests a useful regularisation procedure for loop contributions, different from the one used in \cite{Senatore:2009cf,Melville:2021lst,Lee:2023jby}. These relations crucially depart from the unitary theory case. Yet, let us stress it would be misleading to derive the power spectrum simply by squaring these mode functions. Indeed, in an open theory, dissipation is inextricably linked to the fluctuations generated by the environment \cite{breuerTheoryOpenQuantum2002}. Instead, the power spectrum is derived as follow.

\paragraph{Retarded Green function} \Eq{eq:DRexp} determines the equations of motion obeyed by the Green function  
\begin{align}\label{eq:Greenfct}
\left(\partial_{\eta_1}^2 - \frac{2 + \frac{\gamma}{H}}{\eta_1}\partial_{\eta_1} + k^2  \right) G^{R} (k; \eta_1, \eta_2) = H^2 \eta_1^2 \delta(\eta_1-\eta_2).
\end{align}			
The normalisation is fixed by the continuity of the retarded Green function at coincident time and its first derivative discontinuity is controlled by the time-dependent prefactor of the $\partial_{\eta_1}^2$ term in \Eq{eq:Greenfct}, that is 
\begin{align}
G^R(k; \eta_2,\eta_2) &= 0, \qquad \mathrm{and} \qquad
\left. \partial_{\eta_1}G^R(k; \eta_1,\eta_2) \right|_{\eta_2 = \eta_1} = H^2 \eta^2_2.
\end{align}
This imposes
\begin{align}\label{eq:GRfinsimp}
G^R(k; \eta_1,\eta_2) =  \frac{\pi}{2} H^2 (\eta_1 \eta_2)^{\frac{3}{2}}\left(\frac{\eta_1}{\eta_2} \right)^{\frac{\gamma}{2H}}  \Ima\left[H^{(1)}_{\frac{3}{2}+ \frac{\gamma}{2H}}(-k \eta_1) H^{(2)}_{\frac{3}{2}+ \frac{\gamma}{2H}}(-k\eta_2)   \right] \theta(\eta_1-\eta_2) 
\end{align}
which is consistent with \cite{LopezNacir:2011kk, Creminelli:2023aly}. It turns out to be sometime convenient to express the retarded Green function in terms of Bessel functions of the first kind
\begin{align}\label{eq:GRfinsimpv2}
G^R(k;\eta_1,\eta_2) = \frac{\pi}{2} \frac{H^2}{k^3} \left(\frac{z_1}{z_2}\right)^{\nu_\gamma} z_2^{3}\left[Y_{\nu_\gamma}(z_1)J_{\nu_\gamma}(z_2) -J_{\nu_\gamma}(z_1)Y_{\nu_\gamma}(z_2) \right]
\end{align}
where $z_i \equiv - k \eta_i$. 

\paragraph{Keldysh propagator} We now turn our attention to the computation of the Keldysh propagator of the theory from which the observables such as the power spectrum can be derived.\footnote{Note that we do not deduce a \textit{noise-free} power spectrum out of the mode functions derived above contrarily to \cite{LopezNacir:2011kk, Creminelli:2023aly}. Even in non-equilibrium settings where the fluctuation-dissipation relations do not strictly apply, the two effects are indissociable \cite{Calzetta:2008iqa,Kamenev:2009jj}, which is captured in the formalism below.} 
The Keldysh function is obtained from 
\begin{align}
     G^K(k; \eta_1,\eta_2) &= i\int \dd \eta' G^R(k; \eta_1,\eta') \widehat{D}_K(\eta' ) G^A(k; \eta',\eta_2) + (\eta_1 \leftrightarrow \eta_2) \\
     &= i \int \dd \eta' G^R(k; \eta_1,\eta') \widehat{D}_K(\eta' ) G^R(k; \eta_2,\eta') + (\eta_1 \leftrightarrow \eta_2).
\end{align}
Notice that this equation has precisely the structure one would expect for the power spectrum of a field obeying the Langevin equation with source fluctuations possessing a power spectrum $\widehat{D}_K(\eta' )$, as we will see in \Sec{subsec:Langevin}. Injecting \Eq{eq:GKexp} into the above expressions, we obtain three contributions
 \begin{align}
     G^K_1(k; \eta_1,\eta_2) &= i \frac{\beta_1}{H^4}\int_{-\infty}^{\eta_2} \frac{\dd \eta'}{\eta^{\prime4}} G^R(k; \eta_1,\eta') G^R(k; \eta_2,\eta') + (\eta_1 \leftrightarrow \eta_2) \label{eq:GK1}\\
     G^K_2(k; \eta_1,\eta_2) &= i \frac{\beta_4 - \beta_2}{H^2}\int_{-\infty}^{\eta_2} \frac{\dd \eta'}{\eta^{\prime2}} G^R(k; \eta_1,\eta')\left( \partial_{\eta'}^2 - \frac{2}{\eta'}\partial_{\eta'}\right) G^R(k; \eta_2,\eta') + (\eta_1 \leftrightarrow \eta_2) \label{eq:GK2} \\
     G^K_3(k; \eta_1,\eta_2) &= i \frac{\beta_2 k^2}{H^2}\int_{-\infty}^{\eta_2} \frac{\dd \eta'}{\eta^{\prime2}} G^R(k; \eta_1,\eta') G^R(k; \eta_2,\eta') + (\eta_1 \leftrightarrow \eta_2) \,,\label{eq:GK3}
\end{align}
which correspond to the three noise terms appearing in \Eq{eq:GKexp}. Here we assumed $\eta_2 \leq \eta_1$ without loss of generality. The power spectrum is obtained from the coincident time limit $\eta_1 = \eta_2 = \eta_0$.

Let us compute the first contribution \Eq{eq:GK1}, which corresponds to the noise term of \cite{LopezNacir:2011kk, Creminelli:2023aly} as we will make explicit in \Sec{sec:matching}. Injecting \Eq{eq:GRfinsimpv2} in \Eq{eq:GK1}, we obtain the Keldysh propagator 
\begin{align}\label{eq:GKeq}
    G^K_1(k;\eta_1,\eta_2) = i\frac{\pi^2 \beta_1}{4 k^3}  (z_1 z_2)^{\nu_\gamma} \bigg\{& Y_{\nu_\gamma}(z_1) Y_{\nu_\gamma}(z_2) A^{(1)}_{\nu_\gamma}(z_2) + J_{\nu_\gamma}(z_1) J_{\nu_\gamma}(z_2) C^{(1)}_{\nu_\gamma}(z_2) \\
    -& \left[J_{\nu_\gamma}(z_1)Y_{\nu_\gamma}(z_2) + J_{\nu_\gamma}(z_2)Y_{\nu_\gamma}(z_1)\right]B^{(1)}_{\nu_\gamma}(z_2) \bigg\} + (1 \leftrightarrow 2)\nonumber
\end{align}
where 
\begin{align}
        A^{(1)}_{\nu_\gamma}(z) &\equiv \int_{z}^{\infty} \dd z' z^{\prime2 - 2{\nu_\gamma}} J^2_{\nu_\gamma}(z')  \label{eq:Fbetaref}\\
        B^{(1)}_{\nu_\gamma}(z) &\equiv \int_{z}^{\infty} \dd z' z^{\prime2 - 2{\nu_\gamma}} J_{\nu_\gamma}(z') Y_{\nu_\gamma}(z') \label{eq:Gbetaref} \\
        C^{(1)}_{\nu_\gamma}(z)&\equiv \int_{z}^{\infty} \dd z' z^{\prime2 - 2{\nu_\gamma}} Y^2_{\nu_\gamma}(z') \label{eq:Hbetaref}
\end{align} 
are complicated functions given explicitly in \Eqs{eq:Fbeta}, \eqref{eq:Gbeta} and \eqref{eq:Hbeta} respectively. 

\paragraph{Power spectrum} Considering that $\zeta = - H\pi/ f_\pi^2$ where $\pi$ is the canonically normalised field, the reduced power spectrum 
\begin{align}
\Delta^2_\zeta(k) \equiv \frac{k^3}{2\pi^2}P_\zeta(k) \qquad \mathrm{with} \qquad \langle \zeta_\bmk \zeta_{-\bmk}\rangle = (2\pi)^3 \delta(\bmk + \bmk') P_\zeta(k)
\end{align}
is obtained in the coincident time limit of the Keldysh propagator given in \Eq{eq:GKeq}, such that
\begin{align}\label{eq:PKref}
P_\zeta(k) = \frac{\pi^2\beta_1}{8 k^3}  \frac{H^2}{f_\pi^4} z^{2{\nu_\gamma}} \Big[& Y^2_{\nu_\gamma}(z)  A^{(1)}_{\nu_\gamma}(z) + J^2_{\nu_\gamma}(z) C^{(1)}_{\nu_\gamma}(z) -2J_{\nu_\gamma}(z)Y_{\nu_\gamma}(z) B^{(1)}_{\nu_\gamma}(z) \Big]\,.
\end{align}
In the super-Hubble regime $z\ll 1$, the power spectrum freezes and we recover the result from \cite{Creminelli:2023aly}, that is 
\begin{align}
\Delta^2_\zeta(k) &= \frac{1}{4} \frac{\beta_1}{H^2}   \frac{H^4}{f_\pi^4}  2^{2\nu_\gamma} \frac{\Gamma\left(\nu_\gamma-1\right)\Gamma\left(\nu_\gamma\right)^2}{\Gamma\left(\nu_\gamma - \frac{1}{2}\right)\Gamma\left(2\nu_\gamma - \frac{1}{2}\right)}\label{eq:PK1SH}
\end{align}
which is indeed dimensionless. Keeping in mind that $\nu_\gamma \equiv \frac{3}{2}+ \frac{\gamma}{2H}$, one can expand this result in the small and large dissipation regime leading to 
\begin{tcolorbox}[%
enhanced, 
breakable,
skin first=enhanced,
skin middle=enhanced,
skin last=enhanced,
before upper={\parindent15pt},
]{}

\paragraph{Dissipative power-spectrum}
 \begin{align}\label{eq:dissipPk}
    \Delta^2_\zeta(k)  &\propto  \begin{dcases}
 \frac{\beta_1}{H^2} \frac{H^4}{f_\pi^4}  + \mathcal{O}\left(\frac{\gamma}{H}\right), & \gamma \ll H,\\
 \frac{\beta_1}{H^2} \frac{H^4}{f_\pi^4}   \sqrt{\frac{H}{\gamma}} \left[1 + \mathcal{O}\left(\frac{H}{\gamma}\right) \right], & \gamma \gg H.
\end{dcases}
\end{align}
The observational constraint $\Delta^2_\zeta = 10^{-9}$ is easily obtained by imposing hierarchies between the various scales of the problem. Note that if one further imposes thermal equilibrium of the environment such that the fluctuation-dissipation relation holds, the dynamical KMS symmetry imposes  $\beta_1 =2\pi \gamma T$ where $T$ is the environment temperature \cite{Hongo:2018ant}, such that in the large dissipation regime ($\gamma \gg H$)
\begin{align}
    \Delta^2_\zeta \propto \frac{T}{H}\frac{H^4}{f_\pi^4} \sqrt{\frac{\gamma}{H}}
\end{align}
which reproduces the warm inflation expectation \cite{Berera:1995ie, Berera:1995wh, Berera:1999ws, Berera:2008ar, Ballesteros:2023dno, Montefalcone:2023pvh}.
\end{tcolorbox}

The two other contributions given in \Eqs{eq:GK2} and \eqref{eq:GK3} follow accordingly, the details of which are deferred in \App{app:Pk}. They also lead to equally valid scale invariant power spectra on super-Hubble scales
\begin{align}
   \Delta^2_\zeta(k) &\supset \begin{dcases}
\frac{15}{32}(\beta_4- \beta_2)\frac{H^4}{f_\pi^4} 2^{2\nu_\gamma} \frac{\Gamma\left(\nu_\gamma-2\right)\Gamma\left(\nu_\gamma\right)^2}{\Gamma\left(\nu_\gamma - \frac{3}{2}\right)\Gamma\left(2\nu_\gamma - \frac{1}{2}\right)}, & \text{\Eq{eq:GK2}},\\
\frac{3}{16} \beta_2\frac{H^4}{f_\pi^4} 2^{2\nu_\gamma} \frac{\Gamma\left(\nu_\gamma-2\right)\Gamma\left(\nu_\gamma\right)^2}{\Gamma\left(\nu_\gamma - \frac{3}{2}\right)\Gamma\left(2\nu_\gamma - \frac{3}{2}\right)}, & \text{\Eq{eq:GK3}},
\end{dcases}
\end{align}
which expand in the large dissipation limit ($\gamma \gg H$) to 
\begin{align}
   \Delta^2_\zeta(k) &\supset \begin{dcases}
 (\beta_4- \beta_2)  \frac{H^4}{f_\pi^4}  \sqrt{\frac{H}{\gamma}}  \left[1 + \mathcal{O}\left(\frac{H}{\gamma}\right) \right], & \text{\Eq{eq:GK2}},\\
\beta_2 \frac{H^4}{f_\pi^4}  \sqrt{\frac{\gamma}{H}} \left[1 + \mathcal{O}\left(\frac{H}{\gamma}\right) \right], & \text{\Eq{eq:GK3}}.
\end{dcases}
\end{align}
The obtained contributions may again satisfy the observational constraint $\Delta^2_\zeta = 10^{-9}$ by imposing some hierarchy between the various scales (more stringent for the last contribution due to the $\sqrt{\gamma/H}$ enhancement in the spatial derivative case).
\paragraph{Inclusion of the speed of sound} As we focused on the decoupling limit, we were able to choose units such that $c_{s}=1$. However, if we aim in the future to include coupling to gravity, or to look in details at non-linearly realised boosts, we need to include a speed of sound for the Goldstone mode that may differ from unity. The speed of sound will appear in various observables computed in this work. For the power spectrum we find       
\begin{align}
   \Delta^2_\zeta(k) &\supset \begin{dcases}
   \frac{1}{4 c_{s}^{3}} \frac{\beta_1}{H^2}   \frac{H^4}{f_\pi^4}  2^{2\nu_\gamma} \frac{\Gamma\left(\nu_\gamma-1\right)\Gamma\left(\nu_\gamma\right)^2}{\Gamma\left(\nu_\gamma - \frac{1}{2}\right)\Gamma\left(2\nu_\gamma - \frac{1}{2}\right)},\\
\frac{15}{32 c_{s}^{3}}(\beta_4- \beta_2)\frac{H^4}{f_\pi^4} 2^{2\nu_\gamma} \frac{\Gamma\left(\nu_\gamma-2\right)\Gamma\left(\nu_\gamma\right)^2}{\Gamma\left(\nu_\gamma - \frac{3}{2}\right)\Gamma\left(2\nu_\gamma - \frac{1}{2}\right)} & ,\\
\frac{3}{16c_{s}^{5}} \beta_2\frac{H^4}{f_\pi^4} 2^{2\nu_\gamma} \frac{\Gamma\left(\nu_\gamma-2\right)\Gamma\left(\nu_\gamma\right)^2}{\Gamma\left(\nu_\gamma - \frac{3}{2}\right)\Gamma\left(2\nu_\gamma - \frac{3}{2}\right)}, & 
\end{dcases}
\end{align}
the different powers on $c_{s}$ differentiate between time and spatial derivatives. The speed of sound also enters the discussion of horizon exit, as it takes $z=-k\eta$ to $z=-c_{s}k\eta$.

\paragraph{Imprint of the perturbations} Before closing the discussion on the power spectrum, let us further comment on the freezing of the adiabatic fluctuations on large scales. The early time behaviour of the power spectrum \eqref{eq:PKref} is given in the sub-Hubble regime $z\gg1$ by
  
\begin{align}\label{eq:subHPk1}
    P_\zeta(k;\eta) &=  \frac{1}{2k^3} \frac{\beta_{1}}{H^2} \frac{H^4}{f_\pi^4} \frac{z}{\nu_\gamma-1},
\end{align}
which is $\mathcal{O}(z)$ suppressed at early time compared to the usual closed system Bunch-Davies vacuum solution. In the heuristic estimates of \Sec{subsec:energy}, we made use of the fact that most of the perturbations are imprinted on the statistics around a particular number of e-foldings characterised by $z_{*} \equiv - c_{s}k\eta_{*} = c_s k /(a_* H)$. The characteristic scale of $z_{*}$ given in \Eq{eq:rule3}, can be derived  by matching the early and late time behaviours of the power spectrum given in \Eqs{eq:subHPk1} and \eqref{eq:PK1SH} respectively. From this analysis, we conclude the curvature perturbations freeze around
\begin{equation}
z_{*}=2^{2\nu_\gamma-2}\frac{\Gamma^{3}(\nu_\gamma)}{\Gamma(\nu_\gamma-\frac{1}{2})\Gamma(2\nu_\gamma-\frac{1}{2})} \approx \begin{dcases}
	1 & \gamma \ll H,\\
	\frac{\sqrt{\pi}}{2}\sqrt{\frac{\gamma}{H}} \quad & \gamma \gg H,
\end{dcases}
\end{equation}
summarized through
\begin{equation}\label{eq:zstar}
    z_{*}\sim\sqrt{\frac{\gamma+H}{H}}. 
\end{equation}
While the freezing still occurs around (sound) horizon crossing when dissipation is small, we recover the known fact that freezing occurs sooner in the large dissipation regime, on sub (sound) horizon scales \cite{LopezNacir:2011kk, Creminelli:2023aly}. 

Once $z_*$ is known, we can inject it in the equation of motion for the Goldstone mode $\pir$ to estimate the characteristic frequency\footnote{As discussed around \Eq{eq:rule4} in heuristic estimate of \Sec{subsec:energy}, consistency under integration by part implies that $\pir$ and $\pia$ have the same characteristic frequency.} and the difference in amplitude compared to the advanced component $\pia$. We consider the equation of motion for the retarded component $\pir$ obtained from the quadratic action in \Eq{eq:canonorm}, that is
\begin{align}
	\pir''+2aH\left(1+\frac{\gamma}{2H}\right)\pir'+c_{s}^{2}k^{2}\pir&=2ia^{2}\beta_{1}\pia.\label{eq:eompir}
\end{align}
\Eq{eq:eompir} is sourced term on the right-hand side by the environment noise controlled by $\beta_1$. Let us first estimate the characteristic frequency of the system, $\pi_{r}' \sim a\omega\pi_{r}$ where $\omega$ is to be determined from \Eq{eq:eompir}. We can exploit \Eq{eq:zstar} to estimate spatial derivatives at freezing time, leading to $c_s^2 k^2 \sim a^2 H (H+\gamma)$. It implies that 
\begin{equation}\label{eq:omegareom}
    \left[\omega^{2}+\left(3H+\gamma\right)\omega+H\left(H+\gamma\right)\right] \pir \sim 2i\beta_{1}\pia.
\end{equation}
At small dissipation, there is no doubt that $\omega$ is controlled by $H$ as one recovers the usual closed system dynamics (up to the peculiarity of being sourced by the noise $\beta_1$). At large dissipation, the interplay between the two characteristic timescales $H$ and $\gamma$ renders the estimate less straightforward. \Eq{eq:omegareom} describes a forced damped harmonic oscillator. In this situation, any solution has two contributions: a transient one, with a lifetime $T\sim1/\gamma$ after which it decays away and a stationary one, with a much lower characteristic frequency. In this case, the lowest frequency is given by the Hubble parameter $H$. Therefore, $\omega \sim H$ in both regimes, as stated in \Eq{eq:rule4}. Combining both estimates of $c_s^2 k^2$ and $\omega$ with \Eq{eq:eompir}, we finally obtain the ratio between $\pir$ and $\pia$ stated in \Eq{eq:rule5}, which completes the derivation of the heuristic rules used in \Sec{subsec:energy}. 


\section{Bispectrum}\label{sec:bispec}

Beyond the Gaussian statistics, higher-point functions are computed following the usual perturbative approach. Once the propagators are known, we can derive a new set of Feynman rules from which we construct correlators order by order in perturbation theory. In this section, after reviewing the standard in-in treatment of interactions, we study the structure of the bispectrum (three-point function in Fourier space) both in flat space where analytic results are easily obtained and in de Sitter, which exhibits a specific phenomenology. 


\subsection{Interactions}

\begin{figure}[tbp]
\centering
\includegraphics[width=0.5\textwidth]{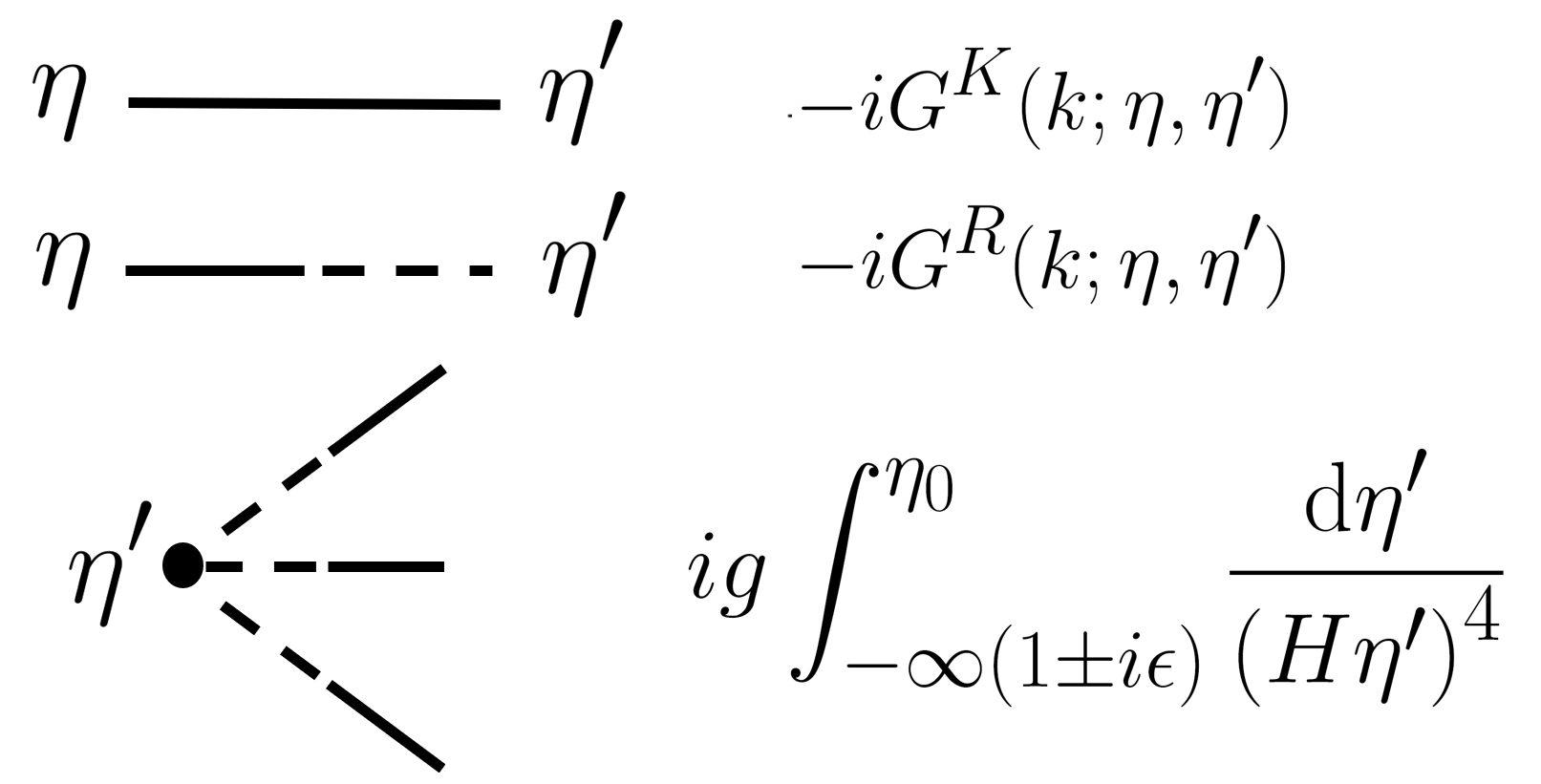}
\caption{Feynman rules in the Keldysh basis.}
\label{fig:rules}
\end{figure}

Interactions are treated as in the familiar in-in approach, see \cite{Chen:2017ryl} for a review. This provides a comforting unified treatment for the cases of an open and closed system. The only small difference from some references is that we find it convenient to work in the Keldysh basis, $\pi_{r,a}$ instead of the $\pi_\pm$ basis. Expectation values of $Q(\eta) \equiv \pi(\eta, \bmx_1) \cdots \pi(\eta, \bmx_n)$ are defined through 
\begin{align}\label{eq:expvalQ}
    \langle Q \rangle = \int \mathcal{D}\pir  \mathcal{D}\pia  \left[\pir(\eta, \bmx_1) \cdots \pir(\eta, \bmx_n) \right] \left.\ee^{i S_{\mathrm{eff}}\left[ \pir, \pia\right]}\right|_{\pia(\eta_0)=0}
\end{align}
where initial conditions lie on the boundaries of the path integral. Notice that since for cosmology we are only interested in the product of fields rather than their momentum conjugate, we are effectively only probing the diagonal of the density matrix prepared by the Schwinger-Keldysh path integral. Once the generating functional is known,  $\langle Q \rangle$ is extracted out of functional derivatives
\begin{align}
    \langle Q \rangle =  \left.\frac{\delta}{\delta J_{r}(\eta, \bmx_{1})} \cdots \frac{\delta}{\delta J_{r}(\eta, \bmx_{n})}\mathcal{Z}[J_{r},J_{a}]\right|_{J_{r,a}=0}
\end{align}
just as we did in \Eq{eq:extract} for the two-point function. The Feynman rules are derived as in \cite{Chen:2017ryl} and lead to \Fig{fig:rules}. There are two propagators that are $-iG^K(k; \eta, \eta')$ (continuous line) which connects $\pir(k;\eta)$ to $\pir(k;\eta')$ and $-iG^R(k; \eta, \eta')$ (continuous-to-dashed line), which connects $\pir(k;\eta)$ to $\pia(k;\eta')$. Notice that the latter is directional, with the continuous line being attached to the $\pir(k;\eta)$ insertion and the dashed part to the $\pia(k;\eta')$.\footnote{There is no propagator connecting $\pia(k;\eta)$ to $\pia(k;\eta')$ which is a consequence of the causality structure of the closed time contour \cite{kamenev_2011}.} If there is an operator $g \pi^m_r \pi^n_a$ appearing in $\mathcal{L}_{\mathrm{eff}}$, it leads to a $ i g$ contribution per vertex with $m$ continuous legs and $n$ dashed legs. Then, diagrams evaluation follows the exact same rules as in \cite{Chen:2017ryl}. As seen from \Eq{eq:expvalQ}, external legs connecting to the conformal boundary $\eta_0 \rightarrow 0^-$ are continuous, corresponding to $\pir$ insertions. An example is given in \Fig{fig:Btree} for a contact bispectrum. One can easily be convinced of these Feynman rules by recovering some known results in flat space as we do in \App{app:unit}. 

\subsection{Flat space intuition}\label{subsec:flatB}

Before turning our attention to primordial cosmology, it is instructive to discuss the generic structure of the contact three-point functions in Minkowski in the presence of dissipation. We derive these results following the Feynman rules enumerated in \Fig{fig:rules}, the propagators being given in \Eqs{eq:retardedM4dissip} and \eqref{eq:distribM4dissip}. We now need to consider both unitary and non-unitary operators presented in \Eqs{eq:L13}, \eqref{eq:L23} and \eqref{eq:L33}. For the sake of clarity, we consider the case where $\beta_2 = \beta_4 = 0$ but the results can be extended to accommodate for non-zero $\beta_2$ and $\beta_4$ if desired. We also set $c_s = 1$ for simplicity. In \App{app:unit}, we further discuss how to recover the unitary results in the absence of dissipation and noise. 

\paragraph{$\dot{\pi}^3$ interactions} Let us first consider 
\begin{align}\label{eq:cubicdot}
\mathcal{L}_{\mathrm{int}} = -\frac{\alpha}{3!}\left(\dot{\pi}^3_+ - \dot{\pi}^3_-\right) = -\frac{\alpha}{2} \left(\dot{\pi}_r^2 \dot{\pi}_a + \frac{1}{12} \dot{\pi}_a^3\right).
\end{align}
where the minus sign in front comes from the Lorentzian signature, assuming $\alpha>0$.\footnote{In the EFT construction given in \Eqs{eq:L13}, \eqref{eq:L23} and \eqref{eq:L33}, this case corresponds to $(4 \alpha_2 -3 \alpha_1)/f_\pi^6 = - \alpha/2$ and $(\delta_4 - \delta_6)/f_\pi^6 = - \alpha/24$ which indeed lies in the unitary direction. } In the unitary shift symmetric case, the operators $\dot{\pi}^3$ and $(\partial_i \pi)^2 \dot{\pi}$ do not generate any contact bispectrum in Minkowski due to the time reversal symmetry $t\rightarrow-t$ and $\pi \rightarrow -\pi$ (one can check explicitly that there is zero contribution to the bispectrum, each diagram vanishing independently). Dissipation spontaneously breaks this symmetry and we observe that the diagrams now lead to a non-zero contribution to the bispectrum. 

\begin{figure}[tbp]
\centering
\includegraphics[width=0.5\textwidth]{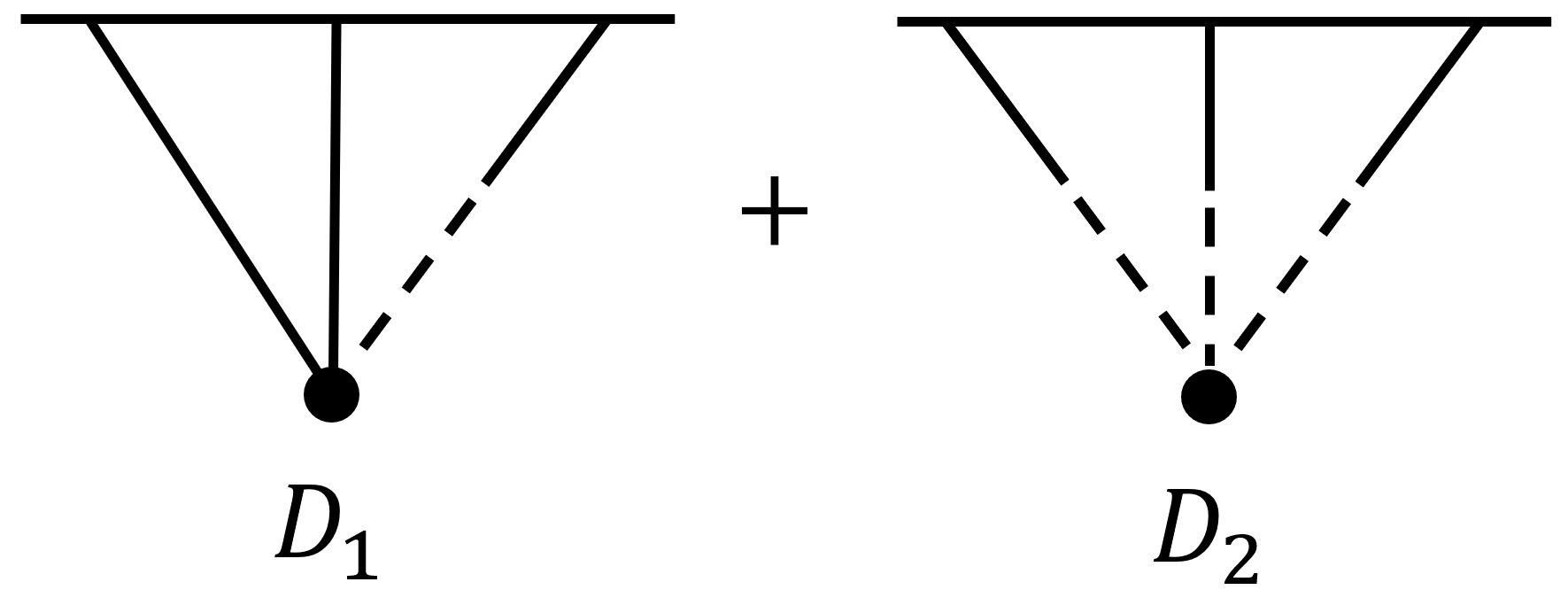}
\caption{The two diagrams to compute the interaction given in \Eq{eq:cubicdot}. \textit{Left}: diagram corresponding to the vertex $\dot{\pi}_r^2 \dot{\pi}_a$, resulting in \Eqs{eq:D1dot}. \textit{Right}: diagram corresponding to the vertex $\dot{\pi}_a^3$, resulting in \eqref{eq:D2dissip}.}
\label{fig:Btree}
\end{figure} 		

Let us consider the bispectrum 
\begin{align}\label{eq:bispecdef}
\langle \pi_{\bmk_1}  \pi_{\bmk_2}  \pi_{\bmk_3} \rangle \equiv (2\pi)^3 \delta(\bmk_1 + \bmk_2 + \bmk_3) B_\pi(k_1,k_2,k_3).
\end{align}
We have two contact diagrams to consider presented in \Fig{fig:Btree}, $B_\pi(k_1,k_2,k_3) = D^{\dot{\pi}^3}_1 + D^{\dot{\pi}^3}_2$. The first one leads to
\begin{align}\label{eq:D1dot}
D^{\dot{\pi}^3}_1 &=  \frac{\alpha}{2} \int_{-\infty(1\pm i \epsilon)}^{t_0} \dd t \left[\partial_t G^K(k_1; t_0,t)\right] \left[\partial_t G^K(k_2; t_0,t)\right] \left[\partial_tG^R(k_3; t_0,t)\right] + 5~\mathrm{perms}. 
\end{align}
Upon injecting \Eqs{eq:retardedM4dissip} and \eqref{eq:distribM4dissip} in \Eq{eq:D1dot} and  summing over the six possible permutations, we observe that $D^{\dot{\pi}^3}_1 = 0$. The second diagram is made of the $\pia$ components only, and reads 
\begin{align}\label{eq:D2dissip}
D^{\dot{\pi}^3}_2 = \frac{\alpha}{24} \int_{-\infty(1\pm i \epsilon)}^{t_0} \dd t \left[\partial_t G^R(k_1; t_0,t) \right] \left[\partial_t G^R(k_2; t_0,t)\right] \left[ \partial_t G^R(k_3; t_0,t)\right] + 5~\mathrm{perms}.
\end{align}
Under the same procedure, this diagram leads to a non-zero contribution to the bispectrum such that
\begin{align}\label{eq:D2resultdissip}
 B_\pi(k_1,k_2,k_3) &=  \frac{\alpha \gamma}{2} \frac{  \mathrm{Poly}_6\left(e_1^\gamma, e_2^\gamma,e_3^\gamma \right)}{ \mathrm{Sing}_\gamma} ,
\end{align}
where we defined the singularity structure
\begin{align}\label{eq:singdissip}
 \mathrm{Sing}_\gamma =& \left| E_1^{\gamma} + E_2^{\gamma} + E_3^{\gamma} + \frac{3}{2}i \gamma\right|^2 \left| -E_1^{\gamma} + E_2^{\gamma} + E_3^{\gamma} + \frac{3}{2}i \gamma\right|^2 \nonumber\\
 &\times \left| E_1^{\gamma} - E_2^{\gamma} + E_3^{\gamma} + \frac{3}{2}i \gamma\right|^2 \left| E_1^{\gamma} + E_2^{\gamma} - E_3^{\gamma} + \frac{3}{2}i \gamma\right|^2, 
\end{align}
remembering that $E_k^{\gamma} = \sqrt{c_s^2 k^2 - \gamma^2/4}$. This singularity structure captures most of the specificities of the non-unitary dynamics. It emerges from time integrals of the form
\begin{align}\label{eq:M4int}
\int_{-\infty(1\pm i \epsilon)}^{t_0} \dd t \ee^{ \pm i E_1^{\gamma} (t_0 - t) } \ee^{\pm i E_2^{\gamma} (t_0 - t) } \ee^{\pm i E_3^{\gamma} (t_0 - t) } \ee^{- \frac{3}{2}\gamma(t_0-t)}\,,
\end{align}
which follow from the structure of the propagators given in \Eqs{eq:retardedM4dissip} and \eqref{eq:distribM4dissip}.
Physically, it represents $3 \leftrightarrow 0$ and $2 \leftrightarrow 1$ interactions mediated by the environment. Fluctuations generate folded singularities while dissipation displaces the pole and regularises the divergence. Consequently, the singularity is not located in the physical plane and the bispectrum remains under perturbative control over the whole kinematical space. As we will see below, this singularity structure is generic and does not depend on the details on the interactions. The details of the particular interaction are imprinted into $\mathrm{Poly}_n$, which is a $n^{\mathrm{th}}$-order polynomial of the elementary symmetric polynomials
\begin{align}\label{eq:var}
e_1^\gamma = E_1^{\gamma} + E_2^{\gamma} + E_3^{\gamma}, \quad e_2^\gamma = E_1^{\gamma} E_2^{\gamma} + E_2^{\gamma}  E_3^{\gamma} + E_1^{\gamma} E_3^{\gamma} \quad  e_3^\gamma = E_1^{\gamma} E_2^{\gamma} E_3^{\gamma}.
\end{align}
For this specific case, 
\begin{align}
\mathrm{Poly}_6\left(e_1^\gamma, e_2^\gamma,e_3^\gamma \right) &= 243\gamma^6 - 792 \gamma^4 e_2^\gamma + 396 \gamma^4 \left(e_1^\gamma\right)^2 + 576 \gamma^2 \left(e_2^\gamma\right)^2 -1088 \gamma^2 e_2^\gamma\left(e_1^\gamma\right)^2 \nonumber \\
&+272 \gamma^2 \left(e_1^\gamma\right)^4 + 1024 \gamma^2e_1^\gamma e_3^\gamma + 512 \left(e_1^\gamma\right)^2 \left(e_2^\gamma\right)^2 - 384 \left(e_1^\gamma\right)^4 e_2^\gamma \\
&-1024 e_1^\gamma e_2^\gamma e_3^\gamma + 64 \left(e_1^\gamma\right)^6 + 512 e_3^\gamma \left(e_1^\gamma\right)^3 + 768 \left(e_3^\gamma\right)^2. \nonumber
\end{align}

\begin{figure}[tbp]
\centering
\includegraphics[width=0.7\textwidth]{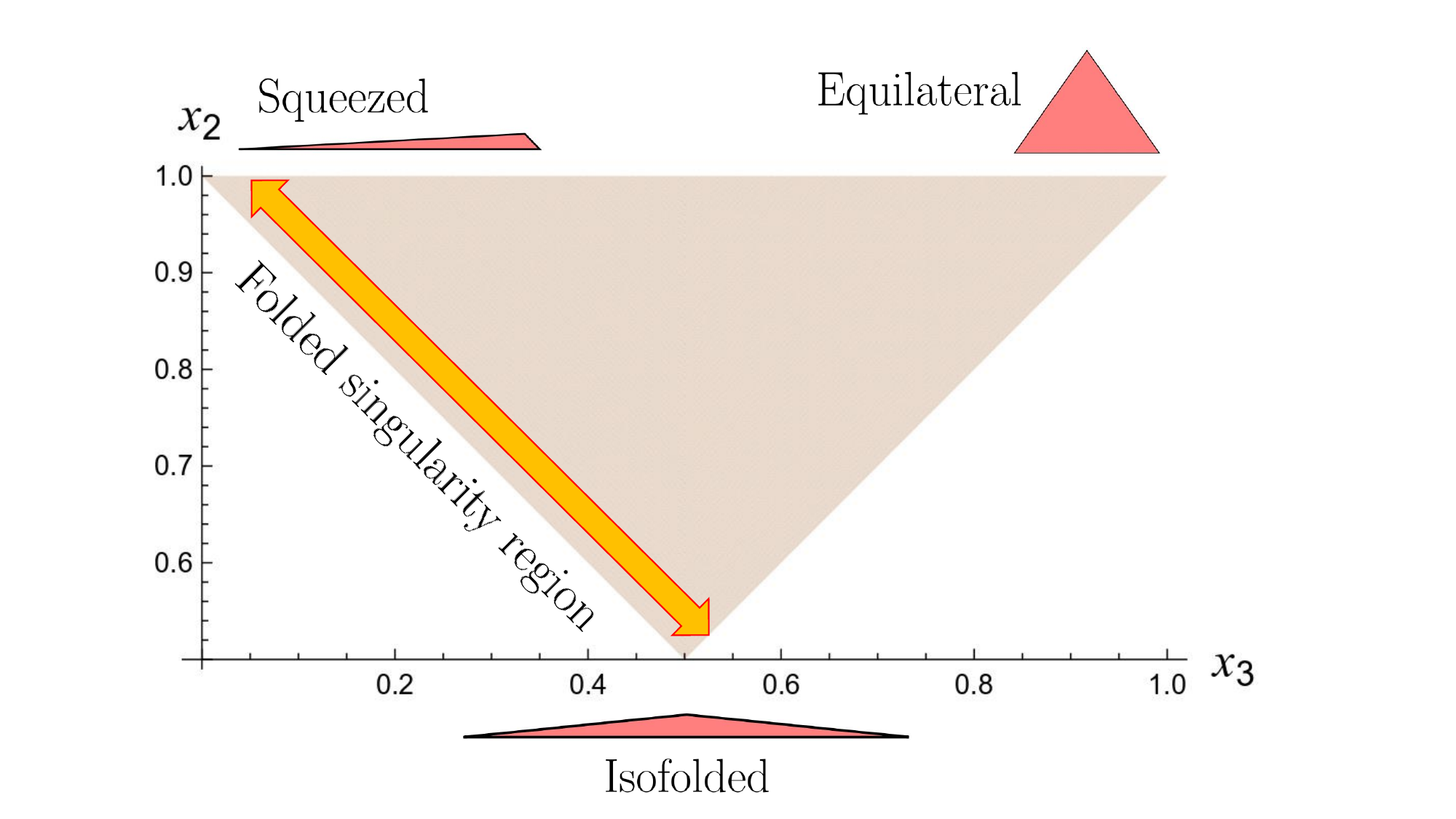}
\caption{The shapes of accessible triangles fulfilling the momentum conservation $\delta(\bmk_1 + \bmk_2 + \bmk_3)$. The triangles are parameterised along $x_2 \equiv k_2/k_1$ and $x_3 \equiv k_3/k_1$ such that $x_3 < x_2 < 1$ and $x_2 + x_3 >1$, the two conditions to construct closed triangles. A region of particular interest for the scope of this article is the folded region where $x_2 + x_3 \simeq 1$, which interpolates between the squeezed and isofolded points. The singularity structure discussed in \Eq{eq:singdissip} peaks close to the folded region, and provides a smoking gun of dissipative dynamics.}
\label{fig:triangles}
\end{figure} 

\begin{figure}[tbp]
\centering 
\makebox[\textwidth][c]{\includegraphics[width=.5\textwidth]{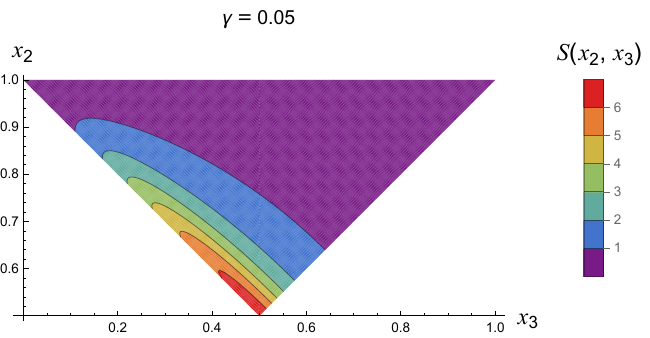}
    \includegraphics[width=.5\textwidth]{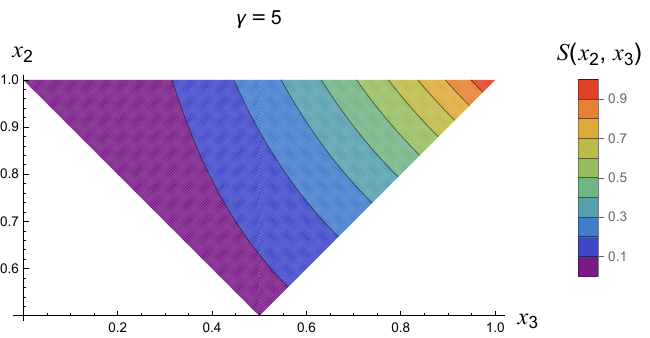}}
\caption{\label{fig:shapeMink} Shape function of the contact bispectrum generated by $\dot{\pi}_a^3$ in Minkowski given in \Eq{eq:D2resultdissip}. \textit{Left:} In the small dissipation regime, the singularity structure $\mathrm{Sing}_\gamma$ given in \Eq{eq:singdissip} becomes small in the folded region $x_2 + x_3 \simeq 1$ which enhances the bispectrum. \textit{Right:} In large dissipation regime, the imaginary contributions from the dissipation in dominates $\mathrm{Sing}_\gamma$ such that no bispectrum enhancement is observed in the $x_2 + x_3 \simeq 1$ folded region.}
\end{figure}

It is instructive to consider the shape of the bispectrum generated by this interaction. The momentum conservation in $\delta(\bmk_1 + \bmk_2 + \bmk_3)$ in \Eq{eq:bispecdef} forces the momenta to form a close triangle. As different inflationary models predict maximal signals in different triangular configurations (see e.g. \cite{Babich:2004gb, Chen:2006nt}), the shape function
\begin{align}\label{eq:shaperef}
S(x_2,x_3) \equiv (x_2 x_3)^2 \frac{B(k_1, x_2 k_1, x_3 k_1)}{B(k_1,k_1,k_1)}
\end{align}
is an informative probe of the mechanism generating primordial non-Gaussianities. The variables $x_2 \equiv k_2/k_1$ and $x_3 \equiv k_3/k_1$ control the shape of the triangles and are restricted by $\delta(\bmk_1 + \bmk_2 + \bmk_3)$ to the region $\max(x_3, 1- x_3) \leq x_2 \leq 1$. In \Fig{fig:triangles}, we present the main shapes of interest in this article. It appears that the singularity structure $\mathrm{Sing}_\gamma$ presented in \Eq{eq:singdissip} exhibits two different behaviour depending the magnitude of the dissipation coefficient $\gamma$. In the strong dissipation regime (\textit{right} panel of \Fig{fig:shapeMink}), the $ \frac{3}{2}i \gamma$ appearing in \Eq{eq:singdissip} always dominates the bispectrum contribution such that the signal peaks in the equilateral shape where $x_2 \simeq x_3 \simeq 1$. On the contrary, in the small dissipation regime, $\mathrm{Sing}_\gamma$ can become small in the folded region where $x_2 + x_3 \simeq 1$ such that the signal predominantly peaks near the isofolded configuration where $x_2 \simeq x_3 \simeq 1/2$ (\textit{left} panel of \Fig{fig:shapeMink}).

\begin{figure}[tbp]
\begin{minipage}{6in}
    \centering
    \raisebox{-0.5\height}{\includegraphics[width=.45\textwidth]{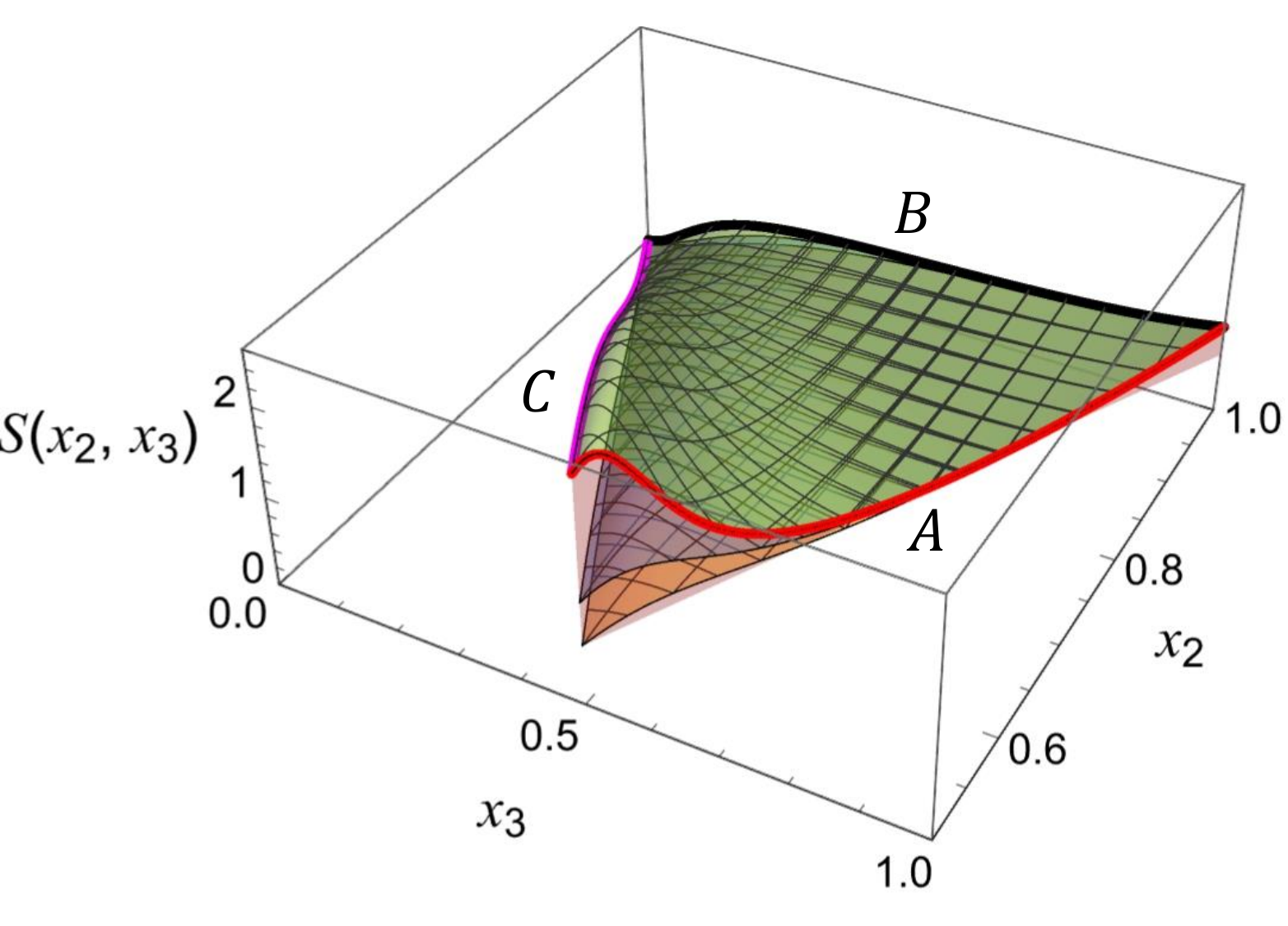}}
    \raisebox{-0.5\height}{\includegraphics[width=.45\textwidth]{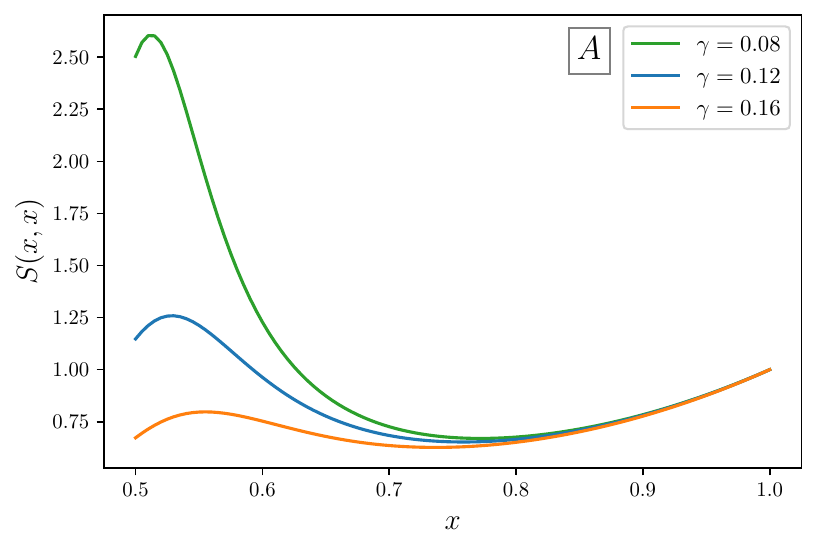}}
        \centering
    \raisebox{-0.5\height}{\includegraphics[width=.45\textwidth]{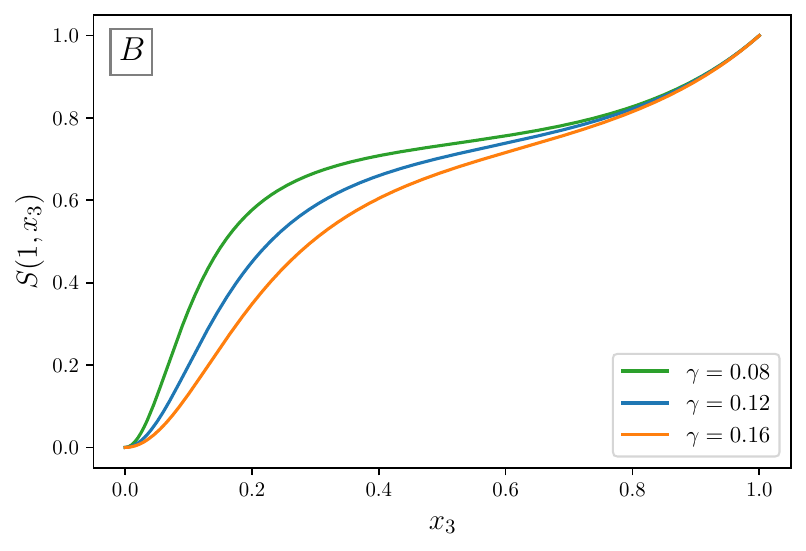}}
    \raisebox{-0.5\height}{\includegraphics[width=.45\textwidth]{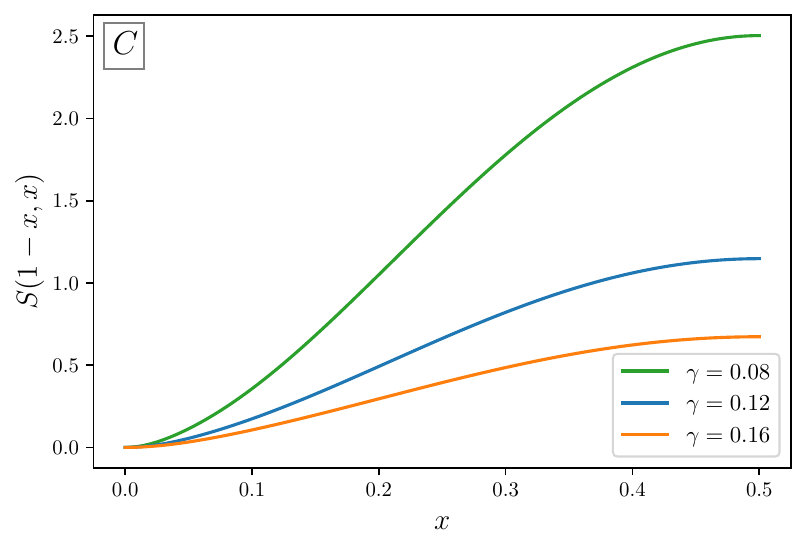}}
\end{minipage}
\caption{\label{fig:foldedMink} \textit{Top left:} $3\mathrm{d}$ shape function of the contact bispectrum generated by $\dot{\pi}_a^3$ in Minkowski given in \Eq{eq:D2resultdissip} for three different values of the dissipation parameter $\gamma \in [0.08; 0.12; 0.16]$. We observe the equilateral-to-folded transition of the shape function as the dissipation parameter decreases. \textit{Top right:} $2\mathrm{d}$ cut along the direction $x_2 = x_3 = x$ appearing in red in the $3\mathrm{d}$ plot. The singularity is resolved such that the bispectrum remains well defined for any triangular configuration and any value of the dissipation parameter $\gamma$. \textit{Bottom left:} $2\mathrm{d}$ cut along the direction $x_2 = 1$ appearing in black in the $3\mathrm{d}$ plot. Consistency relations ensure the signal vanishes in the squeezed limit $x_3 \ll 1$. \textit{Bottom right:} $2\mathrm{d}$ cut along the direction $x_2 = 1 - x_3$ appearing in purple in the $3\mathrm{d}$ plot. Consistency relations are again observed in the squeezed limit $x_3 \ll 1$.}
\end{figure}

This type of singularities have already been encountered in cosmology, mostly in the context of non-Bunch-Davies initial states \cite{Holman:2007na, Chen:2006nt, Meerburg:2009ys, Agullo:2010ws, Ashoorioon:2010xg, Agarwal:2012mq, Ashoorioon:2013eia,  Albrecht:2014aga, Green:2020whw}. The main difference with the current investigation is that, due to the presence of the dissipative environment, the would-be folded singularity is regularised, \ie $\mathrm{Sing}_\gamma \neq 0$ whenever $\gamma \neq 0$. This clearly appears in \Fig{fig:foldedMink} where the peak of the shape function as one approaches the folded singularity $x_2 + x_3 \simeq 1$ is plotted on the \textit{top-right} panel. The resolution of the singularity is a useful feature of the formalism as it allows one to keep perturbative control over all configurations. In particular, one does not have to introduce an artificial cutoff to handle the dissipative interactions. 

\paragraph{Other interactions} Other interactions follow the same structure. In Table \ref{tab:cubicdissip}, we summarize the non-vanishing contact bispectra generated from the cubic interactions presented in \Eqs{eq:L13}, \eqref{eq:L23} and \eqref{eq:L33}.\footnote{Note that while $\left(\partial_i \pir \right)^2 \dot{\pi}_a $ and $\dot{\pi}_{r}	\partial_i \pir  \partial^i \pia$ lead to non-vanishing contributions of the generic form given in \Eq{eq:gen}, the combination $\alpha_1 \left(\partial_i \pir \right)^2 \dot{\pi}_a + 2 \alpha_1  \dot{\pi}_{r}	\partial_i \pir  \partial^i \pia$ appearing in \Eq{eq:L13} leads to a vanishing bispectrum, the two contributions perfectly cancelling.} The generic structure is the following
\begin{tcolorbox}[%
enhanced, 
breakable,
skin first=enhanced,
skin middle=enhanced,
skin last=enhanced,
before upper={\parindent15pt},
]{}

\paragraph{Dissipative bispectrum in Minkowski}
 \begin{align}\label{eq:gen}
    B_\pi(k_1,k_2,k_3) =f(\mathrm{EFT}) \frac{  \mathrm{Poly}_n\left(e_1^\gamma, e_2^\gamma,e_3^\gamma \right)}{ \mathrm{Sing}_\gamma} 
\end{align}
where $f(\mathrm{EFT})$ a rational function of the EFT coefficients (and possibly the kinematics for spatial derivative interactions), $\mathrm{Poly}_n$ are polynomials of the variables given in \Eq{eq:var} and $\mathrm{Sing}_\gamma$ is the singularity structure expressed in \Eq{eq:singdissip}.

\end{tcolorbox}

   The simplicity of the structure, which originates from integrals of the form of \Eq{eq:M4int}, might suggest the future development of bootstrap techniques for this kind of local dissipative dynamics. It also suggests the physics is well captured from the interpretation of $\mathrm{Sing}_\gamma$ controlling the amplitude of $3 \leftrightarrow 0$ and $2 \leftrightarrow 1$ interactions mediated by the environment.

\begin{table}
\centering
\begin{tabular}{cc}
     $\left(4 \gamma_2 -\frac{\gamma}{2} \right)  \dot{\pi}^2_{r} \pia$ & $B_\pi =  -  \frac{3\left(8 \gamma_2 - \gamma  \right)}{2\gamma^2} \frac{\beta_1^2}{f_\pi^6 } \frac{  \mathrm{Poly}_4\left(e_1^\gamma, e_2^\gamma,e_3^\gamma \right)}{ \mathrm{Sing}_\gamma} \Bigg.$ \\
    $\frac{\gamma}{2} 	\left(\partial_i \pir \right)^2 \pia $ & $B_\pi = - \frac{8\gamma}{\gamma^2} \frac{\beta_1^2}{f_\pi^6} \frac{K^2}{ k_1^2 k_2^2 k_3^2} \Bigg.$\\
    $i\beta_3 \partial_i \pir  \partial^i \pia \pia$ & $B_\pi = \frac{\beta_3}{16} \frac{\beta_1}{f_\pi^6}  \frac{K^2}{k_1^2 k_2^2 k_3^2} \frac{  \mathrm{Poly}_8\left(e_1^\gamma, e_2^\gamma,e_3^\gamma \right)}{ \mathrm{Sing}_\gamma}  \Bigg.$ \\
    $2 i\left(\beta_4 + \beta_6 - \beta_8 \right) \dot{\pi}_{r} \dot{\pi}_a^2$ & $B_\pi =   \frac{\beta_4 + \beta_6 - \beta_8}{8\gamma}  \frac{\beta_1}{f_\pi^6} \frac{  \mathrm{Poly}_6\left(e_1^\gamma, e_2^\gamma,e_3^\gamma \right)}{ \mathrm{Sing}_\gamma}  \Bigg.$  \\
    $- 2i \left[\beta_4   \partial_i \pir  \partial^i \pia \dot{\pi}_{a} +  \beta_5   \dot{\pi}_{r}  \pia^2 + \beta_6   \dot{\pi}_{r} (\partial_i \pia)^2\right]$ & $B_\pi = \frac{3\left[\left(2\beta_6 - \beta_4\right)K^2 - \beta_5\right] }{\gamma} \frac{\beta_1}{f_\pi^6}  \frac{  \mathrm{Poly}_4\left(e_1^\gamma, e_2^\gamma,e_3^\gamma \right)}{ \mathrm{Sing}_\gamma}  \Bigg.$  \\ 
    $\delta_1 \pi^3_a + \delta_2 (\partial_i \pia)^2\pia$ & $B_\pi = - \frac{3}{4} \frac{\delta_1 - \delta_2 K^2}{f_\pi^6} \frac{  \mathrm{Poly}_4\left(e_1^\gamma, e_2^\gamma,e_3^\gamma \right)}{ \mathrm{Sing}_\gamma} \Bigg.$\\
    $\left(\delta_5-\delta_2 \right)\dot{\pi}_a^2 \pia $ & $B_\pi = \frac{1}{32} \frac{\delta_2-\delta_5}{f_\pi^6} \frac{  \mathrm{Poly}_6\left(e_1^\gamma, e_2^\gamma,e_3^\gamma \right)}{ \mathrm{Sing}_\gamma} \Bigg.$\\
    $\left(\delta_4-\delta_6 \right)\dot{\pi}_a^3$ & $B_\pi = \frac{3\gamma}{16} \frac{\delta_6-\delta_4}{f_\pi^6} \frac{  \mathrm{Poly}_6\left(e_1^\gamma, e_2^\gamma,e_3^\gamma \right)}{ \mathrm{Sing}_\gamma} \Bigg.$\\
\end{tabular}
\caption{Non-vanishing contributions to the contact bispectrum from the cubic interactions of \Eqs{eq:L13}, \eqref{eq:L23} and \eqref{eq:L33}. $\mathrm{Sing}_\gamma$ is given in \Eq{eq:singdissip} and we defined $K^2 \equiv (\bmk_1.\bmk_2 + \bmk_1.\bmk_3+ \bmk_2.\bmk_3) $. The $\mathrm{Poly}_n$ appearing in the expressions are polynomials of the variables given in \Eq{eq:var} whose details are not given for readability reasons.}
\label{tab:cubicdissip}
\end{table}

\subsection{Non-Gaussianities from dissipation and noises}\label{subsec:bispecdS}

Now that we have gained insight on the properties of dissipative dynamics compared to its unitary counterpart, it is time to apply these techniques to primordial cosmology. Remembering that the curvature perturbations are defined though $\zeta = - H \pi/f_\pi^2$ in the spatially flat gauge, the primordial bispectrum reads 
\begin{align}
    \langle \zeta_{\bmk_1}  \zeta_{\bmk_2}  \zeta_{\bmk_3} \rangle  = - \frac{H^3}{f_\pi^6}	\langle \pi_{\bmk_1}  \pi_{\bmk_2}  \pi_{\bmk_3} \rangle \equiv (2\pi)^3 \delta(\bmk_1 + \bmk_2 + \bmk_3) B(k_1,k_2,k_3).
\end{align} 
Quantities of interest to characterise the non-Gaussian signatures are the amplitude of the signal
\begin{align}
    f_{\mathrm{NL}}(k_1,k_2,k_3) \equiv \frac{5}{6} \frac{B(k_1,k_2,k_3)}{P(k_1) P(k_2) +  P(k_1) P(k_3)  + P(k_2) P(k_3) }\,,
\end{align}
discussed for specific configurations below, and the shape function, already defined in \Eq{eq:shaperef} which we reproduce here for clarity 
\begin{align}
    S(x_2,x_3) \equiv (x_2 x_3)^2 \frac{B(k_1, x_2 k_1, x_3 k_1)}{B(k_1,k_1,k_1)}\,.
\end{align}
Here $x_2 \equiv k_2/k_1$ and $x_3 \equiv k_3/k_1$ are restricted to the region $\max(x_3, 1- x_3) \leq x_2 \leq 1$. One can proceed just as in the flat space case presented in \Sec{subsec:flatB} to evaluate contact bispectra. The generic structure of the integrals is
\begin{align}\label{eq:Bexpref}
    &B(k_1,k_2,k_3) = (-i)^{n_K + n_R + 1}\frac{H^3}{f_\pi^6} \frac{g}{H^{4-n_d}} \int_{-\infty(1 \pm i \epsilon)}^{0^-} \frac{\dd \eta}{\eta^{4-n_d}} \nonumber \\
    &\qquad \widehat{\mathcal{D}}(\{\bmk_i\}, \partial_\eta)\left[ G^{K/R}(k_1, 0, \eta) G^{K/R}(k_2, 0, \eta) G^R(k_3, 0, \eta) + 5~\mathrm{perms}. \right]
\end{align}
where $n_K$ counts the number of Keldysh progagators, $n_R$ the number of retarded ones and $n_d$ the number of (spatial and temporal) derivatives. $\widehat{\mathcal{D}}(\{\bmk_i\}, \partial_\eta)$ is a differential operator schematically representing the $n_d^{\mathrm{th}}$ spatial and temporal derivatives acting on the propagators. Note that there is always at least one $G^R$ due to the at least linearity in $\pia$ inherited from \Eq{eq:norm}. The bulk-to-boundary propagators are 
\begin{align}\label{eq:GRB2b}
    G^R(k, 0, \eta) = -\frac{H^2}{2k^3}  z^{3} \left(\frac{z}{2} \right)^{ - \nu_\gamma}\Gamma(z) J_{\nu_\gamma}(z)
\end{align}
and 
\begin{align}
    G^K(k, 0,\eta) &= - i\frac{\pi}{4k^3} \beta_1 \left(2 z \right)^{\nu_\gamma} \Gamma(\nu_\gamma) \left[ Y_{\nu_\gamma}(z)A^{(1)}_{\nu_\gamma}(z)- J_{\nu_\gamma}(z)B^{(1)}_{\nu_\gamma}(z)\right]\,,
\end{align} 
where we expressed the quantities in terms of $z = - k \eta$ and  $\nu_\gamma = \frac{3}{2} + \frac{\gamma}{2H}$, and $A^{(1)}_{\nu_\gamma}(z)$ and $B^{(1)}_{\nu_\gamma}(z)$ are defined in \Eqs{eq:Fbetaref} and \eqref{eq:Gbetaref} respectively. It is also useful to consider 
\begin{align}
    \partial_\eta G^R(k, 0,\eta) = \frac{H^2}{2k^2} z^2 \left(\frac{z}{2} \right)^{-\nu_\gamma} \Gamma (\nu_\gamma) \left[ z J_{\nu_\gamma -1}(z)- (2 \nu_\gamma-3) J_{\nu_\gamma}(z)\right]
\end{align}
and
\begin{align}
    \partial_\eta G^K(k, 0,\eta)&= i\frac{\pi}{4k^2}\beta_1\left(2 z \right)^{\nu_\gamma-1} \Gamma(\nu_\gamma)  \bigg\{\left[-z Y_{\nu_\gamma+1}(z)+2\nu_\gamma Y_{\nu_\gamma}(z) + z Y_{\nu_\gamma-1}(z)\right]A^{(1)}_{\nu_\gamma}(z) \nonumber \\
    & \left[-z J_{\nu_\gamma+1}(z)+2\nu_\gamma J_{\nu_\gamma}(z) + z J_{\nu_\gamma-1}(z)\right]B^{(1)}_{\nu_\gamma}(z)\bigg\}.
\end{align}
It is now a matter of evaluating the time integral of \Eq{eq:Bexpref} injecting the above expressions for the propagators. This task is analytically hard in full generality, therefore, below we mostly rely on numerical integration and only derive analytical results in some specific regimes. 


\subsubsection*{Numerical results}

In this section, we summarise the main phenomenological implications of the contact bispectra generated by the cubic operators of \Eqs{eq:L13}, \eqref{eq:L23} and \eqref{eq:L33}. Rather than an exhaustive investigation, which we postpone for future work, we here highlight the main features one can expect from the bispectrum signal for this class of models. The \textit{Mathematica} code used to perform the analysis is available upon request.

\begin{figure}[tbp]
\sidesubfloat[]{\includegraphics[width=0.4\textwidth]{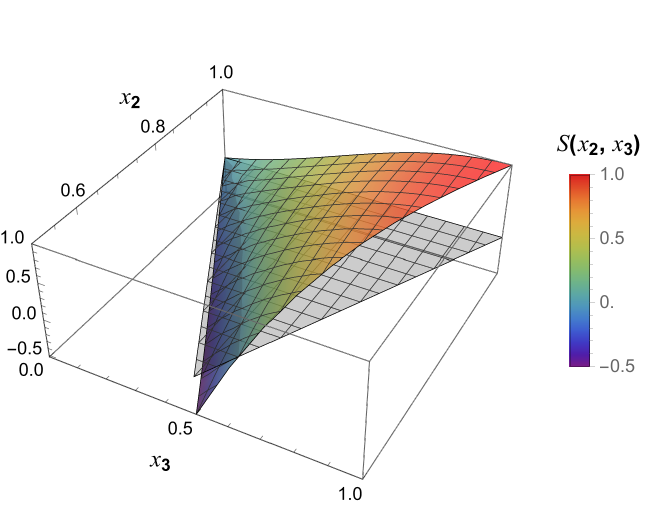}\label{fig:c}}
\hfil
\sidesubfloat[]{\includegraphics[width=0.4\textwidth]{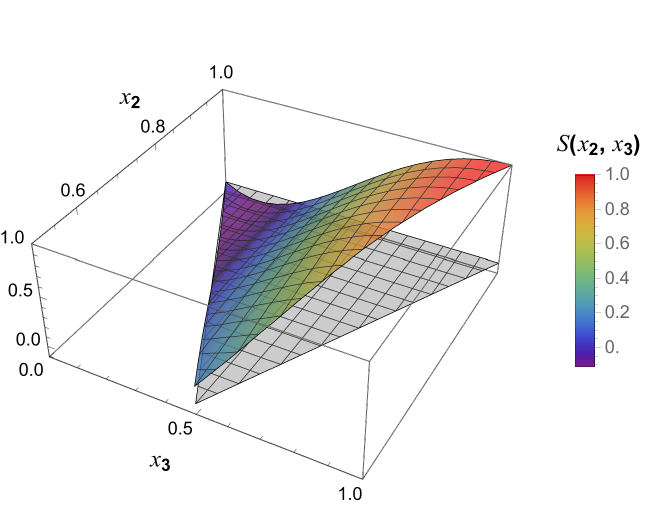}\label{fig:d}}

\medskip
\sidesubfloat[]{\includegraphics[width=0.4\textwidth]{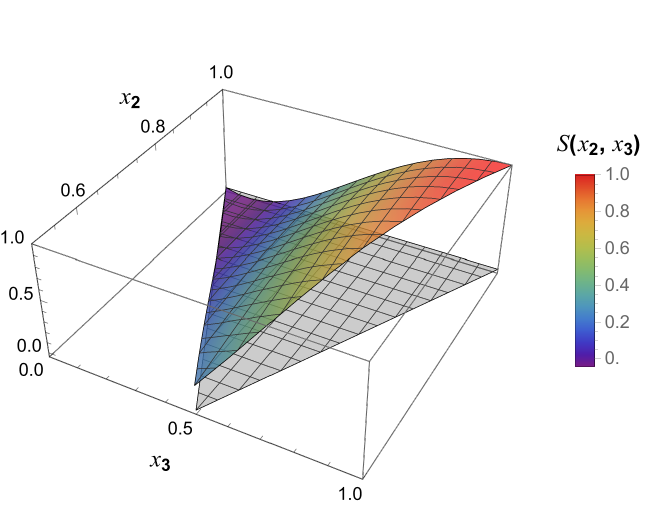}\label{fig:e}}
\hfil
\sidesubfloat[]{\includegraphics[width=0.4\textwidth]{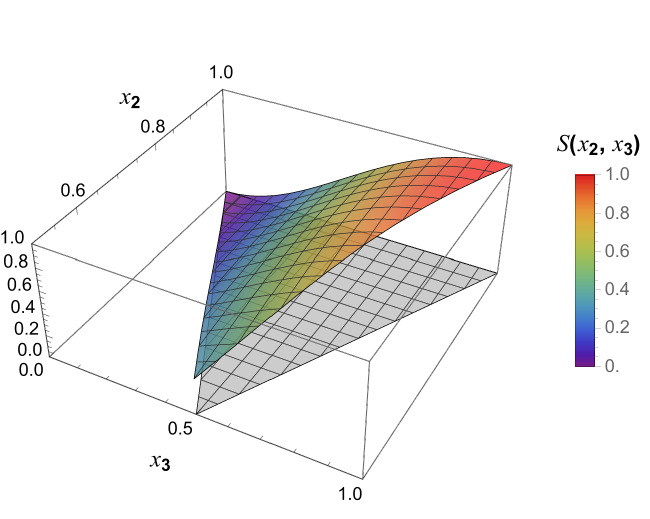}\label{fig:f}}
\caption{Shapes of the bispectrum at large dissipation $\gamma = 5H$. In this regime, the signal peaks in the equilateral configuration $x_2 = x_3 = 1$. Consistency relations hold in the squeezed limit $x_3 \ll x_2 = 1$. \textbf{a.} $\left(\partial_i \pir \right)^2 \pia$ operator; \textbf{b.} $\dot{\pi}^2_{r} \pia$ operator; \textbf{c.} $\dot{\pi}_{r} \pia^2$ operator; \textbf{d.} $\pia^3$ operator.}
    \label{fig:shapes_high}
    \end{figure}

\begin{figure}[tbp]
    \centering
\sidesubfloat[]{\includegraphics[width=0.4\textwidth]{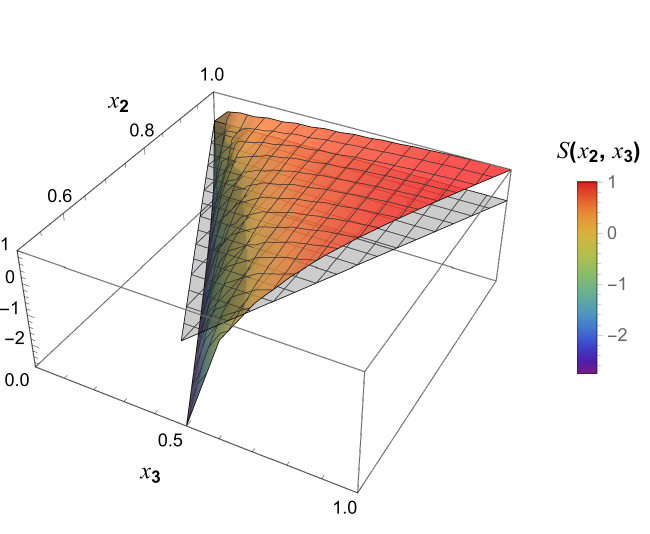}\label{fig:cbis}}
\hfil
\sidesubfloat[]{\includegraphics[width=0.4\textwidth]{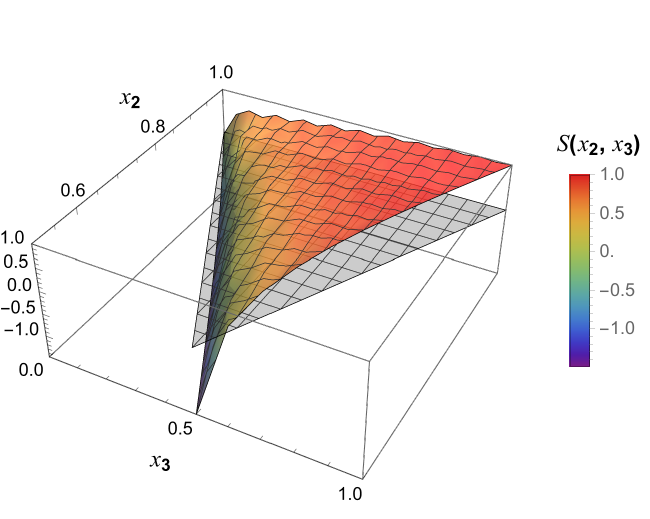}\label{fig:dbis}}

\medskip
\sidesubfloat[]{\includegraphics[width=0.4\textwidth]{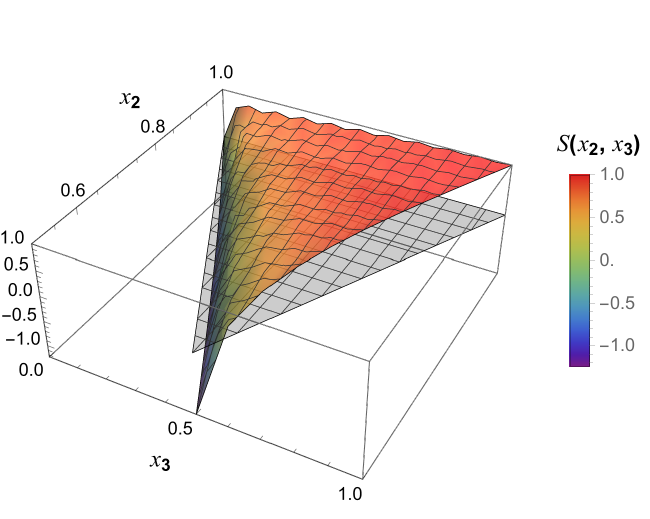}\label{fig:ebis}}
\hfil
\sidesubfloat[]{\includegraphics[width=0.4\textwidth]{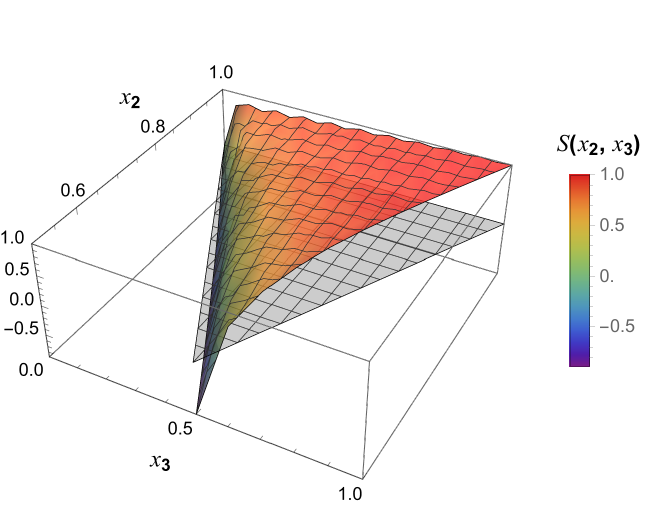}\label{fig:fbis}}
\caption{Shapes of the bispectrum at low dissipation $\gamma= 0.001H$. In this regime, the signal peaks near folded configurations $x_2 + x_3 = 1$. Consistency relations still hold in the squeezed limit $x_3 \ll x_2 = 1$. The tiny oscillations are artefacts of the numerical integration over a finite range. 
\textbf{a.} $\left(\partial_i \pir \right)^2 \pia$ operator; \textbf{b.} $\dot{\pi}^2_{r} \pia$ operator; \textbf{c.} $\dot{\pi}_{r} \pia^2$ operator; \textbf{d.} $\pia^3$ operator.}
    \label{fig:shapes_low}
    \end{figure}

Many of the sixteen cubic operators appearing in \Eqs{eq:L13}, \eqref{eq:L23} and \eqref{eq:L33} lead to similar signatures for the contact bispectrum. Hence, we only display the results for a subset of these operators chosen in the following manner:
\begin{itemize}
    \item From $\mathcal{L}_1$ given in \Eq{eq:L13}, we first consider $\dot{\pi}^2_{r} \dot{\pi}_{a}$ and $\left(\partial_i \pir \right)^2 \dot{\pi}_{a}$, which are two operators appearing in the unitary limit where one recovers the usual EFToI. We highlight how their signature is modified in the presence of dissipation due to the modified structure of the propagators. 
    \item From $\mathcal{L}_1$ given in \Eq{eq:L13}, we also consider $\dot{\pi}^2_{r} \pia$ and $\left(\partial_i \pir \right)^2 \pia$ which are related to the quadratic dissipation $  \dot{\pi}_{r} \pia$ by the non-linear realisation of boosts. These operators have been discussed in \cite{LopezNacir:2011kk}, mostly in the large dissipation regime. Here, we complement that discussion with results at the small dissipation.
    \item From $\mathcal{L}_2$ given in \Eq{eq:L23}, we consider $\partial_i \pir  \partial^i \pia \pia$ and $\dot{\pi}_{r} \pia^2$. The former arises as the non-linear realisation of boosts acting on a noise term $(P^\mu \partial_\mu \pia)\pia$, which is a total derivative at linear order, that is $\dot{\pi}_{a} \pia$. The latter has an interesting signature in $f_{\mathrm{NL}}^{\mathrm{eq}}$ as a function $\gamma/H$ as first noted in \cite{Creminelli:2023aly} and discussed in \Sec{sec:matching}. 
    \item From $\mathcal{L}_3$ given in \Eq{eq:L33}, we consider $\dot{\pi}_{a}^3$ which also appears in the unitary limit where one recovers the usual EFToI. We also consider $\pia^3$ as people might \textit{a priori} worry it behaves differently to the others due to the absence of derivatives. Instead, this term leads to a mostly similar signatures as the other interactions due to the modified propagators compared to the unitary case.
\end{itemize}
Moreover, $\pir^{\prime2} \pia$, $\left(\partial_i \pir \right)^2 \pia$, $\pir^{\prime} \pia^2$ and $\pia^3$ are also the four cubic operators appearing in the matching with the UV completion of \cite{Creminelli:2023aly} discussed in \Sec{sec:matching}. The shapes of the contact bispectrum generated by these four operators are displayed in \Figs{fig:shapes_high} and \ref{fig:shapes_low} (the other operators essentially follow the same trend). Just as for the flat space case, different behaviours emerge in the large ($\gamma \gg H$) and small ($\gamma \ll H$) dissipation regime. While the former peaks in the equilateral configuration as already noted in \cite{LopezNacir:2011kk}, the latter reaches an extremum near the folded region. 

\begin{figure}[tbp]
\centering
\includegraphics[width=0.6\textwidth]{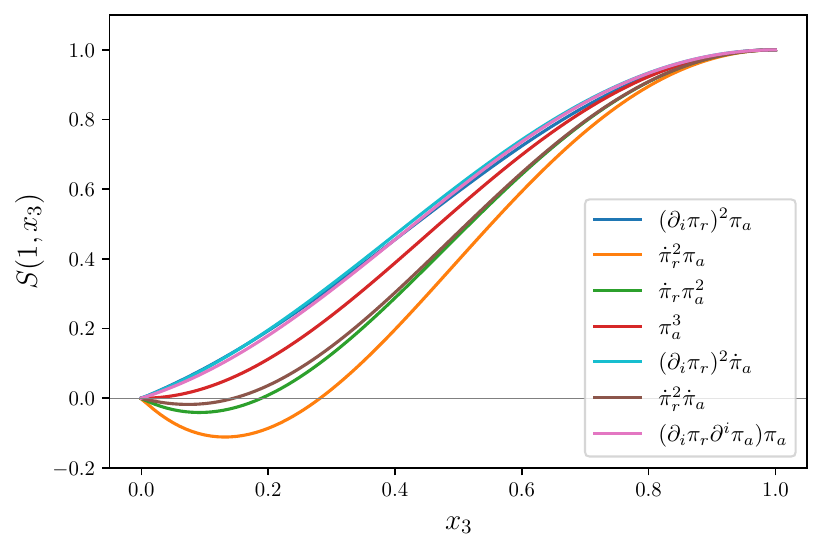}
\caption{Shape function along the direction $x_2 = 1$ for different contributions to the contact bispectrum at large dissipation $\gamma = 5H$. Consistency relations ensure the signal vanishes in the squeezed limit $x_3 \ll 1$. }
\label{fig:squeezed}
\end{figure}

This smoking gun of open dynamics might seem degenerate with other classes of models that also lead to a signal in the folded triangles such as non-Bunch Davies initial states \cite{Holman:2007na, Chen:2006nt, Meerburg:2009ys, Agullo:2010ws, Ashoorioon:2010xg, Agarwal:2012mq, Ashoorioon:2013eia,  Albrecht:2014aga, Green:2020whw}. A crucial difference, which appears in our numerical treatment and is confirmed analytically below, is that dissipation regularises the divergence by smoothing the peak and displacing it from the edge of the triangular configurations, leading to finite values of the bispectrum for any physical configuration.   
In particular, it implies no divergence in the squeezed limit of the bispectrum $k_1 \simeq k_2 \gg k_3$, which is displayed in \Fig{fig:squeezed}. Small values of $\gamma/H$ may eventually lead to an intermediate peak due to the regularised folded singularity, yet consistency relations hold \cite{Maldacena:2002vr,Creminelli:2004yq, Cheung:2007sv, Creminelli:2012ed,Hinterbichler:2012nm,Assassi:2012zq,Pajer:2017hmb,Avis:2019eav} and the squeezed limit goes to zero because of the symmetries of the theory. Notice that operators such as $\pia^3$ follow this trend despite what one may have naively thought in the absence of derivatives. This is because of the modified propagators compared to the free theory, which for instance ensure IR convergence.	

\begin{figure}[tbp]
    \begin{minipage}{6in}
    \centering
    \raisebox{-0.5\height}{\includegraphics[width=.48\textwidth]{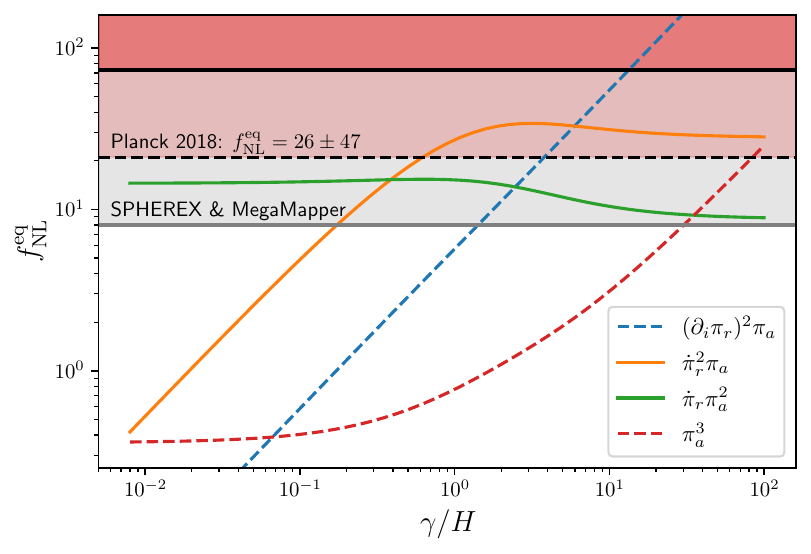}}
    \raisebox{-0.5\height}{\includegraphics[width=.48\textwidth]{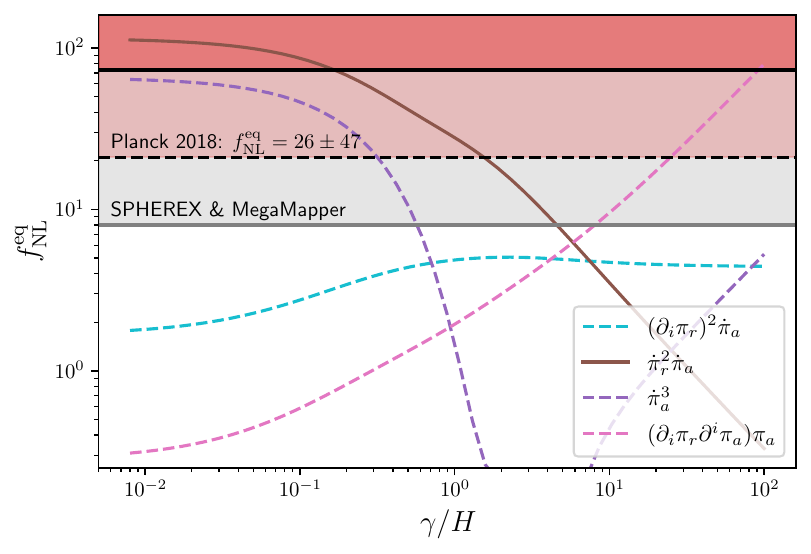}}
\end{minipage}
\caption{Amplitude in the equilateral configuration as a function of the dissipation parameter $\gamma/H$. Continuous (dashed) lines are positive (negative) values. \textit{Left:} The four cubic operators appearing in the matching with \cite{Creminelli:2023aly} performed in \Sec{sec:matching}. All the curves well reproduce the results from \cite{Creminelli:2023aly}. The otherwise arbitrary values of the EFT coefficients were chosen to qualitatively reproduce the results from \cite{Creminelli:2023aly}. \textit{Right:} New operators of the open EFToI that may lead to specific signatures on the non-Gaussian signal. Again, numerical values of the EFT coefficients are arbitrary, chosen to be comparable to the \textit{Left} panel. All scalings with $\gamma$ are consistent with the heuristic estimates of \Sec{subsec:energy}. }
\label{fig:fnleq}
\end{figure}

The amplitude in the equilateral configuration is controlled by 
\begin{align}
    f_{\mathrm{NL}}^{\mathrm{eq}} \equiv \frac{10}{9} \frac{k^6}{(2\pi)^4}\frac{B(k,k,k)}{\Delta_\zeta^4}.
\end{align} 
In \Fig{fig:fnleq}, we display the dependence of $f_{\mathrm{NL}}^{\mathrm{eq}}$ on the dissipation parameter $\gamma/H$ from numerical integration of the bispectrum $B(k,k,k)$ for the operators considered above. One can use observational constraints $f_{\mathrm{NL}}^{\mathrm{eq}} = 26 \pm 47$ from \cite{Planck:2019kim} to place bounds on the EFT parameters in the large dissipation regime. For instance, a numerical fit of the $(\partial_i \pir)^2\pia$ contributions leads to $f_{\mathrm{NL}}^{\mathrm{eq}} \simeq - \gamma/(4H)$ which is consistent with the result from \cite{LopezNacir:2011kk, Creminelli:2023aly} and our heuristic estimate of \Sec{subsec:energy}. It naively implies that $\gamma / H < 80$ at $68\%$ confidence. Of course, this is more of a proof of principle than a realistic estimate due to the cumulative effects of different cubic operators that cannot be disentangled one from another. Yet, it demonstrates how this class of model can be confronted to data. Such observational bounds could tighten thanks to future LSS experiments such as SPHEREX \cite{SPHEREx:2014bgr} or MegaMapper \cite{Cabass:2022epm, Braganca:2023pcp}, further constraining on this class of models.  
One would eventually consider to perform the same kind of analysis in the small dissipation regime for 
\begin{align}
    f_{\mathrm{NL}}^{\mathrm{folded}} \equiv \frac{5}{24} \frac{k^6}{(2\pi)^4} \frac{B(k,k/2,k/2)}{\Delta_\zeta^2(k/2)[4\Delta_\zeta^2(k/2) + \Delta_\zeta^2(k) ]},
\end{align}
for instance using the CMB-BEST pipeline \cite{PhysRevD.108.063504}. We leave it for future work.  

\subsubsection*{Analytical discussion}

An interesting question is the fate of folded singularities in an expanding background. In general, a folded singularity appears when two modes resonate with a third for an infinite amount of time. For this reason, the origin of the divergence can be derived from the early time oscillating behaviour of the propagators. Hence, let us consider the simplest cubic operator $\pia^3$ and the cubic bispectrum it generates
\begin{align}
    &B(k_1,k_2,k_3) = - 6 \frac{H^3}{f_\pi^6} \frac{\delta_1}{f_\pi^2}  \int_{-\infty(1 \pm i \epsilon)}^{0^-} \frac{\dd \eta}{H^4\eta^{4}} G^{R}(k_1, 0, \eta) G^{R}(k_2, 0, \eta) G^R(k_3, 0, \eta). 
\end{align}
Injecting \Eq{eq:GRB2b} into the above expression and expanding at early time in the sub-Hubble regime when $\eta \rightarrow - \infty$, we observe the time integral reduces to a collection of terms of the form
\begin{align}
   \left| \int \dd \eta \frac{\ee^{-i \left(\pm k_1 \pm k_2 \pm k_3 \right)\eta}}{\eta^{1 + \frac{3}{2}\frac{\gamma}{H}}}   \right| < \infty \qquad \mathrm{when} \qquad \frac{\gamma}{H} >0.
\end{align}
This is the analogue of the flat space case discussed in \Eq{eq:M4int} from which we concluded the resolution of the singularity in presence of non-vanishing dissipation. In the vanishing limit, we recover the log-folded singularity familiar to non-Bunch Davies initial states \cite{Holman:2007na, Chen:2006nt, Meerburg:2009ys, Agullo:2010ws, Ashoorioon:2010xg, Agarwal:2012mq, Ashoorioon:2013eia,  Albrecht:2014aga, Green:2020whw}. Contrarily to this dramatically uncontrolled divergence along the folded region of \Fig{fig:triangles}, dissipation tames the peak and displaces it from the edge of the physical region such that the folded singularity remains resolved and under perturbative control for any value of the kinematic variables.

One may wonder if there are implications of the above mentioned singularity on the squeezed limit of the bispectrum $k_1 \simeq k_2 \gg k_3$. As long as $\pir$ transforms as the usual pseudo-Goldstone boson, which non-linearly realises time-translation and boosts, the standard arguments \cite{Creminelli:2004yq, Cheung:2007sv, Creminelli:2012ed} should hold. As a consequence of the singularity being resolved, the curvature perturbation bispectrum still goes to zero in the squeezed limit with an eventual intermediate peak in the small dissipation regime, when $\gamma \ll H$.  

In additions to the sub-Hubble and squeezed regimes discussed above, analytical results can also be obtained for specific values of the dissipation parameter $\gamma$. Indeed, Hankel functions with half-integer order parameter $\nu_\gamma$ are simpler to handle. This happens for $\gamma/H = 2n$ with $n \in \mathbb{N}$. In this case, propagators decompose into polynomials multiplying an oscillatory phase. It offers a window of opportunity into computing analytic expressions for shapes of interest. Yet, beware that the results can significantly depart from the general case of arbitrary $\gamma$ so one must be careful before drawing general conclusions about the behaviour of dissipative dynamics.

\section{Matching}\label{sec:matching}

For an EFT construction to be of interests, it has to encompass a class of physically motivated models. For this reason, we consider in this section the matching of our open EFT to a concrete ``ultra-violet" model recently constructed and studied in \cite{Creminelli:2023aly}. This is not the first model of dissipative dynamics, but it has the distinguishing feature of leading to a \textit{local} low-energy dynamics around and below the Hubble scale. This is related to the fact that the window of instability for particle production involves wavelengths that are parametrically sub-Hubble. This is in constrast with the much studied model based on the coupling $\phi F \widetilde{F} $ \cite{Anber:2009ua}. In that case, gauge fields are produced around horizon crossing and hence mediate interactions at distances and time scales of order Hubble, which appear non-local in the open EFT.\\

The model in \cite{Creminelli:2023aly} provides an example of a partial UV completion and illustrates the scope of our EFT. In addition to the inflaton field $\phi$, the model features a massive scalar field $\chi$ with a softly-broken $U(1)$ symmetry. The action of the model is given by
\begin{align}\label{eq:actionmatch}
S= \int \dd^4 x &\sqrt{- g} \bigg[ \frac{1}{2} M_{\mathrm{Pl}}^2  R - \frac{1}{2} \left(\partial \phi \right)^2 - V(\phi) - \left|\partial \chi \right|^2 + M^2 \left| \chi \right|^2 \nonumber \\
- &\frac{\partial_\mu \phi}{f}\left( \chi \partial^\mu \chi^* - \chi^* \partial^\mu \chi\right) - \frac{1}{2}m^2\left(\chi^2 + \chi^{*2}\right)\bigg].
\end{align}
The last term in \Eq{eq:actionmatch} breaks the $U(1)$ symmetry $\chi \rightarrow \ee^{i \alpha} \chi$ for $\alpha \in \mathbb{R}$. An important parameter for the dynamics of the system is $\rho \equiv \dot{\phi}_0(t)/f$, which controls the mixing of $\phi$ with $\chi$. Here $\dot{\phi}_0(t)$ is the time derivative of the inflaton background $\phi_0(t)$. The hierarchy of scales for which the treatment of \cite{Creminelli:2023aly} is valid is
\begin{align}
f \gg \rho \gtrsim M \gg m \gg H.
\end{align}
The model exhibits a narrow instability band in the sub-Hubble regime, during which particle production occurs. The amount of particle production is controlled by the dimensionless parameter
\begin{align}
\xi \simeq \frac{m^4}{8 H \rho M^2} \gtrsim \mathcal{O}(1).
\end{align}
The inflaton fluctuations experience a dissipative dynamics due to the presence of an environment of $\chi$ particles generated by the instability. The inflaton dynamics is effectively described in terms of a non-linear Langevin equation \cite{Creminelli:2023aly}
\begin{align}\label{eq:Paolo}
\pi'' + \left( 2H + \gamma\right)a\pi' - \partial_i^2 \pi &\simeq \frac{\gamma}{2\rho f} \left[ \left( \partial_i \pi\right)^2 - 2 \pi \xi \pi^{\prime2}\right] - \frac{a^2m^2}{f} \left(1+2\pi\xi \frac{\pi'}{a \rho f} \right) \delta \mathcal{O}_S\,,
\end{align}
where we neglected the inflaton potential contribution $a^2 V'' \pi$ which is slow-roll suppressed. The dissipation parameter $\gamma$ (with units of mass) is determined by the following combination of microphysical parameters
\begin{align}\label{eq:gammaCrem}
\gamma \simeq \frac{\xi m^4}{\pi M f^2} \ee^{2 \pi \xi}.
\end{align}
The effect of noise is captured in terms of its two- and three-point statistics
\begin{align}
\langle  \delta \mathcal{O}_S(\bmk, \eta) \delta \mathcal{O}_S(\bmk',\eta')\rangle &\simeq (2\pi)^3 \delta(\bmk + \bmk') \delta(\eta - \eta')H^4 \eta^4 \nu_{\mathcal{O}}\,, \\
\langle  \delta \mathcal{O}_S(\bmk, \eta) \delta \mathcal{O}_S(\bmk',\eta')  \delta \mathcal{O}_S(\bmk'',\eta'')\rangle &\simeq (2\pi)^3 \delta(\bmk + \bmk' +\bmk'') \delta(\eta - \eta') \delta(\eta - \eta'')H^8 \eta^8 \nu_{\mathcal{O}^3} \,,\label{eq:NGnoiseCrem}
\end{align}
with the amplitudes of the noises being controlled by 
\begin{align}
\nu_{\mathcal{O}} \simeq \frac{M}{m} \frac{\ee^{4\pi \xi}}{4\pi^2} \qquad \mathrm{and} \qquad \nu_{\mathcal{O}^3} \simeq \frac{\ee^{6\pi \xi}}{\pi^2 m^2} .
\end{align}
The parameters at play are summarized in the first line of Table \ref{tab:matching}.

As we demonstrate below, the low-energy dynamics of this model is equivalently described in terms of\footnote{The open EFT of \Eq{eq:Seffmatch} has no derivatives acting on $\pia$ (apart from the standard kinetic term) and up to one derivative acting on $\pir$.} 
\begin{align}\label{eq:Seffmatch}
\quad S_{\mathrm{eff}} &= \int \dd^4 x \Big[ a^2 \pi'_{r} \pi'_{a} - c_{s}^{2} a^2 \partial_i \pir  \partial^i \pia   -  a^3 \gamma \pi'_{r} \pia + i \beta_1 a^4 \pia^2 \nonumber \\
+ &\frac{ \left(8\gamma_2 - \gamma\right)}{2f^{2}_{\pi}}  a^2\pir^{\prime2} \pia	+ \frac{\gamma}{2f_{\pi}^{2}} a^2	\left(\partial_i \pir \right)^2 \pia - 2i\frac{\beta_{5}}{f_\pi^2} a^3 \pir^{\prime} \pia^2 +\frac{\delta_1}{f_\pi^2} a^4 \pia^3  \Big],
\end{align}
upon matching the EFT coefficients to the microphysical parameters of the model through the relations of Table \ref{tab:matching}. The matching can be performed by comparing the power spectrum and bispectrum obtained from the top-down and bottom-up approaches, as we do in \Sec{subsec:obs}. Alternatively, the matching can also be derived at the level of the Langevin equation. In \Sec{subsec:Langevin}, we develop a ``stochastic unravelling" \cite{breuerTheoryOpenQuantum2002} of our EFT\footnote{Indeed, open quantum systems can equivalently be described in terms of a path integral formulation (Feynman-Vernon influence functional), a dynamical equation for the reduced density matrix (master equation - which reduces to Lindblad equation in the Markovian limit) or their stochastic unravelling (Langevin equation or its quantum analogue the stochastic Schr\"odinger equation). Each formulation has its own benefits \cite{breuerTheoryOpenQuantum2002}, the Langevin equation being well-suited for numerical simulations, the influence functional adapted to describe relativistic settings in a manifestly covariant manner and the master equations useful for their ability to implement non-perturbative resummations. See \cite{Colas:2023wxa} for a cosmology oriented review of these techniques.} which holds at non-linear order and matches the top-down construction of \cite{Creminelli:2023aly} upon restricting to \Eq{eq:Seffmatch} and Table \ref{tab:matching}.

\begin{table}
    \begin{center}
    \begin{tabular}{ |c|| P{1.2cm} P{1.2cm} P{1.2cm} P{1.2cm} P{1.2cm} P{1.2cm} P{1.2cm} |  }
    \hline 
    & \multicolumn{7}{|c|}{	\Bigg.
    \textit{Parameters:}} \\               
    \hline 
    \Bigg.UV completion \cite{Creminelli:2023aly} & $M$ & $m$ & $f$ & $\rho$ &  $\xi$ & $\gamma$ & $\nu_{\mathcal{O}}/\nu_{\mathcal{O}^3}$ \\
    \hline
    \Bigg.Open EFT \eqref{eq:Seffmatch} & $f^2_\pi $ & $c_s$ & $\gamma$ & $\gamma_{2}$ &  $\beta_1  $ & $\beta_{5}$ & $\delta_1 $ \\
    \hline
    \end{tabular}
    \vspace{0.2in}

    \begin{tabular}{ |P{4cm} P{4cm} |  }
    \hline 
    \Bigg. \textit{Matching:} & $f^2_\pi = \rho f $ \\
    \Bigg. $c_s = 1$ 	&	$\gamma = \frac{\xi m^4}{\pi M f^2} \ee^{2 \pi \xi}$ \\
    \Bigg. $8\gamma_2 =\left(1-2 \pi \xi\right) \gamma$ & $\beta_1 = \frac{\nu_{\mathcal{O}}}{2 \rho f} \frac{m^4}{f^2} $ \\
    \Bigg. $\beta_{5}=  2\pi\rho f\xi \beta_1$ & 		 
     $6\delta_1 = \rho f \frac{m^6}{f^3} \nu_{\mathcal{O}^3} $  \\
    \hline
\end{tabular}
\end{center}
\caption{Matching of the EFT coefficients appearing in \Eq{eq:Seffmatch} to the microphysical parameters of \cite{Creminelli:2023aly}. The matching is obtained either at the level of the power spectrum and contact bispectra as in \Sec{subsec:obs} or by deriving the Langevin equation for the fluctuations as in \Sec{subsec:Langevin}.}
\label{tab:matching}
\end{table}


\subsection{Power spectrum and bispectrum}\label{subsec:obs}

First, by noticing that in \cite{Creminelli:2023aly} $\zeta = - H \pi/(\rho f) $, and comparing this expression with our linear relation between $\zeta$ and $\pi$ given in \Eq{eq:can}, we obtain $f_\pi^2 = \rho f$. Second, one notices that the speed of sound in this set up is $c_s = 1$ (see \Eq{eq:Paolo}). This provides a matching for our first two EFT coefficients. Then, we turn our attention to the computation of the power spectrum, which in \cite{Creminelli:2023aly} was found to be
\begin{align}\label{eq:PkCrem}
    \langle \zeta_\bmk \zeta_{-\bmk}\rangle = (2\pi)^3 \delta(\bmk + \bmk') \frac{H^2}{\left(\rho f\right)^2} \frac{m^4}{f^2} \nu_{\mathcal{O}} \int_{-\infty}^{\eta} \frac{\dd \eta'}{\left(H\eta'\right)^{4}} \left[ G^R(k; \eta,\eta') \right]^2.
\end{align}
Here we accounted for the rescaling of the Green function by $a^2(\eta')$ in the definition of \cite{Creminelli:2023aly}. Comparing \Eq{eq:PkCrem} with our results from \Sec{subsec:dSPk}, especially the expressions for the retarded Green function in \Eq{eq:GRfinsimpv2} and for the Keldysh function in \Eq{eq:GK1} from which we extract the power spectrum, we can identify
\begin{align}\label{eq:match1}
    \gamma = \frac{\xi m^4}{\pi M f^2} \ee^{2 \pi \xi} \qquad \mathrm{and} \qquad \beta_1 = \frac{\nu_{\mathcal{O}}}{2 \rho f} \frac{m^4}{f^2}\,,
\end{align}
without even performing the integral. Note that we used \Eq{eq:gammaCrem} to relate the derived parameter $\gamma$ to the UV paremeters of \cite{Creminelli:2023aly}.

We now turn our attention the bispectrum expressions found in \cite{Creminelli:2023aly}. We first consider the $\left(\partial_i \pi\right)^2$ term in the right-hand side of the Langevin \Eq{eq:Paolo}. Following \cite{Creminelli:2023aly}, it leads to
\begin{align}\label{eq:bispecCrem}
     B(k_1,k_2,k_3) &= \frac{\gamma}{H} \left(\frac{H}{\rho f} \frac{m^2}{f} \right)^4 \nu_\mathcal{O}^2  \int \frac{\dd \eta_1}{\left(H\eta_1\right)^{2}} \int \frac{\dd \eta_2}{\left(H\eta_2\right)^{4}}  \int \frac{\dd \eta_3} {\left(H\eta_3\right)^{4}}   \nonumber\\
    & \quad \bigg\{G^R(k_1; 0,\eta_1) G^R(k_2; 0,\eta_2) G^R(k_3; 0,\eta_2) \\
    &\times\left[ \left(\bmk_2.\bmk_3\right) G^R(k_2; \eta_1,\eta_2)  G^R(k_3; \eta_3,\eta_2)+ 2~\mathrm{perms.} \right] \bigg\} .  \nonumber
\end{align}
It might not be direct to realise that this expression is strictly equivalent to the contact bispectrum generated by the $\left(\partial_i \pir \right)^2 \pia$ operator of the open EFT spelled in \Eq{eq:Seffmatch}. Following \Eq{eq:Bexpref}, this contribution writes
\begin{align}\label{eq:bspecmatch2}
    &B(k_1,k_2,k_3) = \frac{H^3}{f_\pi^6}  \frac{\gamma}{2f_{\pi}^{2}}  \int \frac{\dd \eta_1}{\left(H\eta_1\right)^{2}} \left[\left(\bmk_1.\bmk_2\right) G^{K}(k_1, 0, \eta_1) G^{K}(k_2, 0, \eta_1) G^R(k_3, 0, \eta_1) + 5~\mathrm{perms}. \right]
\end{align}
which might indeed be non-trivial to match. Yet, if one accounts for the fact that, based on \Eq{eq:GK1}
\begin{align}
   G^{K}(k_1, 0, \eta_1) =  i \beta_1 \int \frac{\dd \eta_2}{\left(H\eta_2\right)^{4}} G^R(k_1; 0, \eta_2) G^R(k_1; \eta_1,\eta_2) 
\end{align}
and inject in \Eq{eq:bspecmatch2} the matching for obtained $ \beta_1$ in \Eq{eq:match1}, one exactly recovers \Eq{eq:bispecCrem} without even having to perform the time integrals.

Similar identification of the functional forms of the integrals for the $\pi^{\prime2}$ term of \Eq{eq:Paolo} with $\pir^{\prime2}\pia$ in the open EFT \eqref{eq:Seffmatch} leads to 
\begin{align}
    8\gamma_2 =\left(1-2 \pi \xi\right) \gamma.
\end{align}
Similarly, the $\pi' \delta \mathcal{O}_S$ term of \Eq{eq:Paolo} is mapped to $\pir'\pia^2$ in the open EFT of \Eq{eq:Seffmatch} such that 
\begin{align}
    \beta_5 = 2\pi\rho f\xi \beta_1.
\end{align}
At last, the non-Gaussian component of the noise in \Eq{eq:Paolo} captured through \Eq{eq:NGnoiseCrem} transforms into the $\pia^3$ term of the open EFT \eqref{eq:Seffmatch}, leading to
\begin{align}
    6\delta_1 = \rho f \frac{m^6}{f^3} \nu_{\mathcal{O}^3} ,
\end{align}
which completes the matching summarised in Table \ref{tab:matching}.

\subsection{Langevin equation}\label{subsec:Langevin}

The formalism presented in this work is not the only way to address open quantum system dynamics \cite{breuerTheoryOpenQuantum2002}. Another commonly employed technique to understand the interaction between system and environment is the Langevin equation. This approach relays on the saddle point approximation for the path integral that defined our open quantum system \eqref{eq:densitymatrix}. We can use it to compute tree-level quantities, such as the contact bispectrum. We will focus on deriving the Langevin equation for the top-down model \cite{Creminelli:2023aly}. We first start with the derivation of the homogeneous stochastic dynamics from the path integral \eqref{eq:densitymatrix}, taking $S_{\text{eff}}$ to be the quadratic functional from \Sec{sec:Pk} with $\widehat{D}^{K}$ set to zero.  We then incorporate the operator $i \beta_{1}\pia^{2}$ on the right-hand side of the Langevin equation through a Hubbard-Stratonovich transformation \cite{Stratonovich1957,Hubbard1959}, that introduces a new field, $\Os$, which is the stochastic noise that appears in \Eq{eq:Paolo}. This allows us to match the Wilsonian coefficient $\beta_{1}$ with the parameters in \cite{Creminelli:2023aly}. Finally, we explain how we introduce the different contact interaction terms derived from the influence functional \eqref{eq:Seffref} into the Langevin equation. This analysis depends on the powers of $\pia$ in each term. Those linear in $\pia$ are easy to incorporate, as the approach is the same as for the $\widehat{D}_{R,A}$ terms in the homogeneous Langevin equation. Those quadratic and cubic in $\pia$ require a modification of the HS trick, yielding a coupling between the system and the noise and self-interactions of the noise (that can be interpreted as non-Gaussian statistics of the noise). 

\subsubsection{Linear stochastic dynamics}

The diagonal of the density matrix is given by a path integral over the $\pir$ component with boundary condition $\pir(\eta_{0})=\pi$ and the $\pia$ component with boundary condition $\pia(\eta_{0})=0$, leading to
\begin{equation}
    \rho[\pi,\pi]=\int_{\text{BD}}^{\pi}\mathcal{D}\pir\int_{\text{BD}}^{0}\mathcal{D}\pia e^{iS_{\text{eff}}[\pir,\pia]}.
\end{equation}

\paragraph{Homogeneous solution} We start by considering the quadratic functional obtained by setting $\widehat{D}_{K}=0$, that is
\begin{equation}\label{eq:noiselessfree}
    S_{\text{eff}}[\pir,\pia] =  \int d^{4}x a^{4}\left(\frac{\pir'\pia'}{a^{2}} - c_{s}^{2}  \frac{\partial_i \pir  \partial^i \pia}{a^{2}} -   \gamma  \frac{\pir' \pia }{a}\right).
 \end{equation}
Integrating by parts, we can write the influence functional as linear in $\pia$ (the boundary terms vanish because they are linear in $\pia(\eta_0)$) such that
\begin{equation}
    S_{\text{eff}}[\pir,\pia] =  \int d^{4}x a^{4}\left[ - \frac{\partial_{\eta}\left(a^{2}\pir'\right)}{a^{4}} + c_{s}^{2}  \frac{\partial_i^{2} \pir}{a^{2}} -   \gamma  \frac{\pir'}{a}\right] \pia.
\end{equation}
We can now consider the path integral over $\pia$ as enforcing the constraint
\begin{equation}
    \rho[\pi,\pi]=\int_{\text{BD}}^{\pi}\mathcal{D}\pir\;\delta\left[ - \frac{\partial_{\eta}\left(a^{2}\pir'\right)}{a^{4}} + c_{s}^{2}  \frac{\partial_i^{2} \pir}{a^{2}} -   \gamma  \frac{\pir'}{a}\right].
\end{equation}
This greatly simplifies the calculation of $n$-point functions of the $\pir$ component. If we want to compute correlation functions of $\pir$, it is then enough to solve the equation of motion defined by the constrain, that is
\begin{equation}
    \pi_{\bfk}^{\prime \prime} + \left( 2H + \gamma\right)a \pi_{\bfk}^{\prime}+c_{s}^{2}k^{2}\pi_{\bfk}=0,
\end{equation}
which defines the differential operator $\widehat{D}_{R}$.
Comparing with \Eq{eq:Paolo}, we see that we were right to call $\gamma$ the same in both equations and that $c_{s}=1$. We can rebuild the retarded propagator from \Sec{sec:Pk} by looking at the Green function for the homogeneous equation of motion
\begin{equation}
    \partial_{\eta}^{2}G^R(k;\eta,\eta')-\frac{1}{\eta}\left(2+\frac{\gamma}{H}\right)\partial_{\eta}G^R(k;\eta,\eta')+c_{s}^{2}k^{2}G^R(k;\eta,\eta')=H^{2}\eta^{2}\delta(\eta-\eta')
\end{equation}
with boundary conditions $G^R(k;\eta<\eta')=0$ and 
\begin{equation}
    G^R(k;\eta,\eta)=0\;, \qquad \;\qquad \partial_{\eta}G^R(k;\eta,\eta')\bigg|_{\eta=\eta'}=H^{2}\eta^{2}.
\end{equation}
It finally leaves to the previously found result given in \Eq{eq:GRfinsimpv2}, which we reproduce here for clarity
\begin{equation}
    G^R(k;\eta_{1},\eta_{2})=\frac{\pi}{2} H^{2} \left(\eta_{1}\eta_{2}\right)^{\frac{3}{2}}\left(\frac{\eta_{1}}{\eta_{2}}\right)^{\frac{\gamma}{2H}}\text{Im}\left[H_{\frac{3}{2}+\frac{\gamma}{2H}}^{(1)}(-c_{s}k\eta_{1})H_{\frac{3}{2}+\frac{\gamma}{2H}}^{(2)}(-c_{s}k\eta_{2})\right].
\end{equation}

\paragraph{Quadratic noise} The derivation of the Langevin equation from the open effective functional relies on rewriting all terms to be linear in $\pia$. When we introduce the quadratic terms in $\pia$ like $ i {\beta}_{1}\pia^{2}$, this rewriting cannot be achieved by integration by parts. Rather, we have to introduce an auxiliary Gaussian field $\Os$ to rewrite quadratic terms in $\pia$ in a linear form.\footnote{This modifies the influence functional by adding terms that depend on $\Os^{2}$ These terms will tell us about the statistics of the noise $\Os$.} This procedure is known as the Hubbard–Stratonovich transformation \cite{Stratonovich1957,Hubbard1959}. We here focus on the term $ i {\beta}_{1}\pia^{2}$, which is the one usually  considered in the literature. We further discuss the inclusions of the derivative noises appearing in \Eq{eq:piq2free} in \App{app:Langevin}. 

Consider the action for the $i \beta_{1}\pia^{2}$ term. This term can be written as the outcome of a path integral over a Gaussian field $\Os$
\begin{equation}\label{eq:HStrick}
    \exp\left(-\int d^{4}x a^{4}\beta_{1}\pia^{2}\right)=\mathcal{N}_{0}\int\mathcal{D}[\Os]\;\exp\left[\int d^{4}x a^{4}\left(-\frac{\Os^{2}}{4\beta_{1}}+i\Os\pia\right)\right],
\end{equation}
with $\mathcal{N}_{0}$ being the normalisation constant. In this way, the path integral that defines the density matrix of the system is linear in $\pia$
\begin{align}
    \rho[\pi,\pi]&=                \mathcal{N}_{0}\int\mathcal{D}[\Os]\int_{\text{BD}}^{\pi}\mathcal{D}\pir\int_{\text{BD}}^{0}\mathcal{D}\pia  \\
    & \qquad \times \exp\left\{\int d^{4}x a^{4}\left[-\frac{\Os^{2}}{4\beta_{1}}+i\left(-\widehat{D}_{R}\pir+\Os\right)\pia\right]\right\},\nonumber
\end{align}
from which we obtain the Langevin equation 
\begin{equation}\label{eq:LinearLangevin}
    \pi''_{\bfk} +\left( 2H + \gamma\right)a\pi_{\bfk}'+c_{s}^{2}k^{2}\pi_{\bfk}=a^2 \Os(\bfk, \eta).
\end{equation}
The new variable $\Os$ behaves as a Gaussian field, getting a non-vanishing two-point function under the path integral
\begin{equation}\label{eq:noisecontact}
    \langle\Os(\bfk,\eta)\Os(\bfk',\eta')\rangle=\frac{2\beta_{1}}{a^{4}(\eta)}\delta(\eta-\eta')(2\pi)^3\delta(\bfk+\bfk').
\end{equation}
A matching of $\beta_1$ with the microphysical parameters of \cite{Creminelli:2023aly} is possible by comparing the Gaussian statistics of the noise, from which we recover the above result. Alternatively, the linear Langevin equation \Eqs{eq:LinearLangevin} contains all the information needed to derive the power spectrum. One can indeed solve the Langevin equation using the convolution of the retarded Green function with the noise and then compute the power spectrum from the inhomogeneous solution
\begin{equation}\label{eq:GRxifree}
    \pi_{\bfk}(\eta)=\int_{-\infty}^{\eta}d\eta'a^{4}(\eta')G^R(k;\eta,\eta')\Os(\bfk,\eta').
\end{equation}
The two-point function of the noise given in \Eq{eq:noisecontact} sources the two-point function of the fluctuations through
\begin{align}
    &\langle\pi_{\bfk_{1}}(\eta_{1})\pi_{\bfk_{2}}(\eta_{2})\rangle=\int_{-\infty}^{\eta_{1}}d\eta_{1}'\int_{-\infty}^{\eta_{2}}d\eta_{2}'a^{4}(\eta'_{1})a^{4}(\eta'_{2})\nonumber \\
    &\qquad \qquad  G^{R}(k_1,\eta_{1},\eta'_{1})G^{R}(k_2, \eta_{2},\eta'_{2})\langle\Os(\bfk_{1},\eta'_{1})\Os(\bfk_{2},\eta'_{2})\rangle,
\end{align}
which obviously reproduces the above result 
\begin{align}
    \langle\pi_{\bfk_{1}}(\eta_{1})\pi_{\bfk_{2}}(\eta_{2})\rangle&=(2\pi)^3\delta(\bfk_{1}+\bfk_{2})\left[\beta_{1}\int_{-\infty}^{\eta_{1}} \frac{d\eta}{H^{4}\eta^{4}}G^{R}(k_1,\eta_{1},\eta)G^{R}(k_2, \eta_{2},\eta) +(\eta_{1}\leftrightarrow\eta_{2})\right].
\end{align}
As shown above, the matching between the results from  \Sec{sec:Pk} and \cite{Creminelli:2023aly} is obtained for
\begin{equation}
    \beta_{1}=\frac{\nu_{\mathcal{O}}}{2\rho f}\frac{m^{4}}{f^{2}}.
\end{equation}

\subsubsection{Interacting theory}

The Langevin equation formalism can be extended to include the interaction terms of the influence functional \eqref{eq:Seffref}. We divide our analysis into three different types of interactions, one for each power of $\pia$. 

\paragraph{Linear terms in $\pia$} The terms of \Eq{eq:Seffmatch} linear in $\pia$ are
\begin{align}
S_{\text{eff}} \supset  \int \dd^4 x \Big[\frac{ \left(8\gamma_2 - \gamma\right)}{2f^{2}_{\pi}}    a^2\pir^{\prime2} \pia	+ \frac{\gamma}{2f_{\pi}^{2}} a^2	\left(\partial_i \pir \right)^2 \pia\Big].
\end{align}
These terms being already linear in $\pia$, their inclusion in the Langevin equation is straightforward, leading to\footnote{If these terms were proportional to derivatives of $\pia$, we would need to integrate by parts, which is further discussed in \App{app:Langevin}.}
\begin{equation}
\pi^{\prime \prime}+\left( 2H + \gamma\right)a\pi'-\partial_{i}^{2}\pi=a^2 \Os+\frac{ \left(8\gamma_2 - \gamma\right)}{2f^{2}_{\pi}}  \pi^{\prime2}+\frac{\gamma}{2f_{\pi}^{2}} 	\left(\partial_i \pi \right)^2.
\end{equation}
We can then compare with \Eq{eq:Paolo}, which leaves
\begin{equation}
\rho f=f_{\pi}^{2}\;,\qquad \mathrm{and}\qquad \;8\gamma_2 =\left(1-2 \pi \xi\right) \gamma.
\end{equation}

\paragraph{Quadratic terms in $\pia$} There exist terms that may source a bispectrum signal that are quadratic in $\pia$. They can be included into the Langevin equation as a coupling between the noise and the system's variable $\pi$. To achieve this task, we first need to rewrite the interactions as being proportional to one of the quadratic terms in $\pia$ in \Eq{eq:S2effparts}. We then modify the Hubbard–Stratonovich trick \cite{Stratonovich1957,Hubbard1959}, allowing for a small deviation in the coupling $\Os\pia$ in \Eq{eq:HStrick}. Then, assuming the quadratic noise to be much stronger than the noise-system coupling, we can work in perturbation theory. This is equivalent to the tree-level approximation at the level of the path integral over the noise $\Os$. Therefore, the Wilsonian coefficients of the open effective functional \eqref{eq:Seffref} and the coefficients appearing in the Langevin equation \Eq{eq:Paolo} are only equal at first order perturbatively. 

The term we need to consider is     
\begin{equation}
S_{\text{eff}}\supset - 2i\frac{\beta_{5}}{f_\pi^2}  \int d^{4}x a^3 \pir^{\prime} \pia^2,
\end{equation}
which couples to the quadratic noise $i\beta_{1}\pia^{2}$. The modified Hubbard–Stratonovich trick reads
\begin{align}
&\exp\left[-\int d^{4}x a^{4}\left(\beta_{1}-2 \frac{\beta_{5}}{f_\pi^2}  \frac{\pir'}{a}\right)\pia^{2}\right]=
\mathcal{N}(\pir')\int\mathcal{D}[\Os]\;\\
&\qquad \qquad \times \exp\Bigg\{\int d^{4}x a^{4}\left[-\frac{\Os^{2}}{4\beta_{1}}+i\lambda(\pir')\Os\pia\right]\Bigg\}\nonumber
\end{align}
with
\begin{equation}
\lambda^{2}(\pir)=1-2\frac{\beta_{5}}{f_\pi^2 \beta_{1}} \frac{\pir'}{a}.
\end{equation}
Note that the full Hubbard–Stratonovich trick implies that the normalisation of the path integral over $\Os$ depends on $\pir'$. Consequently, to follow this procedure we have to take the tree-level approximation where we drop the dependence on $\pir'$ from the normalisation of the integral and expand the square root in powers of $\lambda(\pir')$, leading to
\begin{align}
&\exp\left[-\int d^{4}x a^{4}\left(\beta_{1}-2 \frac{\beta_{5}}{f_\pi^2}  \frac{\pir'}{a}\right)\pia^{2}\right]=\mathcal{N}\int\mathcal{D}[\Os]\; \\
& \qquad \qquad \times \exp\Bigg[\int d^{4}x a^{4} \left(-\frac{\Os^{2}}{4\beta_{1}}+i\Os\pia-i\frac{\beta_{5}}{f_\pi^2 \beta_{1}} \frac{\pir'}{a}\Os\pia\right)\Bigg].\nonumber
\end{align}
This approach generates a perturbativity condition that is similar to demanding that the three-point function is smaller than the corresponding two-point signal, that is
\begin{align}
\frac{2 \beta_{5}\pir' \pia^2}{f_{\pi}^{2} a \beta_{1} \pia^2}\ll1.
\end{align}
At last, we can include the noise-system coupling obtained on the right-hand side of the Langevin equation, leading to
\begin{equation}
\pi^{\prime \prime}+\left( 2H + \gamma\right)a\pi'-\partial_{i}^{2}\pi=a^2 \Os- \frac{\beta_{5}}{f_\pi^2 \beta_{1}} a \pi'\Os. 
\end{equation}
Comparing with \Eq{eq:Paolo}, we recover
\begin{align}
\beta_5 = 2\pi\rho f\xi \beta_1.
\end{align}

\paragraph{Cubic terms in $\pia$}   
Finally, the terms that are cubic in $\pia$ can also be included at leading order into the Langevin equation by modifying the Hubbard–Stratonovich trick \eqref{eq:HStrick}. These terms lead to quadratic corrections in the noise appearing on the right-hand side of the Langevin equation. As we will see below, these corrections mimic non-Gaussian statistics of the noise found for instance in \cite{Creminelli:2023aly}. In particular, we need to modify the coefficient controlling $\Os^{2}$ in \Eq{eq:HStrick} by allowing it to depend on $\pia$. In order to recover the Langevin equation, we again rely on perturbation theory, assuming the quadratic noise to be much stronger than the cubic operator in $\pia$. Therefore, the Wilsonian coefficients in the open effective functional \eqref{eq:Seffref} and the coefficients in the Langevin equation are again only equal at first order in perturbation theory. 

To recover the non-Gaussian statistics of the noise found in \cite{Creminelli:2023aly}, we consider 
\begin{equation}
\quad S_{\mathrm{eff}} = \frac{\delta_1}{f_\pi^2} \int \dd^4 x  a^4 \pia^3.
\end{equation}
The modification of the Hubbard–Stratonovich trick leads to 
\begin{align}
&\exp\left[-\int d^{4}x a^{4}\left(\beta_{1}+i \frac{\delta_1}{f_\pi^2}\pia\right)\pia^{2}\right]=\mathcal{N}(\pia)\int\mathcal{D}[\Os] \\
& \qquad \qquad \times \;\exp\left[\int d^{4}x a^{4}\left(-\frac{\Os^{2}}{4\beta_{1}+4i \frac{\delta_1}{f_\pi^2}\pia}+i\Os\pia\right)\right].\nonumber
\end{align}
We expand at leading order in $\pia$ both the denominator and the normalisation constant, which leads to a term of the form $\Os^{2}\pia$, that is
\begin{align}
&\exp\left[-\int d^{4}x a^{4}\left(\beta_{1}+i \frac{\delta_1}{f_\pi^2}\pia\right)\pia^{2}\right]\approx\mathcal{N}\int\mathcal{D}[\Os] \\
&\qquad \qquad \times \;\exp\left[\int d^{4}x a^{4}\left(-\frac{\Os^{2}}{4\beta_{1}} + i\Os\pia + i\frac{\delta_{1}}{4f_\pi^2\beta_{1}^{2}} \Os^2 \pia\right) \right]. \nonumber 
\end{align}
Here, we again find a perturbativity condition
\begin{equation}
\frac{\delta_{1}\pia^3}{f_{\pi}^{2}\beta_{1} \pia^2}\ll 1,
\end{equation}
that can be related to the heuristic estimate made in \Eq{eq:5thestimate}. 
The $\Os^{2}\pia$ term enters the Langevin equation as
\begin{equation}
\pi^{\prime \prime}+\left( 2H + \gamma\right)a\pi'-\partial_{i}^{2}\pi=a^2 \Os+ \frac{ \delta_{1}}{4f_\pi^2\beta_{1}^{2}} a^2 \Os^{2}
\end{equation}
There is no direct matching with \Eq{eq:Paolo} yet, the connection can be made manifest if one introduces a field redefinition. Indeed, one can map the Gaussian noise $\Os$  to a noise with a non-Gaussian statistics through
\begin{equation}
\Os^{\text{ng}}=\Os+\frac{\delta_{1}}{4f_\pi^2\beta_{1}^{2}}\Os^{2}\;,
\end{equation}
or equivalently in Fourier space
\begin{align}
\delta\mathcal{O}_{S}^{\text{ng}}(\bfk, \eta)=\delta\mathcal{O}_{S}(\bfk,\eta)+\frac{\delta_{1}}{4f_\pi^2\beta_{1}^{2}}\int_{\bfq}\delta\mathcal{O}_{S}(\bfq, \eta)\delta\mathcal{O}_{S}(\bfk-\bfq, \eta).
\end{align}
Under this field redefinition, the non-Gaussian noise adquires a three-point function of the form
\begin{align}
\langle\Os^{\text{ng}}(\bfk_{1},(\eta_{1})\Os^{\text{ng}}(\bfk_{2},\eta_{2})\Os^{\text{ng}}(\bfk_{3},\eta_{3})\rangle'&= \delta(\eta_{1}-\eta_{2})\delta(\eta_{2}-\eta_{3})\frac{24}{a^{4}(\eta_{1})a^{4}(\eta_{2})}  \frac{\delta_1}{f_\pi^2}.
\end{align}
where we used the notation $\langle \cdot \rangle = (2\pi)^3{\delta}(\bfk_{1}+\bfk_{2}+\bfk_{3}) \langle \cdot \rangle'$. Under this construction, we obtain the matching of the last parameter
\begin{equation}
6\delta_1 = \rho f \frac{m^6}{f^3} \nu_{\mathcal{O}^3},
\end{equation}
which completes the reconstruction of the full non-linear Langevin equation of \cite{Creminelli:2023aly} from the path integral language of the open EFToI.


\section{Discussions}\label{sec:disc}

\paragraph{General summary} When energy is not conserved, imprints of new physics on observable sectors might not follow the rules of unitary EFTs. In this work, we constructed a bottom-up open EFT for the Nambu-Goldstone boson of the spontaneous breaking of time-translation symmetry during inflation in the presence of an unknown environment. The theory recovers the usual EFToI \cite{Cheung:2007st} in the unitary limit, and extends it to account for dissipation and noise generated by a surrounding environment, which is generally out-of-equilibrium. New scales enter the problem such as the dissipation scale or the amplitude of noise fluctuations. Symmetries ensure the existence of a nearly scale invariant power spectrum for the curvature perturbations, which can be compared to cosmological observations \cite{Planck:2018jri, DAmico:2019fhj, Colas:2019ret}. Non-Gaussianities are generated that peak in the equilateral configuration for large dissipation and in the folded configurations for small dissipation. The latter possibility constitutes a distinctive feature of this class of model, which departs from non-Bunch Davies initial states \cite{Holman:2007na, Chen:2006nt, Meerburg:2009ys, Agullo:2010ws, Ashoorioon:2010xg, Agarwal:2012mq, Ashoorioon:2013eia,  Albrecht:2014aga, Green:2020whw} in that it is finite for all physical kinematics. Consistency relations \cite{Maldacena:2002vr,Creminelli:2004yq, Cheung:2007sv, Creminelli:2012ed,Hinterbichler:2012nm,Assassi:2012zq,Pajer:2017hmb,Avis:2019eav} still hold due to the symmetry content of the theory. In short, the open EFToI provides an embedding for local dissipative models of inflation such as \cite{Creminelli:2023aly}.

\paragraph{New results} Some results recovered in this article were already known in the past literature, mostly established in \cite{LopezNacir:2011kk,Hongo:2018ant,Creminelli:2023aly}. Still, our work makes progress compared to the previous literature in many directions. The benefit of the current construction is the import of techniques from non-equilibrium EFTs \cite{Liu:2018kfw}, which provides a path-integral formulation for dissipative hydrodynamics \cite{Crossley:2015evo, Glorioso:2016gsa}. Our approach clarifies the comparison between the particle physics approach of unitary in-out evolution \cite{Donath:2024utn} and non-unitary in-in dynamics that emerges from the coarse-graining of UV degrees of freedom in cosmology \cite{Burgess:2022rdo, Colas:2023wxa}. As found in \cite{Crossley:2015evo, Glorioso:2016gsa, Liu:2018kfw}, imposing unitarity of the UV evolution tightly constrains the IR dynamics through a set of non-perturbative non-equilibrium constraints. Further enforcing symmetries through the in-in coset construction  \cite{Akyuz:2023lsm}, the most generic generating functional can be constructed in a systematic manner. The application of these techniques to time-translation symmetry breaking in \cite{Hongo:2018ant} has been an important milestone in the development of non-equilibrium EFTs for cosmology \cite{Ota:2024mps} at the origin of this work. While in the case of a closed system both $\pi_+$ and $\pi_-$ transform non-linearly under time-translation and boosts, the open dynamics breaks the symmetry group $G_+ \times G_-$ to its diagonal subgroup $G_{r}$ such that only $\pir$ transforms non-linearly. Through this property, the open EFToI still relates operators at different orders, not only in the unitary direction (such as the well-studied example of the speed of sound \cite{Cheung:2007st}) but also in the dissipative \cite{LopezNacir:2011kk} and diffusive directions. At last, locality ensures the existence of a radiatively-stable power counting scheme that can be truncated to a finite number of interactions for any desired precision. 

\paragraph{Further investigation} One might look to further constrain the EFT parameters. When the environment reaches thermal equilibrium, the fluctuation-dissipation relations \cite{Kubo:1966fyg} impose non-trivial constraints among the operators of the EFT \cite{Hongo:2018ant, Ota:2024mps} through the Kubo-Martin-Schwinger symmetry \cite{Crossley:2015evo, Glorioso:2016gsa, Glorioso:2017fpd}. One may eventually desire to enforce this symmetry as in \cite{Ota:2024mps} to recover warm inflation results.\footnote{Another direction would consists in using the full out-of-equilibrium treatment to investigate the equilibration properties through the emergence of the KMS symmetry in cosmological settings.} Yet, many situations of cosmological interest evade the thermal assumption \cite{Colas:2022hlq} and require the complete treatment of this work. In the opposite direction, one might also relax some of the underlying assumptions such as the locality of our EFT. Indeed, locality is not necessary \textit{per se} because the environment can mediate interactions at distance and this induces non-locality in time (and/or non-Markovianity \cite{Colas:2022kfu}). Nevertheless, we restricted ourselves to EFTs that are local in time and space as implied by an appropriate underlying separation of scales. This is an important assumption leading to a major simplification, a power counting scheme that can be truncated to a finite number of operators. The radiative stability of the power counting scheme under loop corrections and the renormalization procedure should be further studied along the line of \cite{Adshead:2017srh,Grall:2020tqc}. The non-equilibrium constraints \Eqs{eq:norm}, \eqref{eq:herm} and \eqref{eq:pos} should provide non-perturbative structure ensuring the stability at any order, yet it would be reassuring to control this in a specific example. 

\paragraph{Caveats} The most manifest blind spot of the current analysis is the focus on observables that are not dominated by the mixing with gravity. Working in the decoupling limit of \cite{Cheung:2007st} along the lines of \cite{LopezNacir:2011kk}, here the physics of the Goldstone $\pi$ was studied neglecting metric fluctuations. It would be desirable to go beyond the decoupling limit, which would require an improved understanding of diffeomorphism invariance in the Schwinger-Keldysh formalism \cite{Akyuz:2023lsm}.\footnote{Exploring corrections beyond the leading slow-roll contribution might also be relevant \cite{Green:2024hbw}, as well as resonant setups \cite{Flauger:2009ab,Flauger:2010ja,Behbahani:2011it,DuasoPueyo:2023kyh}.} More generally, a better control over global and gauge symmetries in the in-in formalism in the presence of entangling dynamics is desirable. It would help us to improve expectations for dissipative open dynamics in flat \cite{Matsumura:2023cni, Kashiwagi:2023zwj} and de Sitter \cite{Lotkov:2023vqt} spacetimes. In light of the no-go theorems \cite{Liu:2019fag,Cabass:2022rhr}, it would also be interesting to identify parity-violating operators which might be triggered when integrating out spin-$1$ particles \cite{Anber:2009ua, Peloso:2022ovc}. In addition to symmetries, causality might also play a role in restricting the range of available dynamics. In the flat space case, it imposes $\gamma>0$ at the level of the retarded Green function to avoid divergences \cite{kamenev_2011}. A more systematic understanding of the causality structure of the theory might help us to further constrain the open EFToI, for example constraining loop contributions as studied recently in \cite{Agui-Salcedo:2023wlq}.

\paragraph{Interesting regimes} Two limits of interest of the current construction are the unitary and semiclassical limits. The former allows us to recover the usual EFToI construction \cite{Cheung:2007st}. If the limit is conceptually clear at the level of the open effective functional, the complexity of the de Sitter propagators obscures the comparison whenever $\gamma \neq 0$. Better understanding $\gamma \rightarrow 0$ in a self consistent manner (which would certainly require a double-scaling with $\beta_1 \rightarrow 0$) would be valuable to better control systematic expansions around the unitary result. Notice that a possible direction to improve the computability in the presence of dissipation would consist in working at the level the \textit{transport equations} \cite{Dias:2016rjq, Ronayne:2017qzn, Werth:2023pfl, Pinol:2023oux, Werth:2024aui}, which may greatly simplify the computations of correlators. In this language, the unitary limit might also be more accessible. The second limit of interest is the semiclassical $\hbar \rightarrow 0$ limit. The Schwinger-Keldysh formalism is well-known for its ability to organise a systematic $\hbar$ expansion in powers of $\pia \propto \mathcal{O}(\hbar)$ \cite{Kamenev:2009jj}. Yet, this expansion is valid only around states with large occupation numbers \cite{Radovskaya:2020lns} such that the equivalence should not hold for the flat-space vacuum. In primordial cosmology, the Bunch-Davies vacuum being dynamically populated, a deeper investigation of the semiclassical expansion in this context would be valuable, with eventual connections to hybrid quantum-classical dynamics studied in \cite{Oppenheim:2020wdj, Oppenheim:2022xjr, Oppenheim:2022xjc, Oppenheim:2023izn, Layton:2023wdo, Oppenheim:2024iae, grudka2024renormalisation}. 

\paragraph{Connections with the literature} Systematic bottom-up approaches to open quantum dynamics aims at rendering visible general features following physical principles (e.g. stability, locality, perturbativity, etc.) \cite{breuerTheoryOpenQuantum2002}. It might find applications in the study of particle production in cosmology \cite{Parker:1968mv, Parker:1969au, zeldovichParticleProductionVacuum1972, Birrell:1982ix, Calzetta:2008iqa}, either when the environment thermalizes such as in warm inflation 
\cite{Berera:1995ie, Berera:1995wh, Yokoyama:1998ju, Gupta:2002kn, Moss:2007cv, Berera:2008ar, Moss:2011qc, Bastero-Gil:2011rva, Bastero-Gil:2014raa, deAlwis:2015ioa,  Ballesteros:2023dno, Mirbabayi:2022cbt,Berghaus:2024zfg, tinwala2024thermal,Cheng:2024uvn} or in the larger framework of dissipative models of inflation \cite{Boyanovsky:1994me, Boyanovsky:1996rw,  Boyanovsky:1996ab, Anber:2009ua, Green:2009ds, Barnaby:2011qe, deAlwis:2015ioa, Peloso:2022ovc, Putti:2024uyr, Creminelli:2023aly, Briaud:2023pky}. This construction might also help in fostering formal connections between stochastic inflation \cite{Starobinsky:1986fx, Nambu:1987ef, Nambu:1988je, Kandrup:1988sc, Nakao:1988yi, Nambu:1989uf, Salopek:1990jq, Mollerach:1990zf, Brandenberger:1992sr, Starobinsky:1994bd, Finelli:2008zg, Garbrecht:2013coa, Fujita:2014tja, Vennin:2015hra, Grain:2017dqa, Pinol:2020cdp, Launay:2024qsm, Tinwala:2024wod} and open EFT techniques 
\cite{breuerTheoryOpenQuantum2002, Koks:1996ga, Burgess:2006jn, Wu:2006xp, Lee:2011fj, Anastopoulos:2013zya, Fukuma:2013uxa, Burgess:2014eoa, Burgess:2015ajz, Boyanovsky:2015xoa, Boyanovsky:2015jen, Boyanovsky:2015tba, Nelson:2016kjm, Hollowood:2017bil,Shandera:2017qkg,Boyanovsky:2018fxl,Boyanovsky:2018soy, Martin:2018zbe, Burrage:2018pyg, Bohra:2019wxu, Akhtar:2019qdn, Kaplanek:2019dqu, Brahma:2020zpk, Kaplanek:2020iay, Rai:2020edx, Burgess:2021luo, Kaplanek:2021fnl, Brahma:2021mng, Banerjee:2021lqu, Oppenheim:2022xjr, Brahma:2022yxu, Kaplanek:2022xrr, Kaplanek:2022opa, Colas:2022hlq, Colas:2022kfu, DaddiHammou:2022itk, Burgess:2022nwu, Burgess:2022rdo, Cao:2022kjn, Prudhoe:2022pte, Kading:2022jjl, Colas:2023wxa, Brahma:2023hki, Kading:2023mdk, Sharifian:2023jem, Alicki:2023tfz, Alicki:2023rfv, Ning:2023ybc, Colas:2024xjy, Bhattacharyya:2024duw, Burgess:2024eng}, with an eventual window on quantum diffusion during inflation and its known phenomenology of primordial black holes and associated gravitational waves \cite{Carr:1974nx, Garcia-Bellido:2017mdw, Germani:2017bcs, Pattison:2017mbe, Ezquiaga:2018gbw, Ezquiaga:2019ftu, Kalaja:2019uju, Vennin:2020kng, Ballesteros:2021fsp, Animali:2022otk, Gow:2022jfb, Ezquiaga:2022qpw, Ballesteros:2022hjk, LISACosmologyWorkingGroup:2023njw, Briaud:2023eae, Choudhury:2023jlt, Vennin:2024yzl, Animali:2024jiz, Choudhury:2024jlz}. In the top-down context, both stochastic inflation \cite{Starobinsky:1994bd, Serreau:2013psa, Ezquiaga:2019ftu, Gorbenko:2019rza, Kamenshchik:2021tjh, Miao:2021gic, Cohen:2021fzf, Cohen:2021jbo, Gow:2022jfb, Cohen:2022clv, Woodard:2023rqo, Cespedes:2023aal} and open EFT techniques \cite{Burgess:2014eoa, Burgess:2015ajz, Kaplanek:2019vzj, Kaplanek:2019dqu, Colas:2022hlq, Chaykov:2022zro, Burgess:2024eng} have demonstrated ability to go beyond standard perturbation theory by implementing non-perturbative resummations. It would be interesting to further explore to which extend these resummation techniques can be used in the bottom-up context, in connection with the development of non-perturbative methods for cosmology \cite{Celoria:2021vjw, Creminelli:2024cge}.

\paragraph{Future works} In this context, the construction of bottom-up \textit{open} Effective Field Theories for Dark Energy \cite{Gubitosi:2012hu, Piazza:2013coa, Gleyzes:2013ooa, Raveri:2014cka, Frusciante:2016xoj} and the Large Scale Structure \cite{Baumann:2010tm, Carrasco:2012cv} can be a direct extension of the current work. Applications to the late-time expansion of the universe would explore the remaining parameter space \cite{Peirone:2017lgi, deRham:2018red, deBoe:2024gpf} after GW170817 \cite{LIGOScientific:2017vwq, LIGOScientific:2017zic} and DESI 2024 \cite{DESI:2024mwx} results beyond unitary constructions. The relation between modified gravity and out-of-equilibrium thermodynamics \cite{Giardino:2022sdv, Miranda:2024dhw}, entropic forces \cite{Garcia-Bellido:2021idr, Espinosa-Portales:2021cac, Arjona:2021uxs} and stochastic diffusion \cite{Landau:2022mhm, Oppenheim:2024rcp} is an ever increasing field for which open EFT might provide a systematic framework. The path integral formulation of viscous fluids \cite{Crossley:2015evo, Liu:2018kfw} whose application to cosmology is at its dawn \cite{Ota:2024mps} is a promising avenue, which should reduce in its simplest version to the original formulation of the EFTofLSS \cite{Baumann:2010tm, Carrasco:2012cv}. At last, our formalism is a starting point to study quantum information aspects of inflation (entanglement growth \cite{Campo:2005sv,Lello:2013bva, Lello:2013qza, Maldacena:2015bha, Choudhury:2016cso,Martin:2016tbd,Martin:2017zxs,Bolis:2019fmq, Ando:2020kdz,Espinosa-Portales:2022yok, Adil:2022rgt,Tejerina-Perez:2024opu}, purity and decoherence \cite{Colas:2021llj, Burgess:2022nwu, Colas:2022kfu, Colas:2024xjy, Burgess:2024eng, Colas:2024ysu}, \textit{etc.}). Connections between positivity bounds and entanglement entropy were recently unravelled \cite{Cheung:2023hkq, Aoude:2024xpx}, rising the hope of overcoming obstructions related to the lack of well-defined amplitudes in primordial cosmology (despite the efforts into this direction \cite{Melville:2023kgd,Donath:2024utn,Melville:2024ove}). Accounting for the singularity of cosmology, such as the lack of IR unitary description, might by the route forward to the extension of particle physics techniques to quantum field theory in curved spacetime. 

\subsection*{Acknowledgements:} We thank C. P. Burgess, S. Cespedes, P. Creminelli, J. Grain, R. Holman, G. Kaplanek, S. Melville, A. Nicolis, R. Penco, L. Santoni, B. Salehian, D. Stefanyszyn, V. Vennin and G. Villa for the insightful discussions. T.C. warmly thanks D. Comelli, F. Piazza, A. Tolley and F. Vernizzi for inspiring this research direction. This work has been supported by STFC consolidated grant ST/X001113/1, ST/T000694/1, ST/X000664/1 and EP/V048422/1. S.A.S. is supported by a Harding Distinguished Postgraduate Scholarship. 


\appendix

\section{Derivation of $S_{\mathrm{eff}}$ constraints}\label{app:deriv}

The derivation follows from \cite{Glorioso:2016gsa}. Let us consider a UV evolution operator $\widehat{\mathcal{U}}(t,t_0)$ under which the UV state evolves according to
\begin{align}
    \widehat{\rho}(t) = \widehat{\mathcal{U}}(t,t_0)  \ket{\Omega} \bra{\Omega}\widehat{\mathcal{U}}^\dag(t,t_0),
\end{align}
starting from an initial vacuum state $\ket{\Omega}$. The reduced density matrix is obtained by tracing out the environmental degrees of freedom  denoted $\sigma$ such that 
\begin{align}
    \widehat{\rho}_{\mathrm{red}} (t) &= \mathrm{Tr}_\sigma  \left[  \widehat{\mathcal{U}}(t,t_0)  \ket{\Omega} \bra{\Omega}\widehat{\mathcal{U}}^\dag(t,t_0) \right],\\
    &=\int \dd \sigma \bra{\sigma}  \widehat{\mathcal{U}}(t,t_0)  \ket{\Omega} \bra{\Omega} \widehat{\mathcal{U}}^\dag(t,t_0) \ket{\sigma}.
\end{align}
Let us consider the field-basis matrix element of reduced density matrix
\begin{align}
    \rho_{\pi \pi'} (t) &\equiv \bra{\pi} \widehat{\rho}_{\mathrm{red}} (t)\ket{\pi'} \\
    &= \int \dd \sigma  \bra{\pi} \otimes \bra{\sigma}  \widehat{\mathcal{U}}(t,t_0)  \ket{\Omega} \bra{\Omega} \widehat{\mathcal{U}}^\dag(t,t_0)  \ket{\pi'} \otimes \ket{\sigma}\\
    &=\int \dd \pi_i \dd \pi_i^\prime \int \dd \sigma \int \dd \sigma_i \dd \sigma_i^\prime \bra{\pi} \otimes \bra{\sigma}  \widehat{\mathcal{U}}(t,t_0) \ket{\pi_i} \otimes \ket{\sigma_i} \nonumber \\
    &\qquad \qquad  \rho^{(0)}_{\pi_i \pi_i^\prime} \rho^{(0)}_{\sigma_i \sigma_i^\prime}\bra{\pi_i^\prime} \otimes \bra{\sigma_i^\prime} \widehat{\mathcal{U}}^\dag(t,t_0)\ket{\pi'} \otimes \ket{\sigma},
\end{align}
where $\ket{\pi}$, $\ket{\sigma}$ are eigenstates of the position operators $\widehat{\pi}$, $\widehat{\sigma}$ and we used four representations of the identity, two on each side of the vacuum density matrix. The initial matrix elements are
\begin{align}
    \rho^{(0)}_{\pi_i \pi_i^\prime} &\equiv \braket{\pi_i|\Omega_\pi} \braket{\Omega_\pi|\pi_i^\prime}, \\
    \rho^{(0)}_{\sigma_i \sigma_i^\prime} &\equiv \braket{\sigma_i|\Omega_\sigma} \braket{\Omega_\sigma|\sigma_i^\prime},
\end{align}
where we consider $\ket{\Omega} = \ket{\Omega_\pi} \otimes \ket{\Omega_\sigma}$. The path integral representation of the evolution operator is 
\begin{align}
    \bra{\pi} \otimes \bra{\sigma}  \widehat{\mathcal{U}}(t,t_0) \ket{\pi_i} \otimes \ket{\sigma_i} &= \int_{\pi_i}^\pi \mathcal{D}\pi_+ \int_{\sigma_i}^\sigma \mathcal{D}\sigma_+  \ee^{i S_0\left[\pi_+,\sigma_+ \right]}, \\
    \bra{\pi_i^\prime} \otimes \bra{\sigma_i^\prime} \widehat{\mathcal{U}}^\dag(t,t_0)\ket{\pi'} \otimes \ket{\sigma}  &= \int_{\pi_i^\prime}^{\pi'} \mathcal{D}\pi_- \int_{\sigma_i^\prime}^\sigma \mathcal{D}\sigma_-  \ee^{-i S_0\left[\pi_-,\sigma_- \right]},
\end{align}
such that we obtain 
\begin{align}
    \rho_{\pi \pi'} (t) = \int \dd \pi_i \dd \pi_i^\prime \int_{\pi_i}^\pi \mathcal{D}\pi_+ \int_{\pi_i^\prime}^{\pi'} \mathcal{D}\pi_- \ee^{i S_{\mathrm{eff}}\left[ \pi_+, \pi_-\right]} \rho^{(0)}_{\pi_i \pi_i^\prime},
\end{align}
with the influence functional 
\begin{align}
     \ee^{i S_{\mathrm{eff}}\left[ \pi_+, \pi_-\right]} = \int \dd \sigma \int \dd \sigma_i \dd \sigma_i^\prime  \int_{\sigma_i}^\sigma \mathcal{D}\sigma_+ \int_{\sigma_i^\prime}^\sigma \mathcal{D}\sigma_-  \ee^{i S_0\left[\pi_+,\sigma_+ \right]-i S_0\left[\pi_-,\sigma_- \right]} \rho^{(0)}_{\sigma_i \sigma_i^\prime}.
\end{align}

The central step of the proof is to consider $\sigma$ evolves as if $\pi$ is a background (external source) see Appendix A of \cite{Glorioso:2016gsa}. In this case, we can consider the sourced evolution 
\begin{align}
     \int_{\sigma_i}^\sigma \mathcal{D}\sigma_+  \ee^{i S_0\left[\pi_+,\sigma_+ \right]} &= \bra{\sigma}   \widehat{\mathcal{U}}(t,t_0;\{\pi_+\})\ket{\sigma_i}, \\ 
     \int_{\sigma_i^\prime}^\sigma \mathcal{D}\sigma_-  \ee^{-i S_0\left[\pi_-,\sigma_- \right]} &= \left[\bra{\sigma}   \widehat{\mathcal{U}}(t,t_0;\{\pi_-\})\ket{\sigma_i^\prime}\right]^\dag,
\end{align}
where $\widehat{\mathcal{U}}(t,t_0;\{\pi_+\})$ is the sourced evolution acting on $\mathcal{H}_\sigma$ only. One can the reconsider the influence functional as being 
\begin{align}
    \ee^{i S_{\mathrm{eff}}\left[ \pi_+, \pi_-\right]} &= \int \dd\sigma \bra{\sigma} \widehat{\mathcal{U}}(t,t_0;\{\pi_+\}) \overbrace{\left[\int \dd \sigma_i  \ket{\sigma_i} \bra{\sigma_i} \right]}^{\mathbb{I}}\ket{\Omega_\sigma} \\
    &\bra{\Omega_\sigma} \underbrace{\left[ \int \dd \sigma_i^\prime  \ket{\sigma_i^\prime}  \bra{\sigma_i^\prime} \right]}_{\mathbb{I}}\widehat{\mathcal{U}}^\dag(t,t_0;\{\pi_-\}) \ket{\sigma} \nonumber 
\end{align}
which finally reduces to 
\begin{align}
    \ee^{i S_{\mathrm{eff}}\left[ \pi_+, \pi_-\right]} &= \bra{\Omega_\sigma} \widehat{\mathcal{U}}^\dag(t,t_0;\{\pi_-\})
    \overbrace{\left[\int \dd \sigma \ket{\sigma} \bra{\sigma} \right]}^{\mathbb{I}}\widehat{\mathcal{U}}(t,t_0;\{\pi_+\}) \ket{\Omega_\sigma},
\end{align}
that is 
\begin{tcolorbox}[%
enhanced, 
breakable,
skin first=enhanced,
skin middle=enhanced,
skin last=enhanced
]
\paragraph{Influence functional as a transition probability}
\begin{align}\label{eq:rate}
    \ee^{i S_{\mathrm{eff}}\left[ \pi_+, \pi_-\right]}|  &= \braket{\Omega^{ \{\pi_- \} }_\sigma (t) | \Omega^{ \{\pi_+ \} }_\sigma (t) } 
\end{align}
\end{tcolorbox}
\noindent where we defined the sourced-evolved states
\begin{align}
   \ket{ \Omega^{ \{\pi_+ \} }_\sigma (t)} &=                \widehat{\mathcal{U}}(t,t_0;\{\pi_+\}) \ket{\Omega_\sigma}, \\
    \bra{ \Omega^{ \{\pi_- \} }_\sigma (t)}&= \bra{\Omega_\sigma}\widehat{\mathcal{U}}^\dag(t,t_0;\{\pi_-\}).
\end{align}
In this sense, the influence functional can be interpreted as a probability of obtaining a configuration $(\pi_+,\pi_-)$ given the unitary evolution of the UV theory and taking into consideration the lack of knowledge about the environment. The transition rate is physical if
\begin{align}
    ||\braket{\Omega^{ \{\pi_- \} }_\sigma (t) | \Omega^{ \{\pi_+ \} }_\sigma (t) } ||^2 \leq 1
\end{align}
such that 
\begin{align}
    |\ee^{i S_{\mathrm{eff}}\left[ \pi_+, \pi_-\right]}| \leq 1 \qquad \Rightarrow \qquad \Ima S_{\mathrm{eff}} \left[\pi_+,\pi_-\right] &\geq 0. 
\end{align}
Then, from \Eq{eq:rate}, one can easily see that 
\begin{align}
    \ee^{-i S^*_{\mathrm{eff}}\left[ \pi_+, \pi_-\right]} = \braket{\Omega^{ \{\pi_+ \} }_\sigma (t) | \Omega^{ \{\pi_- \} }_\sigma (t) } = \ee^{i S_{\mathrm{eff}}\left[ \pi_-, \pi_+\right]}
\end{align}
from which we deduce
\begin{align}
    S_{\mathrm{eff}} \left[\pi_+,\pi_-\right] &= - 	S^*_{\mathrm{eff}} \left[\pi_-,\pi_+\right].
\end{align}
Lastly, the causality structure which is obvious in the unitary theory, $S_0[\pi_+] - S_0[\pi_-] = 0$ if $\pi_+ = \pi_-$, is much less straightforward in the non-unitary case yet still holds as
\begin{align}
    \ee^{i S_{\mathrm{eff}}\left[ \pi_+, \pi_+\right]}  &= \braket{\Omega^{ \{\pi_+ \} }_\sigma (t) | \Omega^{ \{\pi_+ \} }_\sigma (t) }  =1 \qquad \Rightarrow \qquad S_{\mathrm{eff}}\left[ \pi_+, \pi_+\right] = 0.
\end{align}

\section{de Sitter Keldysh functions}\label{app:dSKG}

This appendix gathers the outcome of the lenghty integrals appearing in the main text in the expression of \Eq{eq:GK1}. Indeed, to analytically access the Keldysh function given in \Eq{eq:GK1}, we have three integrals to compute
\begin{align}
A^{(1)}_{\nu_\gamma}(z) &\equiv \int_{z}^{\infty} \dd z' z^{\prime2 - 2{\nu_\gamma}} J^2_{\nu_\gamma}(z'), \\
        B^{(1)}_{\nu_\gamma}(z) &\equiv \int_{z}^{\infty} \dd z' z^{\prime2 - 2{\nu_\gamma}} J_{\nu_\gamma}(z') Y_{\nu_\gamma}(z'), \\
        C^{(1)}_{\nu_\gamma}(z)&\equiv \int_{z}^{\infty} \dd z' z^{\prime2 - 2{\nu_\gamma}} Y^2_{\nu_\gamma}(z'),
\end{align}
that lead for the first contribution to 
\begin{align}
    A^{(1)}_{\nu_\gamma}(z) &= \frac{1}{4} \bigg[\frac{\Gamma ({\nu_\gamma} -1)}{\Gamma \left({\nu_\gamma} -\frac{1}{2}\right) \Gamma \left(2 {\nu_\gamma} -\frac{1}{2}\right)} \label{eq:Fbeta} \nonumber\\
&\qquad \qquad -z^3 \Gamma \left({\nu_\gamma} +\frac{1}{2}\right) \, _2\tilde{F}_3\left(\frac{3}{2},{\nu_\gamma} +\frac{1}{2};\frac{5}{2},{\nu_\gamma} +1,2 {\nu_\gamma} +1;-z^2\right)\Bigg], 
\end{align}
for the second contribution to 
\begin{align} 
B^{(1)}_{\nu_\gamma}(z) 
 &=\frac{z^{3-2 {\nu_\gamma} } \, _2F_3\left(\frac{1}{2},\frac{3}{2}-{\nu_\gamma} ;1-{\nu_\gamma} ,\frac{5}{2}-{\nu_\gamma} ,{\nu_\gamma} +1;-z^2\right)}{3 \pi  {\nu_\gamma} -2 \pi  {\nu_\gamma} ^2}  \label{eq:Gbeta} \nonumber \\
 +&\frac{z^3 \Gamma (-{\nu_\gamma} ) \, _2F_3\left(\frac{3}{2},{\nu_\gamma} +\frac{1}{2};\frac{5}{2},{\nu_\gamma} +1,2 {\nu_\gamma} +1;-z^2\right)}{3 \sqrt{\pi } \Gamma \left(\frac{1}{2}-{\nu_\gamma} \right) \Gamma (2 {\nu_\gamma} +1)}-\frac{\Gamma \left(\frac{3}{2}-{\nu_\gamma} \right)}{4 \Gamma (2-{\nu_\gamma} ) \Gamma \left(2 {\nu_\gamma} -\frac{1}{2}\right)},
\end{align}
and for the third contribution to
\begin{align}
 \label{eq:Hbeta}
C^{(1)}_{\nu_\gamma}(z) &= \frac{2 \cot (\pi  {\nu_\gamma} ) z^{3-2 {\nu_\gamma} } \, _2F_3\left(\frac{1}{2},\frac{3}{2}-{\nu_\gamma} ;1-{\nu_\gamma} ,\frac{5}{2}-{\nu_\gamma} ,{\nu_\gamma} +1;-z^2\right)}{3 \pi  {\nu_\gamma} -2 \pi  {\nu_\gamma} ^2} \nonumber \\
+&\frac{4^{{\nu_\gamma} } \Gamma ({\nu_\gamma} )^2 z^{3-4 {\nu_\gamma} } \, _2F_3\left(\frac{3}{2}-2 {\nu_\gamma} ,\frac{1}{2}-{\nu_\gamma} ;1-2 {\nu_\gamma} ,\frac{5}{2}-2 {\nu_\gamma} ,1-{\nu_\gamma} ;-z^2\right)}{\pi ^2 (4 {\nu_\gamma} -3)}\nonumber\\
-&\frac{4^{-{\nu_\gamma} } z^3 \cot ^2(\pi  {\nu_\gamma} ) \, _2F_3\left(\frac{3}{2},{\nu_\gamma} +\frac{1}{2};\frac{5}{2},{\nu_\gamma} +1,2 {\nu_\gamma} +1;-z^2\right)}{3 \Gamma ({\nu_\gamma} +1)^2}\nonumber \\
+&\frac{\Gamma ({\nu_\gamma} -1)}{4 \Gamma \left({\nu_\gamma} -\frac{1}{2}\right) \Gamma \left(2 {\nu_\gamma} -\frac{1}{2}\right)}-\frac{4^{3 {\nu_\gamma} -2} \sin (2 \pi  {\nu_\gamma} ) \Gamma (3-4 {\nu_\gamma} )}{\Gamma (1-{\nu_\gamma} ) \Gamma (2-{\nu_\gamma} )}.
\end{align}

\section{de Sitter power spectrum with derivative noises}\label{app:Pk}

In this appendix, we compute the inflationary power spectrum in the presence of derivatives noises $\pia^{\prime2}$ and $\left(\partial_i \pia \right)^2$. It amounts to explicitly evaluate \Eq{eq:GK2} and \Eq{eq:GK3} in the coincident time limit. We proceed as we did in the main text in \Sec{subsec:dSPk} for the $\pia^2$ noise. 

\subsection*{Time-derivative noise} 

Injecting \Eq{eq:GRfinsimpv2} in \Eq{eq:GK2}, we obtain the Keldysh propagator 
\begin{align}
    G^K_2(k;\eta_1,\eta_2) &= i\frac{\pi^2  H^2}{4 k^3} (\beta_4 - \beta_2 ) (z_1 z_2)^{\nu_\gamma} \bigg[ Y_{\nu_\gamma}(z_1) Y_{\nu_\gamma}(z_2) F^{(2)}_{\nu_\gamma}(z_2) + J_{\nu_\gamma}(z_1) J_{\nu_\gamma}(z_2) H^{(2)}_{\nu_\gamma}(z_2) \nonumber \\
    +& J_{\nu_\gamma}(z_1)Y_{\nu_\gamma}(z_2)G^{(2)}_{\nu_\gamma}(z_2) + J_{\nu_\gamma}(z_2)Y_{\nu_\gamma}(z_1)\tilde{G}^{(2)}_{\nu_\gamma}(z_2) \bigg] + (1 \leftrightarrow 2)
\end{align}
where 
\begin{align}
F^{(2)}_{\nu_\gamma}(z) &\equiv \int_{z}^{\infty} \dd z'       z^{\prime2 - 2{\nu_\gamma}} J_{\nu_\gamma}(z') \bigg\{\left[z^{\prime2}-2 (\nu_\gamma -1) \nu_\gamma \right] J_{\nu_\gamma }(z') \nonumber \\
        +&(\nu_\gamma -1) z' J_{\nu_\gamma -1}(z')-(\nu_\gamma -2) z' J_{\nu_\gamma +1}(z') \bigg\}\\
        G^{(2)}_{\nu_\gamma}(z) &\equiv \int_{z}^{\infty} \dd z' z^{\prime2 - 2{\nu_\gamma}} Y_{\nu_\gamma}(z') \bigg\{-\left[z^{\prime2}-2 (\nu_\gamma -1) \nu_\gamma \right] J_{\nu_\gamma }(z') \nonumber \\
        -&(\nu_\gamma -1) z' J_{\nu_\gamma -1}(z')+(\nu_\gamma -2) z' J_{\nu_\gamma +1}(z') \bigg\} \\
        \tilde{G}^{(2)}_{\nu_\gamma}(z) &\equiv \int_{z}^{\infty} \dd z' z^{\prime2 - 2{\nu_\gamma}} J_{\nu_\gamma}(z') \bigg\{ -\left[z^{\prime2}-2 (\nu_\gamma -1) \nu_\gamma \right] Y_{\nu_\gamma }(z')\nonumber \\
        +&(1-\nu_\gamma )z' Y_{\nu_\gamma -1}(z')+(\nu_\gamma -2) z' Y_{\nu_\gamma +1}(z') \bigg\} \\
        H^{(2)}_{\nu_\gamma}(z)&\equiv \int_{z}^{\infty} \dd z' z^{\prime2 - 2{\nu_\gamma}} Y_{\nu_\gamma}(z') \bigg\{ \left[z^{\prime2}-2 (\nu_\gamma-1) \nu_\gamma\right] Y_{\nu_\gamma}(z') \nonumber \\
        -&(1-\nu_\gamma) z'Y_{\nu_\gamma-1}(z')-(\nu_\gamma-2) z' Y_{\nu_\gamma+1}(z')\bigg\}
\end{align} are complicated functions easily obtained from Mathematica (the integrals being convergent) yet too lengthy to have any practical interest being explicitly spelled here. Just as above, the power spectrum is obtained in the coincident time limit
\begin{align}
P_\zeta(k) &= \frac{\pi^2}{8 k^3} (\beta_4 - \beta_2 ) \frac{H^4}{f_\pi^4}z^{2\nu_\gamma} \bigg[Y_{\nu_\gamma}(z) Y_{\nu_\gamma}(z) F^{(2)}_{\nu_\gamma}(z) + J_{\nu_\gamma}(z) J_{\nu_\gamma}(z) H^{(2)}_{\nu_\gamma}(z) \nonumber \\
+& J_{\nu_\gamma}(z)Y_{\nu_\gamma}(z)G^{(2)}_{\nu_\gamma}(z) + J_{\nu_\gamma}(z)Y_{\nu_\gamma}(z)\tilde{G}^{(2)}_{\nu_\gamma}(z) \bigg]
\end{align}
In the super-Hubble regime $z\ll 1$,\footnote{In the sub-Hubble regime $z \gg 1$, the power spectrum reads 
\begin{align}\label{eq:subHPk2}
    P_\zeta(k;\eta) &=  \frac{1}{2k^3} (\beta_4- \beta_2)\frac{H^4}{f_\pi^4} \frac{z^3}{\nu_\gamma-2}
\end{align}
which turns negative for $\nu_\gamma < 2$ (or equivalently $\gamma < H$), indicating a positivity violation in small dissipation regime. Moreover, the sub-Hubble power spectrum scale as $z^3$ which might violate Hadamard condition in the asymptotic past.} the power spectrum again freezes and we obtain
\begin{align}
\Delta^2_\zeta(k) &=   \frac{15}{32}(\beta_4- \beta_2)\frac{H^4}{f_\pi^4} 2^{2\nu_\gamma} \frac{\Gamma\left(\nu_\gamma-2\right)\Gamma\left(\nu_\gamma\right)^2}{\Gamma\left(\nu_\gamma - \frac{3}{2}\right)\Gamma\left(2\nu_\gamma - \frac{1}{2}\right)}. \label{eq:PK2SH}
\end{align}
One can expand this result in the large dissipation regime, leading to 
\begin{align}
    \Delta^2_\zeta(k)  &\propto (\beta_4- \beta_2)  \frac{H^4}{f_\pi^4}  \sqrt{\frac{H}{\gamma}}\left[1 + \mathcal{O}\left(\frac{H}{\gamma}\right) \right],
\end{align}
which also easily fulfils the observational constraint $\Delta^2_\zeta = 10^{-9}$ under imposing some hierarchy between the various scales. The small dissipation regime, on the contrary, leads to a positivity violation with an unphysical negative power spectrum for $\beta_4- \beta_2 > 0$, which is consistent with the sub-Hubble behaviour discussed below \Eq{eq:subHPk2}. It appears $\pia^{\prime 2}$ generates an unphysical regime whenever $\nu_\gamma < 2$ (or equivalently $\gamma < H$).

\subsection*{Spatial-derivative noise} 
 
Injecting \Eq{eq:GRfinsimpv2} in \Eq{eq:GK3}, we obtain the Keldysh propagator 
\begin{align}
    G^K_3(k;\eta_1,\eta_2) = i\frac{\pi^2 \beta_2 H^2}{4 k^3}  &(z_1 z_2)^{\nu_\gamma} \bigg\{ Y_{\nu_\gamma}(z_1) Y_{\nu_\gamma}(z_2) F^{(3)}_{\nu_\gamma}(z_2) + J_{\nu_\gamma}(z_1) J_{\nu_\gamma}(z_2) H^{(3)}_{\nu_\gamma}(z_2) \nonumber \\
    -& \left[J_{\nu_\gamma}(z_1)Y_{\nu_\gamma}(z_2) + J_{\nu_\gamma}(z_2)Y_{\nu_\gamma}(z_1)\right]G^{(3)}_{\nu_\gamma}(z_2) \bigg\} + (1 \leftrightarrow 2)
\end{align}
where 
\begin{align}
F^{(3)}_{\nu_\gamma}(z) &\equiv \int_{z}^{\infty} \dd z' z^{\prime4 - 2{\nu_\gamma}} J^2_{\nu_\gamma}(z') \\
        G^{(3)}_{\nu_\gamma}(z) &\equiv \int_{z}^{\infty} \dd z' z^{\prime4 - 2{\nu_\gamma}} J_{\nu_\gamma}(z') Y_{\nu_\gamma}(z') \\
        H^{(3)}_{\nu_\gamma}(z)&\equiv \int_{z}^{\infty} \dd z' z^{\prime4 - 2{\nu_\gamma}} Y^2_{\nu_\gamma}(z')
\end{align} are complicated functions which are explicitly given for the first contribution by    
\begin{align} \label{eq:F3beta}
     F^{(3)}_{\nu_\gamma}(z) &= \frac{3}{8} \left[\frac{\Gamma (\nu_\gamma -2)}{\Gamma \left(\nu_\gamma -\frac{3}{2}\right) \Gamma \left(2 \nu_\gamma -\frac{3}{2}\right)}-z^5 \Gamma \left(\nu_\gamma +\frac{1}{2}\right) \, _2\tilde{F}_3\left(\frac{5}{2},\nu_\gamma +\frac{1}{2};\frac{7}{2},\nu_\gamma +1,2 \nu_\gamma +1;-z^2\right)\right],
\end{align}
for the second contribution by
\begin{align}\label{eq:G3beta}
    G^{(3)}_{\nu_\gamma}(z) &= \frac{z^{5-2 \nu_\gamma } \, _2F_3\left(\frac{1}{2},\frac{5}{2}-\nu_\gamma ;1-\nu_\gamma ,\frac{7}{2}-\nu_\gamma ,\nu_\gamma +1;-z^2\right)}{5 \pi  \nu_\gamma -2 \pi  \nu_\gamma ^2} \nonumber \\
    +&\frac{z^5 \Gamma (-\nu_\gamma ) \, _2F_3\left(\frac{5}{2},\nu_\gamma +\frac{1}{2};\frac{7}{2},\nu_\gamma +1,2 \nu_\gamma +1;-z^2\right)}{5 \sqrt{\pi } \Gamma \left(\frac{1}{2}-\nu_\gamma \right) \Gamma (2 \nu_\gamma +1)}-\frac{3 \Gamma \left(\frac{5}{2}-\nu_\gamma \right)}{8 \Gamma (3-\nu_\gamma ) \Gamma \left(2 \nu_\gamma -\frac{3}{2}\right)},
\end{align}
and for the third contribution by 
\begin{align}\label{eq:H3beta}
    H^{(3)}_{\nu_\gamma}(z) &=\frac{2 \cot (\pi  \nu_\gamma ) z^{5-2 \nu_\gamma } \, _2F_3\left(\frac{1}{2},\frac{5}{2}-\nu_\gamma ;1-\nu_\gamma ,\frac{7}{2}-\nu_\gamma ,\nu_\gamma +1;-z^2\right)}{5 \pi  \nu_\gamma -2 \pi  \nu_\gamma ^2} \nonumber \\
    +&\frac{4^{\nu_\gamma } \Gamma (\nu_\gamma )^2 z^{5-4 \nu_\gamma } \, _2F_3\left(\frac{5}{2}-2 \nu_\gamma ,\frac{1}{2}-\nu_\gamma ;1-2 \nu_\gamma ,\frac{7}{2}-2 \nu_\gamma ,1-\nu_\gamma ;-z^2\right)}{\pi ^2 (4 \nu_\gamma -5)} \nonumber \\ 
    -&\frac{4^{-\nu_\gamma } z^5 \cot ^2(\pi  \nu_\gamma ) \, _2F_3\left(\frac{5}{2},\nu_\gamma +\frac{1}{2};\frac{7}{2},\nu_\gamma +1,2 \nu_\gamma +1;-z^2\right)}{5 \Gamma (\nu_\gamma +1)^2}+\frac{3 \Gamma (\nu_\gamma -2)}{8 \Gamma \left(\nu_\gamma -\frac{3}{2}\right) \Gamma \left(2 \nu_\gamma -\frac{3}{2}\right)} \nonumber \\
    +&\frac{3 \sin (2 \pi  \nu_\gamma ) \Gamma \left(\frac{5}{2}-\nu_\gamma \right) \Gamma \left(\frac{5}{2}-2 \nu_\gamma \right)}{4 \pi  \Gamma (3-\nu_\gamma )}.
\end{align}
Just as above, the power spectrum is obtained in the coincident time limit
\begin{align}
P_\zeta(k) &= \frac{\pi^2}{8 k^3} \beta_2 \frac{H^4}{f_\pi^4}z^{2\nu_\gamma} \Big[ Y^2_{\nu_\gamma}(z)  F^{(3)}_{\nu_\gamma}(z) + J^2_{\nu_\gamma}(z) H^{(3)}_{\nu_\gamma}(z) -2J_{\nu_\gamma}(z)Y_{\nu_\gamma}(z) G^{(3)}_{\nu_\gamma}(z) \Big]
\end{align}
In the super-Hubble regime $z\ll 1$,\footnote{In the sub-Hubble regime $z \gg 1$, the power spectrum takes the same form as the one obtained from $\pia^{\prime2}$, that is 
\begin{align}\label{eq:subHPk3}
    P_\zeta(k;\eta) &=   \frac{1}{2k^3} \beta_2\frac{H^4}{f_\pi^4} \frac{z^3}{\nu_\gamma-2} 
\end{align}
which again turns negative for $\nu_\gamma < 2$ (or equivalently $\gamma < H$), indicating a positivity violation in small dissipation regime. } the power spectrum again freezes and we obtain
\begin{align}
\Delta^2_\zeta(k) &=  \frac{3}{16} \beta_2\frac{H^4}{f_\pi^4} 2^{2\nu_\gamma} \frac{\Gamma\left(\nu_\gamma-2\right)\Gamma\left(\nu_\gamma\right)^2}{\Gamma\left(\nu_\gamma - \frac{3}{2}\right)\Gamma\left(2\nu_\gamma - \frac{3}{2}\right)}. \label{eq:PK3SH}
\end{align}
One can expand this result in the large dissipation regime, leading to 
\begin{align}
    \Delta^2_\zeta(k)  &\propto \beta_2 \frac{H^4}{f_\pi^4}  \sqrt{\frac{\gamma}{H}} \left[1 + \mathcal{O}\left(\frac{H}{\gamma}\right) \right],
\end{align}
which is enhanced due to the $\sqrt{\gamma/H}$ yet may still fulfil the observational constraint $\Delta^2_\zeta = 10^{-9}$ under imposing some hierarchy between the various scales. The small dissipation regime, on the contrary, again leads to a positivity violation, consistent with the sub-Hubble behaviour discussed below \Eq{eq:subHPk3}. It appears that $(\partial_i \pia)^2$, just as $\pia^{\prime2}$, generates an unphysical regime whenever $\nu_\gamma < 2$ (or equivalently $\gamma < H$).

\section{Unitary results in Keldysh basis}\label{app:unit}

In this appendix, we derive in flat space some tree level unitary results useful to apprehend to formalism and build physical intuition on use of the Keldysh basis.

\paragraph{Propagators}

For a free bosonic theory, the quadratic open effective functional reads \cite{Kamenev:2009jj}
\begin{align}
S^{(2)}_{\mathrm{eff}} \left[\pir,\pia\right] &= \int \dd^4 x  \begin{pmatrix}
\pir(x) & \pia(x)
\end{pmatrix} \begin{pmatrix}
0 & (\partial_0 + \epsilon)^2 - \partial_i^2 + m^2 \\
(\partial_0 - \epsilon)^2 - \partial_i^2 + m^2  & i f(E_k)\epsilon
\end{pmatrix} \begin{pmatrix}
\pir(x) \\ \pia(x)
\end{pmatrix}
\end{align}
where the $\epsilon$ are here for convergence purpose and $f(E_k)$ is a function of $E_k\equiv \sqrt{k^2+m^2}$ encoding the occupation of the state \cite{Kamenev:2009jj}. It generates the equations of motion for the propagators  
\begin{align}\label{eq:Minkret}
\left[(\partial_0 \pm \epsilon)^2 + k^2 + m^2 \right] G^{R/A} (k; t_1, t_2) =  \delta(t_1-t_2)			 
\end{align}
and
\begin{align}
G^K (k;t_1,t_2) = - i f(E_k) \epsilon \int \dd t' G^{R} (k;t,t') G^{A} (k;t',t).
\end{align}
One can benefit from the flat space frequency decomposition to solve \Eq{eq:Minkret} in temporal Fourier space
\begin{align}
G^{R/A} (k; \omega) &= - \frac{1}{(\omega \mp i \epsilon)^2 - E_k^2}= -\frac{1}{2k} \left[ \frac{1}{\omega- (E_k\pm i\epsilon)} - \frac{1}{\omega- (-E_k\pm i\epsilon)}   \right]
\end{align}
The real-time Green functions are given by
\begin{align}\label{eq:retardedM4}
G^{R/A} (k; \tau) &= \int \frac{\dd \omega}{2\pi} \ee^{i \omega \tau} G^{R/A} (k; \omega) =\pm \frac{\sin \left[E_k \tau\right]}{E_k}\theta\left(\pm \tau\right)
\end{align}
where $\tau \equiv t_1-t_2 $. We can then compute the Keldysh function either in Fourier or real space. For instance, in real space, it writes
\begin{align}
G^K (k;t_1,t_2) &= - i \frac{f(E_k)}{E_k^2} \epsilon \int \dd t'  \left\{\sin[E_k(t_1-t')] \ee^{\epsilon (t_1-t')} \theta(t'-t_1) \right\}\\
&\qquad \qquad \qquad \qquad \times \left\{\sin[E_k(t'-t_2)] \ee^{-\epsilon (t'-t_2)} \theta(t'-t_2) \right\} \nonumber 
\end{align}
which leads to 
\begin{align}\label{eq:distribM4}
G^K (k;\tau)  = i \frac{f(E_k)}{4E_k^2}\cos\left[E_k \tau\right].
\end{align}
The $f(E_k)$ factor characterises the occupation of the system's state. The vacuum result is recovered for $f(E_k) = 2 E_k$ which we assume to be the case below. Note that this textbook result relies on a subtle balance between fluctuation and dissipation controlled by the $\epsilon$ prescription. Indeed, $\epsilon$ appearing in $\widehat{D}_R$ ensures the convergence of the integrals. Taking $\epsilon \rightarrow 0$ makes the field interacts for longer and longer in the asymptotic past. Then, the amplitude of the fluctuations controlled by $\widehat{D}_K$ must be rescaled by $\epsilon$ accordingly such that the contributions equilibrate despite longer interactions and a finite result is reached.

\subsubsection*{Interactions}

Let us now recover some known results from the Feynman rules presented in \Fig{fig:rules}. For this purpose, we work with a vacuum state such that $f(E_k) = 2 E_k$ below.

\paragraph{Polynomial interactions} Let us first consider the unitary interaction
\begin{align}\label{eq:cubic}
\mathcal{L}_{\mathrm{int}} = \frac{\lambda}{3!}\left(\pi^3_+ - \pi^3_-\right) = \frac{\lambda}{2} \left(\pir^2 \pia + \frac{1}{12} \pia^3\right).
\end{align}
We have two diagrams to consider similar to the one presented in \Fig{fig:Btree}. The first one leads to
\begin{align}\label{eq:D1}
D^{\pi^3}_1 &= -\frac{\lambda}{2} \int_{-\infty(1\pm i \epsilon)}^{t_0} \dd t G^K(k_1; t_0,t) G^K(k_2; t_0,t)G^R(k_3; t_0,t) + \mathrm{perms}. 
\end{align}
Upon injecting \Eqs{eq:distribM4} and \eqref{eq:retardedM4} in \Eq{eq:D1}, we obtain
\begin{align}\label{eq:D1result}
D^{\pi^3}_1 &=  \frac{\lambda}{32 E_1 E_2 E_3} \left( \frac{1}{E_T} + \frac{1}{-E_1 + E_2 + E_3} + \frac{1}{E_1 - E_2 + E_3} - \frac{1}{E_1 + E_2 - E_3}\right)+ \mathrm{perms}. 
\end{align}
where we defined $E_T \equiv E_1 + E_2 + E_3$. We observe a collection of naked folded singularities. In \cite{Green:2020whw}, these singularities have been argued to correspond to classical exchanges of momenta through scattering of physical particles (in the case of \Eq{eq:D1result}, $1+2\rightarrow3$). The second diagram is made of the $\pia$ components only and reads
\begin{align}\label{eq:D2}
D^{\pi^3}_2 = - \frac{\lambda}{24} \int_{-\infty(1\pm i \epsilon)}^{t_0} \dd t G^R(k_1; t_0,t) G^R(k_2; t_0,t)G^R(k_3; t_0,t) + \mathrm{perms}. 
\end{align}
Upon injecting \Eqs{eq:distribM4} and \eqref{eq:retardedM4} in \Eq{eq:D2}, we obtain
\begin{align}\label{eq:D2result}
D^{\pi^3}_2 &=   \frac{\lambda}{96 E_1 E_2 E_3} \left( \frac{1}{E_T} + \frac{1}{-E_1 + E_2 + E_3} + \frac{1}{E_1 - E_2 + E_3} + \frac{1}{E_1 + E_2 - E_3}\right)+ \mathrm{perms} .
\end{align}
Upon performing all the six possible permutations and summing \Eqs{eq:D1result} and \eqref{eq:D2result}, we recover the standard result
\begin{align}\label{eq:std}
D^{\pi^3}_1 + D^{\pi^3}_2 &= \frac{\lambda}{4 E_1 E_2 E_3 E_T}.
\end{align}		

\paragraph{Derivative interactions} In order to clarify the use of differential operators, we consider the unitary interactions $\dot{\pi}^2 \pi$ and $(\partial_i \pi)^2 \pi$ which can also be described in terms of the diagrams given in \Fig{fig:Btree}. Yet, one has to be careful about the use of the differential operators. Indeed, $\dot{\pi}^2 \pi$ decomposes into
\begin{align}\label{eq:timeder}
-\frac{\lambda}{2}\left(\dot{\pi}^2_+ \pi_+ - \dot{\pi}^2_- \pi_-\right) = -\frac{\lambda}{2}\left( \dot{\pi}_r^2 \pia + 2 \dot{\pi}_r \dot{\pi}_a \pir + \frac{1}{4} \dot{\pi}_a^2 \pia\right)
\end{align}
where the minus sign in front comes from the Lorentzian signature. It leads to three contributions (two for the diagram $D_1$ and one for the diagram $D_2$). The first one is
\begin{align}\label{eq:D1a}
D^{\dot{\pi}^2 \pi}_{1a} &= \frac{\lambda}{2} \int_{-\infty(1\pm i \epsilon)}^{t_0} \dd t \partial_t G^K(k_1; t_0,t) \partial_t G^K(k_2; t_0,t)G^R(k_3; t_0,t) + \mathrm{perms}. 
\end{align}
Upon injecting \Eqs{eq:distribM4} and \eqref{eq:retardedM4} in \Eq{eq:D1a} 
\begin{align}\label{eq:D1aresult}
D^{\dot{\pi}^2 \pi}_{1a} &=   \frac{\lambda}{32E_3} \left( \frac{1}{E_T} - \frac{1}{-E_1 + E_2 + E_3} - \frac{1}{E_1 - E_2 + E_3} - \frac{1}{E_1 + E_2 - E_3}\right)+ \mathrm{perms}. 
\end{align}
The second contribution is
\begin{align}\label{eq:D1b}
D^{\dot{\pi}^2 \pi}_{1b} &= \lambda \int_{-\infty(1\pm i \epsilon)}^{t_0} \dd t \partial_t G^K(k_1; t_0,t)  G^K(k_2; t_0,t)\partial_t G^R(k_3; t_0,t) + \mathrm{perms}. 
\end{align}
Upon injecting \Eqs{eq:distribM4} and \eqref{eq:retardedM4} in \Eq{eq:D1a}, we obtain
\begin{align}\label{eq:D1bresult}
D^{\dot{\pi}^2 \pi}_{1b} &=  \frac{\lambda}{16E_2} \left( \frac{1}{E_T} - \frac{1}{-E_1 + E_2 + E_3} + \frac{1}{E_1 - E_2 + E_3} + \frac{1}{E_1 + E_2 - E_3}\right)+ \mathrm{perms}.
\end{align}
Lastly, the final contribution is given by \begin{align}\label{eq:D2a}
D^{\dot{\pi}^2 \pi}_{2} &= \frac{\lambda}{8} \int_{-\infty(1\pm i \epsilon)}^{t_0} \dd t \partial_t G^R(k_1; t_0,t) \partial_t G^R(k_2; t_0,t)G^R(k_3; t_0,t) + \mathrm{perms}. 
\end{align}
Upon injecting \Eqs{eq:distribM4} and \eqref{eq:retardedM4} in \Eq{eq:D2a}, we obtain
\begin{align}\label{eq:D2aresult}
D^{\dot{\pi}^2 \pi}_{2} &=   \frac{\lambda}{32E_3} \left( \frac{1}{E_T} + \frac{1}{-E_1 + E_2 + E_3} + \frac{1}{E_1 - E_2 + E_3} - \frac{1}{E_1 + E_2 - E_3}\right)+ \mathrm{perms}.
\end{align}
Upon performing all the six possible permutations and summing the various contributions, we recover the expected result
\begin{align}
D^{\dot{\pi}^2 \pi}_{1a} +D^{\dot{\pi}^2 \pi}_{1b} + D^{\dot{\pi}^2 \pi}_{2} = \frac{2\lambda \left(E_1 E_2 + E_2 E_3 + E_1 E_3\right)}{8 E_1 E_2 E_3 E_T}.
\end{align}

The spatial derivative case $(\partial_i \pi)^2 \pi$  follows accordingly upon decomposing into
\begin{align}\label{eq:spatialder}
\frac{\lambda}{2}\left[(\partial_i \pi_+)^2 \pi_+ - (\partial_i \pi_-)^2 \pi_-\right] = \frac{\lambda}{2}\left[ (\partial_i \pir)^2  \pia + 2 \partial_i\pir \partial^i\pia \pir + \frac{1}{4}  (\partial_i \pia)^2 \pia\right]
\end{align}
Following the same procedure, we recover the standard result
\begin{align}
D^{(\partial_i \pi)^2 \pi}_{1a} +D^{(\partial_i \pi)^2 \pi}_{1b} + D^{(\partial_i \pi)^2 \pi}_{2} = \frac{2\lambda \left(\bmk_1.\bmk_2 + \bmk_2.\bmk_3 + \bmk_1.\bmk_3\right)}{8 E_1 E_2 E_3 E_T}.
\end{align}

\begin{figure}[tbp]
\centering
\includegraphics[width=0.9\textwidth]{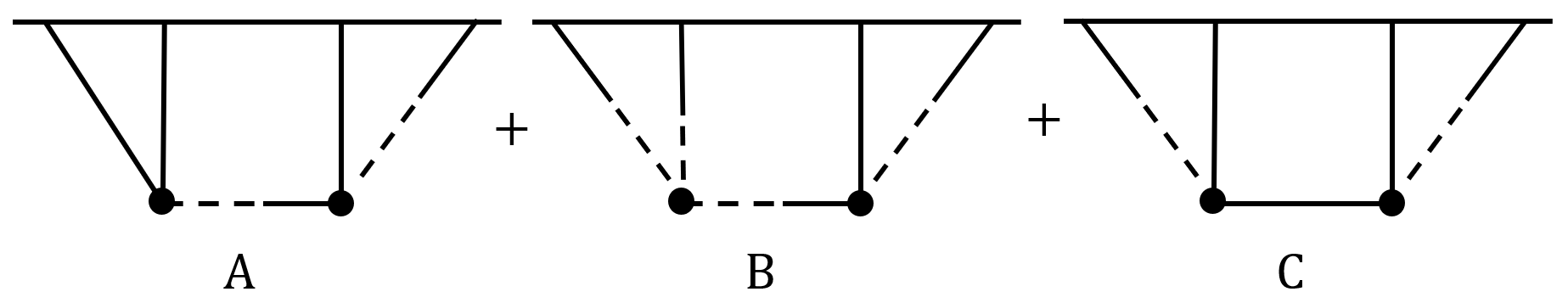}
\caption{The three diagrams to compute given in \Eqs{eq:A}, \eqref{eq:B} and \eqref{eq:C}.}
\label{fig:exchange}
\end{figure}

\paragraph{Exchange diagram} At last, we aim at recovering the tree level exchange in the presence of the cubic interaction specified in \Eq{eq:cubic}. For the two vertices corresponding to the operators $\pir^2 \pia$ and $\pi^3_a$, there are three diagrams to compute that are represented in \Fig{fig:exchange}. For simplicity, we focus on the $s$-channel $(12)\rightarrow(34)$ for which we define $k_s \equiv |\bmk_1+\bmk_2|$ and $E_s \equiv \sqrt{|\bmk_1+\bmk_2|^2+m^2}$. For convenience, we also define $E_{ij\cdots n} \equiv E_i+E_j+\cdots E_n$, $E_L \equiv E_s + E_{12}$ and $E_R \equiv E_s + E_{34}$.

The first diagram in \Fig{fig:exchange} is given by 
\begin{align}\label{eq:A}
A =& 2\left(i \frac{\lambda}{2}\right)^2(-i)\int_{-\infty(1\pm i \epsilon)}^{t_0} \dd t_1 \int_{-\infty(1\pm i \epsilon)}^{t_0} \dd t_2  \\
&\left[ G^K(k_1; t_0,t_1)   G^K(k_2; t_0,t_1)G^R(k_S; t_2,t_1)   G^K(k_3; t_0,t_2)  G^R(k_4; t_0,t_2)+ \mathrm{perms}.\right] \nonumber
\end{align}
which is manifestly real (the Keldysh propagator being pure imaginary) and where the permutations are over each side of the exchange $E_1 \leftrightarrow E_2$ and $E_3 \leftrightarrow E_4$, together with the two partial poles of the exchange $(E_1,E_2) \leftrightarrow (E_3,E_4)$. There is a symmetry factor $2$ in front coming from the freedom of choosing one out of two legs of the left vertex going into the exchange on the A-diagram. It leads to $A \equiv A_1 +A_2$ with 
\begin{align}
    A_1 =&  \frac{\lambda^2}{16 E_1 E_2 E_3 E_4} \frac{1}{E_T E_L E_R} \left[1- \frac{E_T(E_T-E_s)}{(E_{L} - 2E_s)(E_{R} - 2E_s)}  \right] 
\end{align}
and
\begin{align}
A_2 =&-  \frac{\lambda^2 }{16 E_1 E_2 E_3 E_4} \Bigg\{ \frac{E_{34}}{(E_{134}-E_2)(E_{234}-E_1)\left[(E_1-E_2)^2 - E_s^2\right]} \nonumber \\
&\qquad \qquad + \frac{E_{12}}{(E_{123}-E_4)(E_{124}-E_3)\left[(E_3-E_4)^2 - E_s^2\right]}\Bigg\}
\end{align}
The second diagram in \Fig{fig:exchange} is given by 
\begin{align}\label{eq:B}
B =& 6\left(i \frac{\lambda}{2}\right) \left(i \frac{\lambda}{24}\right) (-i) \int_{-\infty(1\pm i \epsilon)}^{t_0} \dd t_1 \int_{-\infty(1\pm i \epsilon)}^{t_0} \dd t_2  \\
&\left[G^R(k_1; t_0,t_1)  G^R(k_2; t_0,t_1)G^R(k_S; t_2,t_1)  G^K(k_3; t_0,t_2)  G^R(k_4; t_0,t_2)+ \mathrm{perms}.\right] \nonumber
\end{align}
where the symmetry factor $6$ in front comes from the freedom of choosing one out of two legs of the left vertex and one out of three legs of the right vertex going into the exchange on the B-diagram. It leads, after performing the exchanges, to $B = A_1 - A_2$ so that the $A_2$ contributions cancel out. Lastly, the third diagram in \Fig{fig:exchange} is given by 
\begin{align}\label{eq:C}
C =& 2\left(i \frac{\lambda}{2}\right)^2 (-i)\int_{-\infty(1\pm i \epsilon)}^{t_0} \dd t_1 \int_{-\infty(1\pm i \epsilon)}^{t_0} \dd t_2  \\
&\left[G^R(k_1; t_0,t_1)  G^K(k_2; t_0,t_1)  G^K(k_S; t_2,t_1)   G^K(k_3; t_0,t_2)  G^R(k_4; t_0,t_2)+ \mathrm{perms}.\right] \nonumber
\end{align}
where the symmetry factor $2$ in front comes from the freedom of choosing one out of two legs of the left vertex, which leads, after performing the exchanges, to 
\begin{align}
C=& \frac{ 2 \lambda^2}{16 E_1 E_2 E_3 E_4} \frac{1}{E_L E_R E_s} \frac{(E_L-E_s)(E_R-E_s)}{(E_L-2E_s)(E_R-2E_s)}.
\end{align}
Summing over the three contributions, we recover the expected result, that is 
\begin{align}
A+B+C = \frac{2 \lambda^2}{16 E_1 E_2 E_3 E_4} \frac{(E_T+E_s)}{E_T E_L E_R E_s}.
\end{align}

\section{Langevin equation with derivative noises}\label{app:Langevin}

In this appendix, we discuss the inclusion of the derivative noises $\pia^{\prime2}$ and $(\partial_{i}\pia)^{2}$ into the Langevin equation through a modification of the Hubbard–Stratonovich trick \eqref{eq:HStrick}. We present two approaches that yield the same power spectrum. 

\subsection*{Single noise} 

In order to account for derivative noises, one can introduce a differential operator $\mathfrak{D}$ in the coupling $\pia\Os$ such that
\begin{equation}
\exp\left(i\int d^{4}x a^{4}\mathcal{L}_{2}^{(2)}\right)=\mathcal{N}\int\mathcal{D}[\Os]\;\exp\left[\int d^{4}x a^{4}\left(-\frac{\Os^{2}}{2\beta_{1}}+i\Os\overset{\rightarrow}{\mathfrak{D}}\pia\right)\right],
\end{equation}
with
\begin{equation}\label{eq:frakDprop}
(\overset{\rightarrow}{\mathfrak{D}}\pia)^{2}=\pia^{2}+\frac{\beta_{4}-\beta_{2}}{\beta_{1}}\frac{\pia^{\prime2}}{a^{2}}+\frac{\beta_{2}}{\beta_{1}}\frac{(\partial_{i}\pia)^{2}}{a^{2}}.
\end{equation}
This way, we can include the derivative noises into the Langevin equation via this differential operator acting on the Gaussian noise $\Os$, leading to
\begin{equation}
\widehat{D}_{R} [\pi(\eta)] =\frac{\left[a^{4}(\eta)\Os(\eta)\right]\overset{\leftarrow}{\mathfrak{D}}}{a^{2}(\eta)}.
\end{equation}
With a non-vanishing right-hand side, we obtain the solution to the differential equation in terms of the Green function for $\widehat{D}_{R}$, leading to
\begin{equation}
\pi_{\bfk}(\eta)=\int_{-\infty}^{0}d\eta'\; G^{R}(k;\eta,\eta')\left[a^{4}(\eta')\delta\mathcal{O}_{S}(\bfk, \eta')\right]\overset{\leftarrow}{\mathfrak{D}},
\end{equation}
where we can then integrate by parts to leave the noise without derivatives\footnote{The boundary terms vanish thanks to the Bunch-Davies vacuum (to the infinite past) and the retarded boundary condition in the coincident time limit for $\eta'=\eta$.}
\begin{equation}
\pi_{\bfk}(\eta)=-\int_{-\infty}^{0}d\eta'\; a^{4}(\eta')\delta\mathcal{O}_{S}(\bfk,\eta')\overset{\rightarrow}{\mathfrak{D}}\left[G^{R}(k;\eta,\eta')\right].
\end{equation}
We can then compute the power spectrum in the usual way, such that 
\begin{equation}
\langle\pi_{\bfk}(\eta)\pi_{\bfk'}(\eta)\rangle=2\beta_{1}(2\pi)^3{\delta}(\bfk+\bfk')\int_{-\infty}^{0} d\eta'\;a^{4}(\eta')\overset{\rightarrow}{\mathfrak{D}}\left[G^{R}(k;\eta,\eta')\right]\overset{\rightarrow}{\mathfrak{D}}\left[G^{R}(k;\eta,\eta')\right].
\end{equation}
Using the property \eqref{eq:frakDprop}, we simplify the integrand to
\begin{align}
2\beta_{1}\overset{\rightarrow}{\mathfrak{D}}\left[G^{R}(k;\eta,\eta')\right]&\overset{\rightarrow}{\mathfrak{D}}\left[G^{R}(k;\eta,\eta')\right]= 2\beta_{1}\left[G^{R}(k;\eta,\eta')\right]^{2} \Bigg.\\
+&2(\beta_{4}-\beta_{2})\frac{\left[\partial_{\eta'}G^{R}(k;\eta,\eta')\right]^{2}}{a^{2}(\eta)}+2\beta_{2}\frac{\left[\partial_{i}G^{R}(k;\eta,\eta')\right]^{2}}{a^{2}(\eta)}.\nonumber
\end{align}
Under this form, it is explicit that it matches the result for the power spectrum \eqref{eq:Keldyshreff} derived in \Sec{sec:Pk}.

\subsection*{Three noises} 

Another way to include derivative noises into the Langevin equation consists in introducing new Gaussian fields $\Os^{(i)}$ that couple derivatively to $\pia$ in the Hubbard–Stratonovich trick \eqref{eq:HStrick}. For the time derivative noise, the Hubbard–Stratonovich tricks takes the form
\begin{align}
\exp\Bigg[-&\int d^{4}x a^{2}(\beta_{4}-\beta_{2})\pia^{\prime2}\Bigg]=\mathcal{N}^{(2)}\int\mathcal{D}[\Os^{(2)}]\\
&\times\;\exp\left\{\int d^{4}x a^{4}\left[-\frac{[\Os^{(2)}]^{2}}{4\beta_{1}}+i\sqrt{\frac{\beta_{4}-\beta_{2}}{\beta_{1}}}\frac{\pia'}{a}\Os^{(2)}\right]\right\}.\nonumber
\end{align}
For the spatial derivative noise, we need to consider the square root of the Laplacian
\begin{align}
\exp\Bigg[-&\int d^{4}x a^{2}\beta_{2}(\partial_{i}\pia)^{2}\Bigg]=\mathcal{N}^{(3)}\int\mathcal{D}[\Os^{(3)}]\\
&\times\;\exp\left\{\int d^{4}x a^{4}\left[-\frac{[\Os^{(3)}]^{2}}{4\beta_{1}}+i\sqrt{\frac{\beta_{2}}{\beta_{1}}}\frac{1}{a}\Os^{(3)}\sqrt{\nabla^{2}}\pia\right]\right\}.\nonumber
\end{align}
We can then modify the Langevin equation which becomes
\begin{equation}
\widehat{D}_{R}[\pi_{\bfk}(\eta)]=a^{2}(\eta)\Os^{(1)}-\frac{1}{a^{2}(\eta)}\partial_{\eta}\left[\sqrt{\frac{\beta_{4}-\beta_{2}}{\beta_{1}}}a^{3}(\eta)\Os^{(2)}\right]-ka(\eta)\sqrt{\frac{\beta_{2}}{\beta_{1}}}\Os^{(3)}.
\end{equation}
The statistics of the noises now have a matrix structure with indices labelling the noise under consideration
\begin{equation}
\langle\delta\mathcal{O}^{(a)}_{S}(\bfk, \eta_{1})\delta\mathcal{O}^{(b)}_{S}(\bfk', \eta_{2})\rangle=2\beta_{1}H^{4}\eta_{1}^{4}(2\pi)^{3}\delta(\bfk+\bfk')\delta(\eta_{1}-\eta_{2})\delta^{ab}.
\end{equation}
Using this technique, we fully recover from the Langevin equation the expression of the path integral Keldysh propagator \eqref{eq:Keldyshreff} from \Sec{sec:Pk}.

\bibliographystyle{JHEP}
\bibliography{Biblio}

\end{document}